\documentclass[twocolumn]{aastex62}
\usepackage{natbib}

\begin{document}

\title{The Type II-Plateau Supernova 2017\lowercase{eaw} in NGC 6946 and Its Red Supergiant Progenitor}

\author{Schuyler D.~Van Dyk}
\affil{Caltech/Spitzer Science Center, Caltech/IPAC, Mailcode 100-22, 
Pasadena, CA 91125, USA 0000-0001-9038-9950}

\author{WeiKang Zheng}
\affil{Department of Astronomy, University of California,
Berkeley, CA 94720-3411, USA}

\author{Justyn R.~Maund}
\affil{Department of Physics and Astronomy, University of 
Sheffield, Hicks Building, Hounsfield Road, Sheffield 
S3 7RH, UK 0000-0003-0733-7215}

\author{Thomas G.~Brink}
\affil{Department of Astronomy, University of California,
Berkeley, CA 94720-3411, USA}

\author{Sundar Srinivasan}
\affil{Institute of Astronomy \& Astrophysics, Academia 
Sinica, 11F, Astronomy-Mathematics Building, No. 1, 
Roosevelt Rd, Sec 4, Taipei 10617, Taiwan, Republic of China}
\affil{Instituto de Radioastronom\'ia y Astrof\'isica,
Universidad Nacional Auton\'oma de M\'exico, Antigua Carretera a P\'atzcuaro \# 8701 Ex-Hda. San Jos\'e de la Huerta, Morelia, Michoac\'an. M\'exico. C.P. 58089}

\author{Jennifer E.~Andrews}
\affil{Steward Observatory, University of Arizona, 933 N.~Cherry Avenue, Tucson, AZ 85721, USA}

\author{Nathan Smith}
\affil{Steward Observatory, University of Arizona, 933 N.~Cherry Avenue,
Tucson, AZ 85721, USA}

\author{Douglas C.~Leonard}
\affil{Department of Astronomy, San Diego State University, San Diego, CA 92182-1221, USA}

\author{Viktoriya Morozova}
\affil{Department of Astrophysical Sciences, Princeton University, Princeton, NJ 08544, USA}

\author{Alexei V.~Filippenko}
\affil{Department of Astronomy, University of California, Berkeley, CA 94720-3411, USA 0000-0003-3460-0103}
\affil{Miller Senior Fellow, Miller Institute for Basic Research in Science, University of California, Berkeley, CA 94720, USA}

\author{Brody Conner}
\affil{Department of Physics and Astronomy, Purdue University, 525 Northwestern Avenue, West Lafayette, IN 
47907, USA}

\author{Dan Milisavljevic}
\affil{Department of Physics and Astronomy, Purdue University, 525 Northwestern Avenue, West Lafayette, IN 
47907, USA 0000-0002-0763-3885}

\author{Thomas de Jaeger}
\affil{Department of Astronomy, University of California, Berkeley, CA 94720-3411, USA}

\author{Knox S.~Long}
\affil{Space Telescope Science Insitute, 3700 San Martin Drive, Baltimore, MD 21218, USA}
\affil{Eureka Scientific, Inc. 2452 Delmer Street, Suite 100, Oakland, CA 94602-3017, USA}

\author{Howard Isaacson}
\affil{Department of Astronomy, University of California, Berkeley, CA 94720-3411, USA 0000-0002-0531-1073}

\author{Ian J.~M.~Crossfield}
\affil{Department of Physics, Massachusetts Institute of Technology, 77 Massachusetts Avenue, Cambridge, MA, USA}

\author{Molly R.~Kosiarek}
\affil{University of California Santa Cruz, Santa Cruz, CA, 95064, USA 0000-0002-6115-4359}
\affil{NSF Graduate Research Fellow}

\author{Andrew W.~Howard}
\affil{California Institute of Technology, Pasadena, CA 91125, USA 0000-0001-8638-0320}

\author{Ori D.~Fox}
\affil{Space Telescope Science Institute, 3700 San Martin Drive,
Baltimore, MD 21218, USA 0000-0002-4924-444X}

\author{Patrick L.~Kelly}
\affil{College of Science \& Engineering, Minnesota Institute for Astrophysics, University of Minnesota, 115 Union St.~SE, Minneapolis, MN 55455 USA 0000-0003-3142-997X}

\author{Anthony L.~Piro}
\affil{The Observatories of the Carnegie Institution for Science, 813 Santa Barbara St., Pasadena, CA 91101, USA 0000-0001-6806-0673}

\author{Stuart P. Littlefair}
\affil{Department of Physics and Astronomy, University of 
Sheffield, Hicks Building, Hounsfield Road, Sheffield S3 7RH, UK}

\author{Vik S. Dhillon}
\affil{Department of Physics and Astronomy, University of 
Sheffield, Hicks Building, Hounsfield Road, Sheffield S3 7RH, UK}
\affil{Instituto de Astrof\'isica de Canarias, E-38205 La Laguna, Tenerife, Spain}

\author{Richard Wilson}
\affil{Center for Advanced Instrumentation, Department of Physics, University of Durham, South Road, Durham DH1 3LE, UK}

\author{Timothy Butterley}
\affil{Center for Advanced Instrumentation, Department of Physics, University of Durham, South Road, Durham DH1 3LE, UK}

\author{Sameen Yunus}
\affil{Department of Astronomy, University of California, Berkeley, CA 94720-3411, USA}

\author{Sanyum Channa}
\affil{Department of Astronomy, University of California, Berkeley, CA 94720-3411, USA}

\author{Benjamin T.~Jeffers}
\affil{Department of Astronomy, University of California, Berkeley, CA 94720-3411, USA}

\author{Edward Falcon}
\affil{Department of Astronomy, University of California, Berkeley, CA 94720-3411, USA}

\author{Timothy W.~Ross}
\affil{Department of Astronomy, University of California, Berkeley, CA 94720-3411, USA}

\author{Julia C.~Hestenes}
\affil{Department of Astronomy, University of California, Berkeley, CA 94720-3411, USA}

\author{Samantha M.~Stegman}
\affil{Department of Astronomy, University of California, Berkeley, CA 94720-3411, USA}

\author{Keto Zhang}
\affil{Department of Astronomy, University of California, Berkeley, CA 94720-3411, USA}

\author{Sahana Kumar}
\affil{Department of Astronomy, University of California, Berkeley, CA 94720-3411, USA}
\affil{Department of Physics, Florida State University, 77 Chieftain Way, Tallahassee, Florida 32306, USA}

\begin{abstract}
We present extensive optical photometric and spectroscopic observations, 
from 4 to 482 days after explosion, of the Type II-plateau (II-P) supernova (SN) 2017eaw in NGC 6946. 
SN 2017eaw is a normal SN II-P intermediate in properties between, for example, SN 1999em and SN 2012aw and
the more luminous SN 2004et, also in NGC 6946. 
We have determined that the extinction to SN 2017eaw is primarily due to the Galactic foreground and 
that the SN site metallicity is likely subsolar.
We have also independently confirmed a tip-of-the-red-giant-branch (TRGB) distance to NGC 6946
of $7.73{\pm}0.78$ Mpc. 
The distances to the SN that we have also estimated via both the standardized candle method and
expanding photosphere method corroborate the TRGB distance.
We confirm the SN progenitor identity in pre-explosion archival {\sl Hubble Space Telescope} 
({\sl HST}) and {\sl Spitzer Space Telescope\/} images,
via imaging of the SN through our {\sl HST\/} Target of Opportunity program.
Detailed modeling of the progenitor's spectral energy distribution indicates that the star was
a dusty, luminous red supergiant consistent with an initial mass of $\sim 15\ M_{\odot}$.

\end{abstract}


\section{Introduction}

\bibpunct[;]{(}{)}{;}{a}{}{;}

Supernovae (SNe) have a profound influence on the host galaxies in which they occur: through 
chemical enrichment, galactic feedback, and the formation of compact neutron star and black hole remnants.
A large fraction of SNe, $\sim 76$\% in the local Universe \citep{Li+2011}, arise from the core collapse of 
massive stars with initial masses $M_{\rm ini} \gtrsim 8$--$10\ M_{\odot}$.
The most common of these core-collapse SNe are the Type II-Plateau 
(SNe II-P, $\sim 48$\% locally; \citealt{Smith+2011}).

Solid evidence has emerged through the direct identification of the progenitor stars of a number of recent,
nearby SNe II-P that these explosions represent the termination of stars in the red supergiant (RSG) 
evolutionary phase 
(e.g., 
\citealt{VanDyk+2003,Smartt+2004,Maund+2009,VanDyk+2012a,VanDyk+2012b,Fraser+2012,Fraser+2014,Maund+2014a,Maund+2014b}).
From a still incomplete sample of about 27 SNe II-P, it has been inferred that the initial mass range 
for RSGs leading to SNe II-P is $\sim 9.5$--$16.5\ M_{\odot}$ \citep{Smartt+2009,Smartt2015}\footnote{Up till 
now, 17 SNe II-P have had their progenitors identified directly through high-resolution pre-SN imaging, with 
the remainder consisting of upper limits on detection (\citealt{VanDyk2017} and also including SN 2018aoq; \citealt{ONeill+2019}).}.
More indirect indicators, such as the ages of the immediate stellar environment 
(e.g., \citealt{Williams+2014,Williams+2018,Maund2017}) and mass measurement 
of the nucleosynthetic products via modeling of SNe II-P nebular spectra \citep{Jerkstrand+2012,Jerkstrand+2014}, tend 
generally to agree with this approximate mass range.
Corroborating evidence for the lack of SN II-P progenitors above an initial mass of $\sim 17\ M_{\odot}$, 
known as ``the RSG problem,'' stems from theoretical modeling of massive star explosions for which models at 
this approximate mass are unable to explode, but would instead collapse to form a black hole directly 
(e.g., \citealt{Sukhbold+2016}). Related to 
this is the possible discovery of an RSG with $M_{\rm ini} \approx 22\ M_{\odot}$ which appears to have 
vanished, rather than terminated as an SN \citep{Adams+2017}.

Several alternative possibilities have been offered to 
explain away the RSG problem, including enhanced wind-driven mass loss in more massive RSGs, stripping much 
of the H envelope \citep{yoon:10,Georgy+2012}; possible self-obscuration by more luminous, more massive RSGs 
possessing dustier envelopes (e.g., \citealt{Walmswell+2012}); and, 
inadequate assumptions of the bolometric correction for RSGs nearing explosion \citep{Davies+2018}. 
In addition, \citet{Davies+2018} have argued that the uncertainty in the mass-luminosity relationship, 
which can be different for various theoretical stellar evolution models, may shift the limiting mass up by 
as much as several solar masses.
Furthermore, we note that some SN II-P/II-linear hybrid cases may have more massive progenitors (e.g., SN 2016X 
with $M_{\rm ini} \approx 19$--$20\ M_{\odot}$; \citealt{Huang+2018}).

The mass function of SN II-P progenitors requires additional development through additional cases of 
directly identified progenitors. Furthermore, the data for existing examples are sparse, with initial 
progenitor masses being precariously inferred from fitting of a spectral energy distribution (SED) based on 
one or two photometric data points. 
Inevitably in the near future, the Large Synoptic Survey Telescope (LSST) and a possible general-observer 
program for nearby galaxies using the {\sl Wide-Field Infrared Survey Telescope\/} ({\sl WFIRST}) would 
provide detailed, multiband pre-explosion images for the progenitors of ever-larger numbers of 
discovered SNe~II-P.
In the meantime, we can continue to build the sample slowly through the smatterings of nearby SNe II-P for 
which sufficient archival ground- and space-based data are available.

To that end, in this paper we discuss the Type II-P SN 2017eaw, which was discovered by 
\citet{Wiggins2017} at an unfiltered brightness of 12.8 mag on
2017 May 14.238 (UT dates are used throughout this paper). 
The discovery was confirmed by \citet{Dong+2017}. The object was found to be a young SN~II-P by 
\citet{Cheng+2017}, \citet{Xiang+2017}, and \citet{Tomasella+2017}.
A progenitor candidate was quickly identified by \citet{Khan2017} in pre-explosion data obtained by the 
{\sl Spitzer Space Telescope\/} and by \citet{VanDyk+2017} in archival {\sl Hubble Space Telescope\/}
({\sl HST}) data. Analyses of the progenitor have been published by \citet{Kilpatrick+2018} and \citet{Rui+2019}.
\citet{Tsvetkov+2018} presented detailed {\it UBVRI\/} light curves of the SN over the first 200 days.
\citet{Rho+2018} analyzed near-infrared spectroscopy of the SN to examine the possibility of dust formation.
The dust properties of the evolving SN also have been explored by \citet{Tinyanont+2019}.
SN 2017eaw is the tenth historical SN in the prodigious NGC 6946, also colloquially
known as the ``Fireworks Galaxy.'' The other events are SN 1917A, SN 1939C, SN 1948B, 
SN 1968D \citep{Hyman+1995}, SN 1969P, the Type II-L SN 1980K (e.g., \citealt{Milisavljevic+2012}), the 
Type II-P SN 2002hh (e.g., \citealt{Pozzo+2006}) and SN 2004et (e.g., \citealt{Sahu+2006,Maguire+2010}),
and the ``SN impostor'' SN 2008S \citep{Prieto+2008,Botticella+2009,Thompson+2009}.
A number of SN remnants are also known to exist in this
host galaxy \citep{MF1997,Bruursema+2014}.

The organization of this paper is as follows. In Section 2 we describe the various observations of SN 
2017eaw, both pre- and post-explosion. We estimate the extinction to the SN in Section 3, and in Section 4 we 
confirm recent estimates of the distance to the SN host galaxy. The metallicity at the SN site is inferred
in Section 5. In Section 6 we provide an analysis of the SN, including an estimate of the date of explosion, 
studies of the absolute light curves, color curves, and bolometric light curve, an estimate of the 
synthesized nickel mass,  and analysis of the spectra. In Section 7 
we present identification and characterization of the SN progenitor. We provide a discussion and summarize  
our conclusions in Section 8.

\section{Observations}

\subsection{SN Photometry}

Multiband {\it BVRI\/} images of SN~2017eaw were obtained with both
the Katzman Automatic Imaging Telescope (KAIT; \citealt{Filippenko+2001})
and the 1 m Nickel telescope at Lick Observatory. Unfiltered images were also obtained with KAIT.
All KAIT and Nickel images were reduced using a custom pipeline \citep{Ganeshalingam+2010}.

We obtained near-daily {\it BVRI\/} coverage with the pt5m, a 0.5 m robotic telescope 
at the Roque de los Muchachos Observatory, La Palma \citep{Hardy+2015}.
Exposure times for these observations were $5 \times 15$ s for both 
$B$ and $I$, and $5 \times 10$ s for both $V$ and $R$, prior to 2017 October 26. From that date onward,
exposure times were $5 \times 45$ s for both $B$ and $I$, and $5 \times 40$ s for both $V$ and $R$.
All pt5m observations were reduced using bias, dark, and flat-field frames acquired on the same night (or as 
close in time as possible) as the science observations.  Image alignment was conducted using 
\texttt{astrometry.net}, and coaddition of the images was performed using \texttt{swarp} \citep{Bertin+2002}.  

\begin{figure}
\plotone{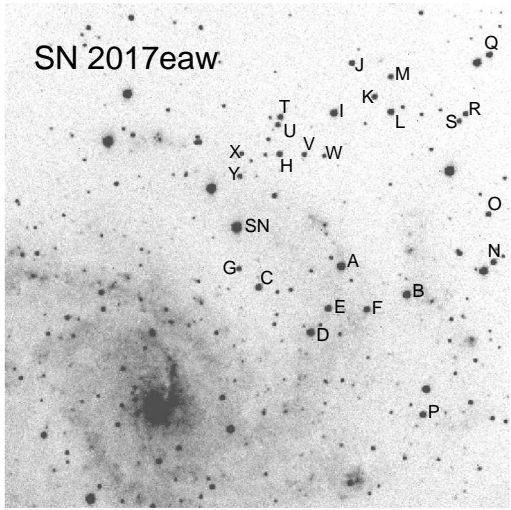}
\caption{KAIT $R$-band image from 2017 June 23 showing a $6{\farcm}7 \times 6{\farcm}7$ field, 
including SN 2017eaw and the local sequence of calibration stars (labeled)
listed in Table~\ref{tabseq}. North is up and east is to the left.\label{figseq}}
\end{figure}

Calibration of the SN 2017eaw photometry was achieved via transformation from Pan-STARRS1 
(PS1)\footnote{http://archive.stsci.edu/panstarrs/search.php} magnitudes for stars in
the field to {\it BVRI\/}, following relations provided by \citet{Tonry+2012}. We show the SN field with 
this local sequence of stars in Figure~\ref{figseq} and list their 
brightnesses in Table~\ref{tabseq}. As a check on the validity of this calibration method, we transformed 
the PS1 magnitudes to {\it BVRI\/} for the local sequences employed by \citet{Sahu+2006} and 
\citet{Misra+2007} for SN 2004et and by \citet{Botticella+2009} for SN 2008S, and we found differences of 
$-0.026 \pm 0.048$, $-0.016 \pm 0.024$, $-0.001 \pm 0.031$, and $-0.029 \pm 0.045$ mag in $B$, $V$, $R$, 
and $I$, respectively. A systematic offset exists, in the sense that the published magnitudes are slightly
brighter than the transformed PS1 photometry. However, the rms in the difference is small
in all bands.

Point-spread-function (PSF) photometry was extracted from the KAIT and Nickel images using \texttt{DAOPHOT} 
\citep{Stetson1987} from the IDL Astronomy User's Library\footnote{http://idlastro.gsfc.nasa.gov/} for the 
SN and the local sequence.
Apparent magnitudes were all measured in the KAIT4/Nickel2 natural system and were transformed to the 
standard system using the local sequence and color terms for KAIT4 from Table 4 of \citet{Ganeshalingam+2010} 
and updated Nickel color terms from \citet{Shivvers+2017}. 
The KAIT clear (unfiltered) photometry was also calibrated using this sequence in the $R$ band.

Aperture photometry of the SN and local sequence 
stars in the pt5m images was conducted using \texttt{SExtractor} \citep{Bertin+1996}.
Comparison with the local sequence was used to determine the zero-point (although no color correction was 
derived), with sigma-clipping ($\sigma = 2$ for 10 iterations) to remove outliers.

\begin{deluxetable}{ccccc}
\tablewidth{0pt}
\tablecolumns{5}
\tablecaption{Photometric Sequence around SN 2017eaw\tablenotemark{a}\label{tabseq}}
\tablehead{
\colhead{Star} & \colhead{$B$ (mag)} & \colhead{$V$ (mag)} &
\colhead{$R$ (mag)} & \colhead{$I$ (mag)}
}
\startdata
A & 14.788 & 14.138 & 13.756 & 13.332 \\ 
B & 15.176 & 14.404 & 13.956 & 13.482 \\
C & 16.006 & 15.140 & 14.640 & 14.135 \\ 
D & 16.032 & 14.836 & 14.154 & 13.507 \\
E & 16.288 & 15.346 & 14.805 & 14.332 \\
F & 15.878 & 15.379 & 15.079 & 14.686 \\
G & 17.444 & 16.490 & 15.942 & 15.367 \\ 
H & 16.830 & 15.923 & 15.400 & 14.843 \\
I & 15.534 & 14.670 & 14.172 & 13.689 \\
J & 17.073 & 15.935 & 15.285 & 14.637 \\
K & 16.923 & 16.015 & 15.492 & 14.955 \\ 
L & 16.345 & 15.401 & 14.859 & 14.315 \\ 
M & 17.081 & 16.191 & 15.677 & 15.152 \\ 
N & 17.632 & 16.078 & 15.195 & 14.330 \\ 
O & 17.044 & 16.080 & 15.526 & 14.982 \\ 
P & 16.824 & 15.637 & 14.960 & 14.330 \\ 
Q & 16.358 & 15.530 & 15.051 & 14.548 \\ 
R & 17.520 & 16.436 & 15.816 & 15.229 \\ 
S & 17.394 & 16.400 & 15.831 & 15.274 \\ 
T & 16.870 & 15.884 & 15.317 & 14.779 \\ 
U & 17.387 & 16.494 & 15.980 & 15.475 \\ 
V & 18.008 & 16.764 & 16.055 & 15.365 \\ 
W & 18.489 & 17.387 & 16.757 & 16.116 \\ 
X & 18.274 & 16.994 & 16.266 & 15.583 \\ 
Y & 17.291 & 16.384 & 15.861 & 15.324 \\ 
\enddata
\tablenotetext{a}{The uncertainties in these magnitudes are those from the PS1-to-Johnson-Cousins transformations from
\citet{Tonry+2012},
i.e., 0.034, 0.012, 0.015, and 0.017 mag in $B$, $V$, $R$, and $I$, respectively (the uncertainties in the observed PS1 
magnitudes for these stars are all $\ll 0.01$ mag).}
\end{deluxetable}

\begin{figure*}
\plotone{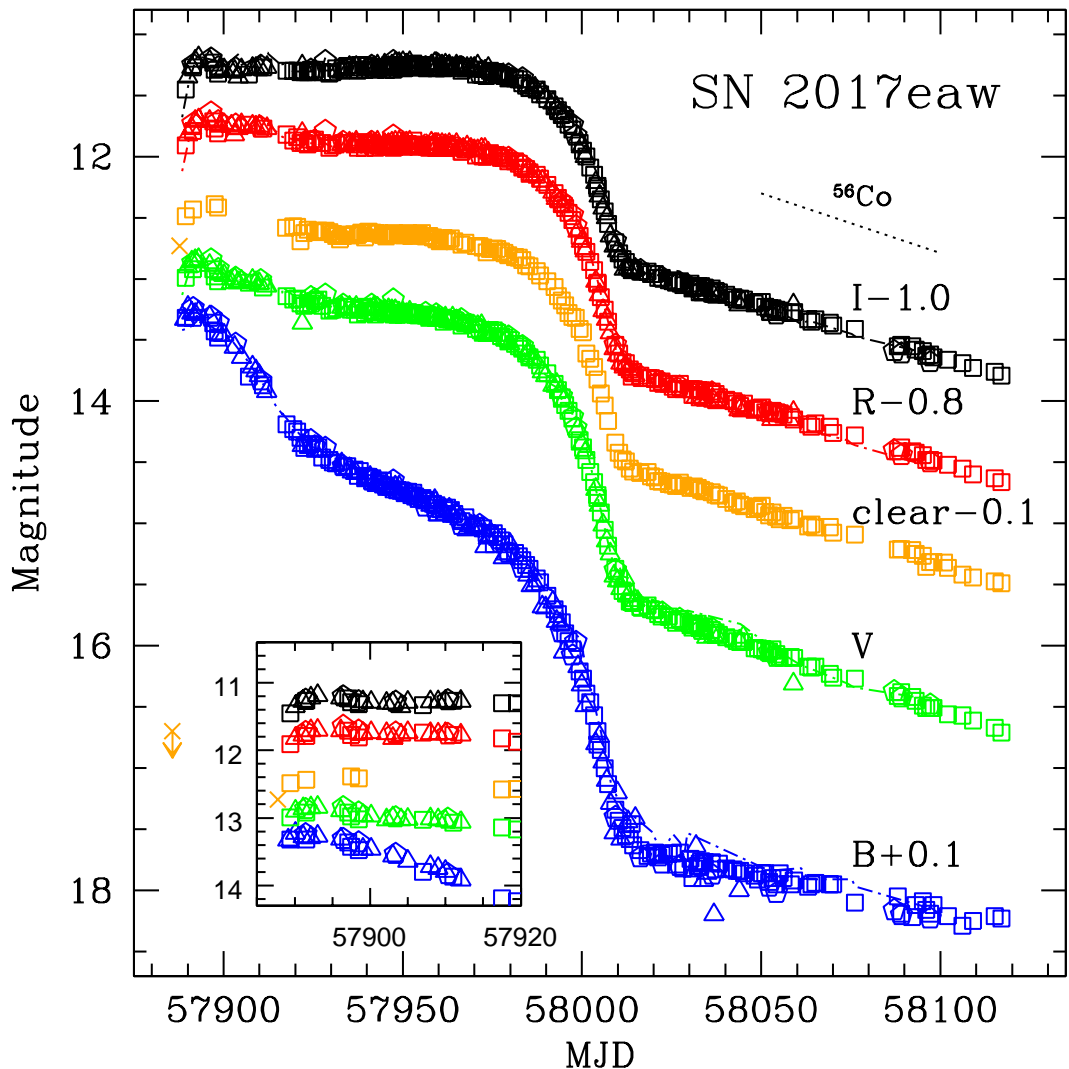}
\caption{The {\it BVRI\/} and unfiltered (clear; orange points) light curves of SN 2017eaw. 
KAIT data are shown as open squares, 
Nickel data as open pentagons, and pt5m data as open triangles. We also include our remeasurements of the 
discovery and upper limit to discovery from \citet{Wiggins2017} as orange crosses (see 
Section~\ref{explosion}). For comparison we include
the {\it BVRI\/} light curves from \citet[][ dotted-dashed curves]{Tsvetkov+2018}. We also show the expected 
decline rate if the exponential tail of the light curve is powered by the decay of $^{56}$Co. In the inset 
we focus on the first 35 days of the light curves; the offsets for the curves are of the same magnitude as 
in the main figure.\label{figlc}}
\end{figure*}

We combined all of the photometry from the three telescopes, covering from about 1 day after the discovery by
\citet{Wiggins2017} through nearly the end of calendar year 2017.
We show the light curves for SN 2017eaw in Figure~\ref{figlc}. The curves seem to exhibit a conspicuous 
initial ``bump'' near maximum brightness (see the figure inset), followed by an extended plateau phase, 
a steady rapid decline 
from the plateau, and an exponential tail (see, e.g., \citealt{Anderson+2014}). 
In the figure we also compare with the previously published
{\it BVRI\/} light curves from \citet{Tsvetkov+2018} at (approximately) matching epochs.
The agreement is quite good, with $\Delta B=0.04 \pm 0.09$, $\Delta V=0.00 \pm 0.05$, $\Delta R=0.00 \pm 0.04$, and
$\Delta I=0.06 \pm 0.04$ mag, in the sense of ``ours$-$\citeauthor{Tsvetkov+2018}''
A similar comparison with the photometry of \cite{Rui+2019} results in
$\Delta B=0.04 \pm 0.05$, $\Delta V=0.04 \pm 0.04$,  $\Delta R=0.03 \pm 0.02$, and $\Delta I=0.05 \pm 0.02$ mag,
again in the sense of ``ours$-$\citeauthor{Rui+2019},'' indicating that their photometry was slightly brighter, particularly
in the redder bands. \citeauthor{Rui+2019} also began monitoring the SN about 1 day before we did.

We have not performed $S$-corrections \citep{Stritzinger+2002} to our photometry. Nevertheless, we have made a 
comparison of all of the photometry obtained between the three telescopes and instruments: $\Delta B=-0.06 \pm 0.09$, 
$\Delta V=-0.02 \pm 0.05$, $\Delta R=-0.01 \pm 0.05$, and $\Delta I=-0.01 \pm 0.04$ mag, ``KAIT$-$pt5m''; 
$\Delta B=0.01 \pm 0.12$, $\Delta V=-0.03 \pm 0.04$, $\Delta R=-0.03 \pm 0.05$, and $\Delta I=-0.02 \pm 0.03$ mag, 
``Nickel$-$pt5m''; and, $\Delta B=-0.04 \pm 0.06$, $\Delta V=0.02 \pm 0.03$, $\Delta R=0.00 \pm 0.03$, and 
$\Delta I=0.00 \pm 0.03$ mag, ``KAIT$-$Nickel.'' As one can see, the differences are quite small.
We also have generated synthetic photometry from the Lick Observatory Kast spectra that we obtained (see 
Section~\ref{specobs}) using \texttt{pysynphot}\footnote{https://github.com/spacetelescope/pysynphot} \citep{STScI2013} with 
the KAIT, Nickel, and pt5m bandpasses, comparing with the photometry at or near contemporaneous epochs (these spectra 
were all first recalibrated to the $V$ magnitude for the nearest epoch), and find the following: 
$\Delta(B-V)=-0.31 \pm 0.14$, $\Delta(V-R)=0.00 \pm 0.05$, $\Delta(V-I)=0.11 \pm 0.04$ mag, ``KAIT$-$phot''; 
$\Delta(B-V)=-0.27 \pm 0.12$, $\Delta(V-R)=0.05 \pm 0.05$, $\Delta(V-I)=-0.10 \pm 0.06$ mag, ``Nickel$-$phot''; and, 
$\Delta(B-V)=0.06 \pm 0.10$, $\Delta(V-R)=0.05 \pm 0.05$, $\Delta(V-I)=0.13 \pm 0.05$ mag, ``pt5m$-$phot.''
One could therefore correct our photometry by these amounts, although we caution that the blue end of the KAIT and Nickel $B$
bandpasses extends shortward of the bluest wavelengths of the spectra, so there is likely flux missing within those bandpasses; 
likewise, the spectra extend redward of the end of the Nickel $I$ bandpass trace, so, again, not all of the flux may be represented 
in the synthetic photometry with that filter. The authors will happily provide the bandpasses to the reader should an inquiry be 
made.

We analyze the photometric properties of SN 2017eaw more extensively in Section~\ref{photanal}.

\begin{deluxetable*}{lccccc}
\tablewidth{0pt}
\tablenum{2}
\tablecolumns{6}
\tablecaption{Log of Optical Spectroscopy of SN 2017eaw\label{tabspec}}
\tablehead{
\colhead{Obs.~Date} & \colhead{MJD} & \colhead{Age\tablenotemark{a}} & \colhead{Instrument} & \colhead{Wavelength} & \colhead{Resolution} \\
\colhead{} & \colhead{} & \colhead{} & \colhead{} & \colhead{Range (\AA)} & \colhead{(\AA)}}
\startdata
17-05-17.402 & 57890.902 & 5.2 & MMT-BC  & 5697--6997 & 0.49 \\
17-05-19.481 & 57892.981 & 7.3 & Kast & 3650--10650 & 2.0 \\
17-05-20.377 & 57893.877 & 8.2 & Kast & 3660--10600 & 2.0 \\
17-05-20.429 & 57893.929 & 8.2 & MMT-BC  & 5785--7090 & 0.49 \\
17-05-21.482 & 57894.982 & 9.3 & MMT-BC  & 5788--7093 & 0.49 \\
17-06-02.231 & 57906.731 & 21.0 & HIRES & 3642.9--7967.1 & 0.02--0.03 \\
17-06-02.478 & 57906.978 & 21.3 & Kast & 3660--10630 & 2.0 \\
17-06-21.480 & 57925.980 & 40.3 & Kast & 3622--10718 & 2.0 \\
17-06-24.418 & 57928.918 & 43.2 & MMT-BC  & 5711--7022 & 0.49 \\
17-06-27.500 & 57932.000 & 46.3 & Kast & 3627--10718 & 2.0 \\
17-06-30.351 & 57934.851 & 49.2 & MMT-BC  & 5709--7020 & 0.49 \\
17-07-01.468 & 57935.968 & 50.3 & Kast & 3638--10710 & 2.0 \\
17-07-17.488 & 57951.968 & 66.3 & Kast & 3614--10690 & 2.0 \\
17-07-26.498 & 57960.998 & 75.3 & Kast & 3622--10680 & 2.0 \\
17-07-30.489 & 57964.989 & 79.3 & Kast & 3620--10708 & 2.0 \\
17-08-01.488 & 57966.988 & 81.3 & Kast & 3620--10710 & 2.0 \\
17-08-27.492 & 57992.992 & 107.3 & Kast & 3620--10716 & 2.0 \\
17-08-29.166 & 57994.666 & 109.0 & Bok-B\&C & 3684--8040 & 3.6 \\
17-09-12.330 & 58008.830 & 123.1 & Bok-B\&C & 4000--8039 & 3.6 \\
17-09-14.191 & 58010.691 & 125.0 & Kast & 3632--10720 & 2.0 \\
17-09-27.155 & 58023.655 & 138.0 & Kast & 3630--10680 & 2.0 \\
17-09-29.249 & 58025.749 & 140.0 & Bok-B\&C & 3799--8029 & 3.6 \\
17-10-08.149 & 58034.649 & 148.9 & MMT-BC  & 3476--8695 & 1.9 \\
17-10-09.100 & 58035.600 & 149.9 & MMT-BC  & 5705--7010 & 0.49 \\
17-10-11.117 & 58037.617 & 151.9 & Bok-B\&C & 4731--9079 & 3.6 \\
17-10-19.346 & 58045.846 & 160.1 & Kast & 3622--10670 & 2.0 \\
17-10-25.330 & 58051.830 & 166.1 & Kast & 3620--10680 & 2.0 \\
17-10-27.080 & 58053.580 & 167.9 & MMT-BC  & 5646--6958 & 0.49 \\
17-10-28.113 & 58054.613 & 168.9 & Bok-B\&C  & 4415--8728 & 3.6 \\
17-10-30.100 & 58056.600 & 170.9 & Kast & 3622--10706 & 2.0 \\
17-11-20.091 & 58077.591 & 191.9 & MMT-BC  & 5657--6965 & 0.49 \\
17-11-26.099 & 58083.599 & 197.9 & Kast & 3630--10700 & 2.0 \\
17-12-12.098 & 58099.598 & 213.9 & Kast & 3632--10680 & 2.0 \\
17-12-18.096 & 58105.596 & 219.9 & Kast & 3632--10712 & 2.0 \\
18-01-13.093 & 58131.593 & 245.9 & Kast & 3630--10680 & 2.0 \\
18-07-01.415 & 58300.915 & 415.2 & MMT-BC  & 5720--7025 & 0.49 \\
18-09-06.316 & 58367.816 & 482.1 & MMT-BC  & 3245--8104 & 1.9 \\
\enddata
\tablenotetext{a}{The age is referenced to our estimate of the date of explosion, JD 2,457,885.7.}
\end{deluxetable*}

\subsection{SN Optical Spectroscopy}\label{specobs}

Over an eight-month period beginning on 2017 May 19, a series of 20 optical spectra of SN 2017eaw were 
obtained with the Kast double spectrograph \citep{Miller+1993} mounted on the 3 m Shane telescope at Lick 
Observatory.  These spectra were taken at or near the parallactic angle \citep{Filippenko1982} to minimize 
slit losses caused by atmospheric dispersion. Data were reduced following standard techniques for CCD 
processing and spectrum extraction \citep{Silverman+2012} utilizing \texttt{IRAF}\footnote{IRAF is distributed by the 
National Optical Astronomy Observatory, which is operated by AURA, Inc., under a cooperative agreement with 
the NSF.} routines and custom Python and IDL codes\footnote{https://github.com/ishivvers/TheKastShiv}.  
Low-order polynomial fits to arc-lamp spectra were used to calibrate the wavelength scale, and small 
adjustments derived from night-sky lines in the target frames were applied.  Observations of appropriate 
spectrophotometric standard stars were used to flux-calibrate the spectra.

With the Blue Channel (BC) spectrograph on the MMT we also obtained 
nine spectra with the 1200 lines mm$^{-1}$ grating, with a central wavelength of 6360\,\AA\ and a 1$\farcs$0 
slit width, and two spectra with the 300 lines mm$^{-1}$ grating.
We obtained five epochs of optical spectroscopy with the Boller $\&$ Chivens (B$\&$C) spectrograph 
mounted on the 2.3~m Bok telescope on Kitt Peak using the 300 line~mm$^{-1}$ grating.

Standard reductions were carried out using \texttt{IRAF}, and wavelength solutions were determined using internal 
He-Ne-Ar lamps. Flux calibration was achieved using spectrophotometric standards at a similar airmass taken 
throughout the night.

Additionally, some of us (H.I., I.J.M.C., D.H., M.R.K.) obtained an optical spectrum of SN 2017eaw with the
HIRES spectrometer \citep{Vogt+1994} on the Keck I 10 m telescope on Maunakea on 2017 June 2.
The spectrum, with an exposure time of 292 s, has a continuum signal-to-noise ratio (S/N) of 40 per pixel at 
5500 \AA. The use of the $0{\farcs}87 \times 14{\farcs}0$ (C2) decker provided a resolution of 50,000; 
while sky subtraction can be performed, the extra on-sky pixels in the spatial direction provide additional 
information about the environment of the primary target.
The spectrum was reduced using the standard California Planet Search pipeline \citep{Howard+2010}.

\begin{figure*}
\plotone{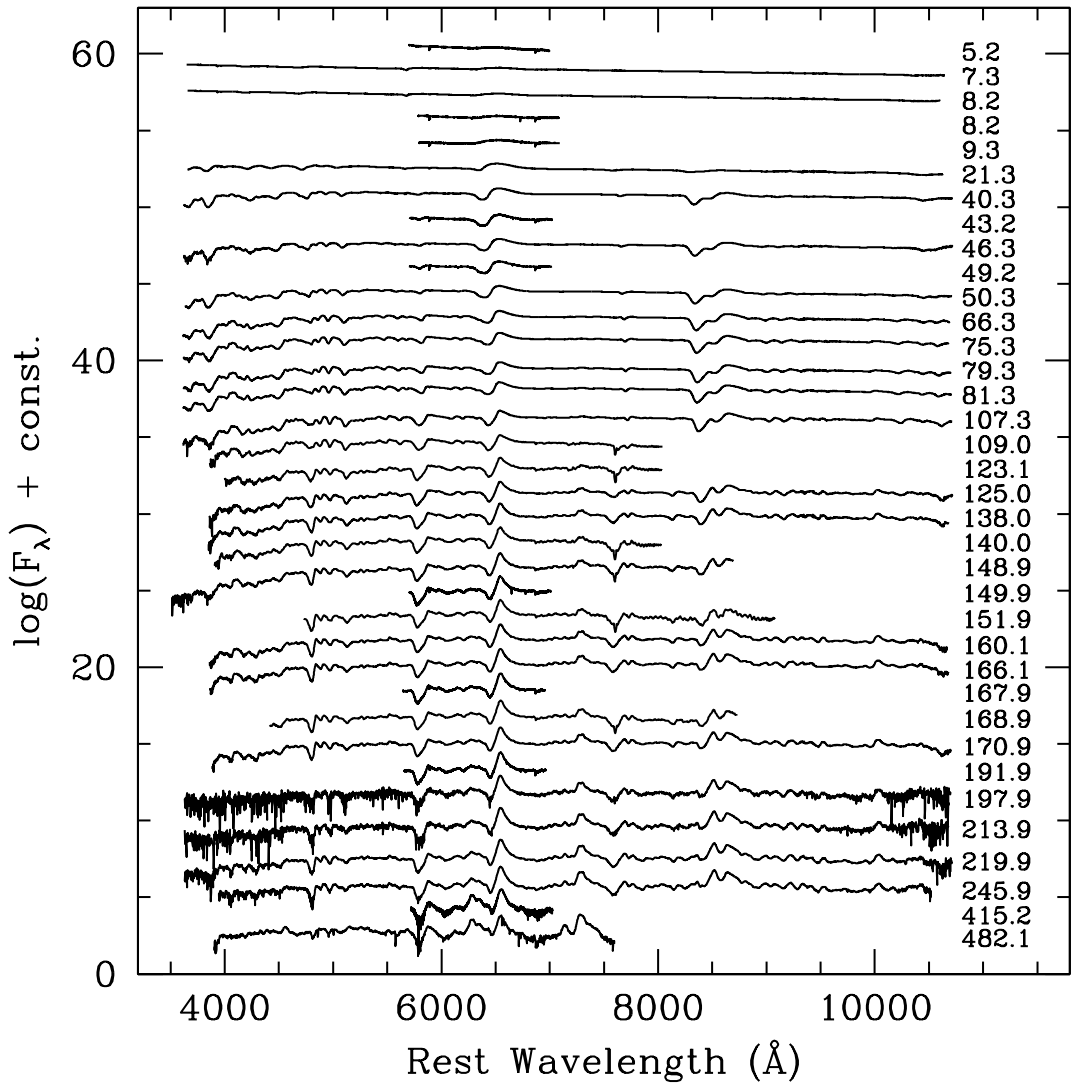}
\caption{Sequence of Kast, MMT, and Bok spectra of SN 2017eaw; see Table~\ref{tabspec}. The spectra have all been 
corrected for the redshift of the host galaxy, but not for the reddening to the SN. The spectra are labeled
by the day since explosion, assumed to be JD 2,457,885.7 (see Section~\ref{explosion}).\label{figspec}}
\end{figure*}

We provide a log of the Kast, MMT, Bok, and Keck observations in Table~\ref{tabspec}.
The sequence of Kast and MMT spectra is shown in Figure~\ref{figspec}.
All of the spectra have been corrected for the redshift of NGC 6946, taken to be 
$z=0.000133$\footnote{From the NASA/IPAC Extragalactic Database (NED), http://ned.ipac.caltech.edu/.}.
We can see from the sequence in Figure~\ref{figspec} that SN 2017eaw appears to be a normal SN II-P.

\subsection{Pre-SN {\sl HST\/} Imaging}\label{HST}

The site of SN 2017eaw was observed serendipitously by several {\sl HST\/} programs prior to explosion.
These include GO-14156 (PI A.~Leroy) with the Wide Field Camera 3
(WFC3) IR channel in bands F110W (total exposure 455.87 s) and F128N
(1411.74 s) on 2016 February 9, and GO-14638 (PI K.~Long) with WFC3/IR F160W (396.93 s) on 2016 October
24; and GO-9788 (PI L.~Ho) with the Advanced Camera for Surveys (ACS) Wide Field Channel (WFC) on
2004 July 29 in F658N (700 s) and F814W (120 s), and GO-14786 (PI B.~Williams) with 
ACS/WFC on 2016 October 26 in F814W (2570 s) and F606W (2430 s). 

\subsection{Pre-SN {\sl Spitzer\/} Imaging}

The SN site was also serendipitously observed pre-explosion by a number of programs on various dates, from 
2004 June 10 to 2017 March 31, using {\sl Spitzer}.
We list all of these observations and the data that we considered in Table~\ref{tabspitzer}.
The data were obtained primarily with the Infrared Array Camera (IRAC; \citealt{Fazio+2004}), but also with 
the Multiband Imaging Photometer for {\sl Spitzer\/} (MIPS; \citealt{Rieke+2004}). 
The IRAC imaging was obtained both during the cryogenic mission in all four bands
(3.6, 4.5, 5.8, and 8.0 $\mu$m), and during the so-called ``Warm'' (post-cryogenic) mission only in the two 
shortest wavelength bands.
For MIPS we consider only the available 24 $\mu$m data.

\begin{deluxetable}{lcc}
\tablewidth{0pt}
\tablecolumns{3}
\tablenum{3}
\tablecaption{Log of {\sl Spitzer\/} Data Covering the SN 2017eaw Site\label{tabspitzer}}
\tablehead{
\colhead{Obs.~Date} & \colhead{AORKEY} & \colhead{Data}}
\startdata
04-06-10 & 5508864 & IRAC 3.6, 4.5, 5.8, 8.0 $\mu$m \\
04-07-10 & 5576704 & MIPS 24, 70, 160 $\mu$m \\
04-07-11 & 5576960 & MIPS 24, 70, 160 $\mu$m \\
04-09-12 & 10550528 & IRAC 3.6, 4.5, 5.8, 8.0 $\mu$m \\
04-11-25 & 5508608 & IRAC 3.6, 4.5, 5.8, 8.0 $\mu$m \\
04-11-25 & 10550784 & IRAC 3.6, 4.5, 5.8, 8.0 $\mu$m \\
05-07-19 & 14526976 & IRAC 3.6, 4.5, 5.8, 8.0 $\mu$m \\
05-07-20 & 14456320 & IRAC 3.6, 4.5, 5.8, 8.0 $\mu$m \\
05-07-20 & 14457856 & IRAC 3.6, 4.5, 5.8, 8.0 $\mu$m \\
05-09-24 & 14528000 & MIPS 24 $\mu$m \\
05-12-30 & 14528256 & IRAC 3.6, 4.5, 5.8, 8.0 $\mu$m \\
06-01-10 & 14528512 & MIPS 24 $\mu$m \\
06-11-26 & 17965568 & IRAC 3.6, 4.5, 5.8, 8.0 $\mu$m \\
06-12-29 & 18277120 & IRAC 3.6, 4.5, 5.8, 8.0 $\mu$m \\
07-01-06 & 18270976 & MIPS 24 $\mu$m \\
07-07-03 & 18277376 & IRAC 3.6, 4.5, 5.8, 8.0 $\mu$m \\
07-07-10 & 18271232 & MIPS 24 $\mu$m \\
07-12-27 & 23508224 & IRAC 3.6, 4.5, 5.8, 8.0 $\mu$m \\
08-01-27 & 24813824 & IRAC 3.6, 4.5, 5.8, 8.0 $\mu$m \\ 
08-07-18 & 27190016 & IRAC 3.6, 4.5, 5.8, 8.0 $\mu$m \\
08-07-29 & 27189248 & MIPS 24 $\mu$m \\
09-08-06 & 34777856 & IRAC 3.6, 4.5 $\mu$m \\
10-01-05 & 34778880 & IRAC 3.6, 4.5 $\mu$m \\
10-08-13 & 39560192 & IRAC 3.6, 4.5 $\mu$m \\
11-07-27 & 42195456 & IRAC 3.6, 4.5 $\mu$m \\
11-08-01 & 42415872 & IRAC 3.6, 4.5 $\mu$m \\
12-02-01 & 42502144 & IRAC 3.6, 4.5 $\mu$m \\
13-08-17 & 48000768 & IRAC 3.6, 4.5 $\mu$m \\
14-01-03 & 49665024 & IRAC 3.6, 4.5 $\mu$m \\
14-02-17 & 48934144 & IRAC 3.6, 4.5 $\mu$m \\
14-03-26 & 50623232 & IRAC 3.6, 4.5 $\mu$m \\
14-09-16 & 50623744 & IRAC 3.6, 4.5 $\mu$m \\
14-10-15 & 50623488 & IRAC 3.6, 4.5 $\mu$m \\
15-01-31 & 53022464 & IRAC 3.6, 4.5 $\mu$m \\
15-09-02 & 52785664 & IRAC 3.6, 4.5 $\mu$m \\
15-09-07 & 52785920 & IRAC 3.6, 4.5 $\mu$m \\
15-09-16 & 52786176 & IRAC 3.6, 4.5 $\mu$m \\
15-09-28 & 52786432 & IRAC 3.6, 4.5 $\mu$m \\
15-10-27 & 52786688 & IRAC 3.6, 4.5 $\mu$m \\
15-11-25 & 52786944 & IRAC 3.6, 4.5 $\mu$m \\
15-12-23 & 52787200 & IRAC 3.6, 4.5 $\mu$m \\
16-09-04 & 52787456 & IRAC 3.6, 4.5 $\mu$m \\
16-10-12 & 60832256 & IRAC 3.6, 4.5 $\mu$m \\
16-12-30 & 60832512 & IRAC 3.6, 4.5 $\mu$m \\
17-03-31 & 60832768 & IRAC 3.6, 4.5 $\mu$m \\
\enddata
\end{deluxetable}

\subsection{Post-explosion {\sl HST\/} Imaging}\label{postimaging}

We observed SN 2017eaw on 2017 May 29.79 with {\sl HST\/} WFC3/UVIS in subarray mode in F814W (270 s total 
exposure), as part of our Target of Opportunity (ToO) program (GO-14645, PI S.~Van Dyk). Although the SN 
was quite bright at the time, we successfully avoided 
saturating the detector (the F814W band was expressly chosen to reduce the probability 
of this happening on the likely observing date), while achieving appreciable S/N on fainter
point-like objects in the
SN environment that could be used as astrometric fiducials (see Section~\ref{progenitorid}).
SN 2017eaw was also observed on 2018 January 5 with WFC3/UVIS in F555W and F814W (710 and 780 s)
as part of the {\sl HST\/} Snapshot program GO-15166 (PI A.~Filippenko); unfortunately, the SN was still quite
bright ($V \approx 16.8$ and $I \approx 14.8$ mag) at the time, and the central pixels of the SN image are saturated, 
further leading to prominent detector row and column bleeding.

\section{Extinction to SN 2017\lowercase{eaw}}\label{reddening}

\begin{figure}
\plottwo{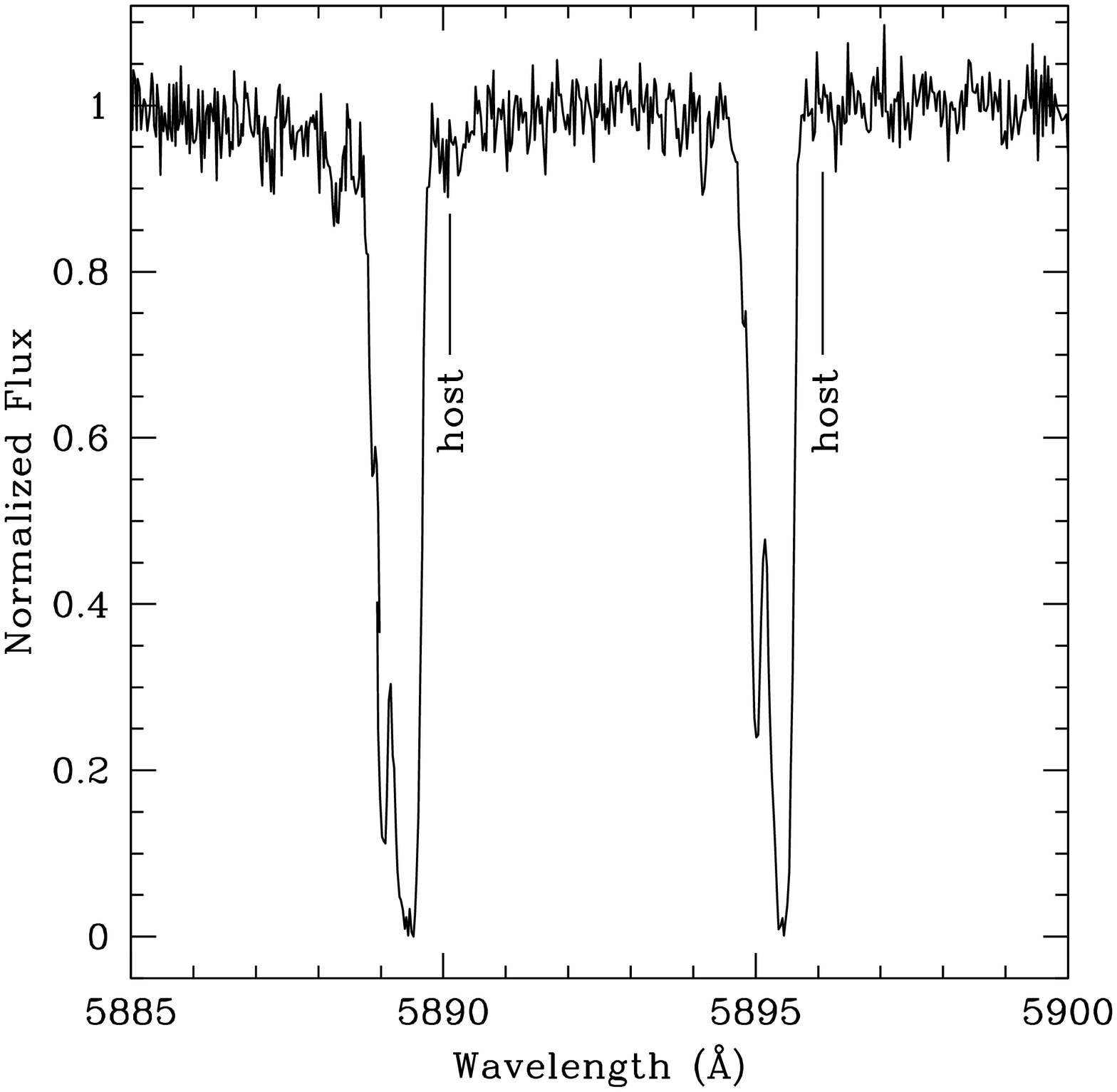}{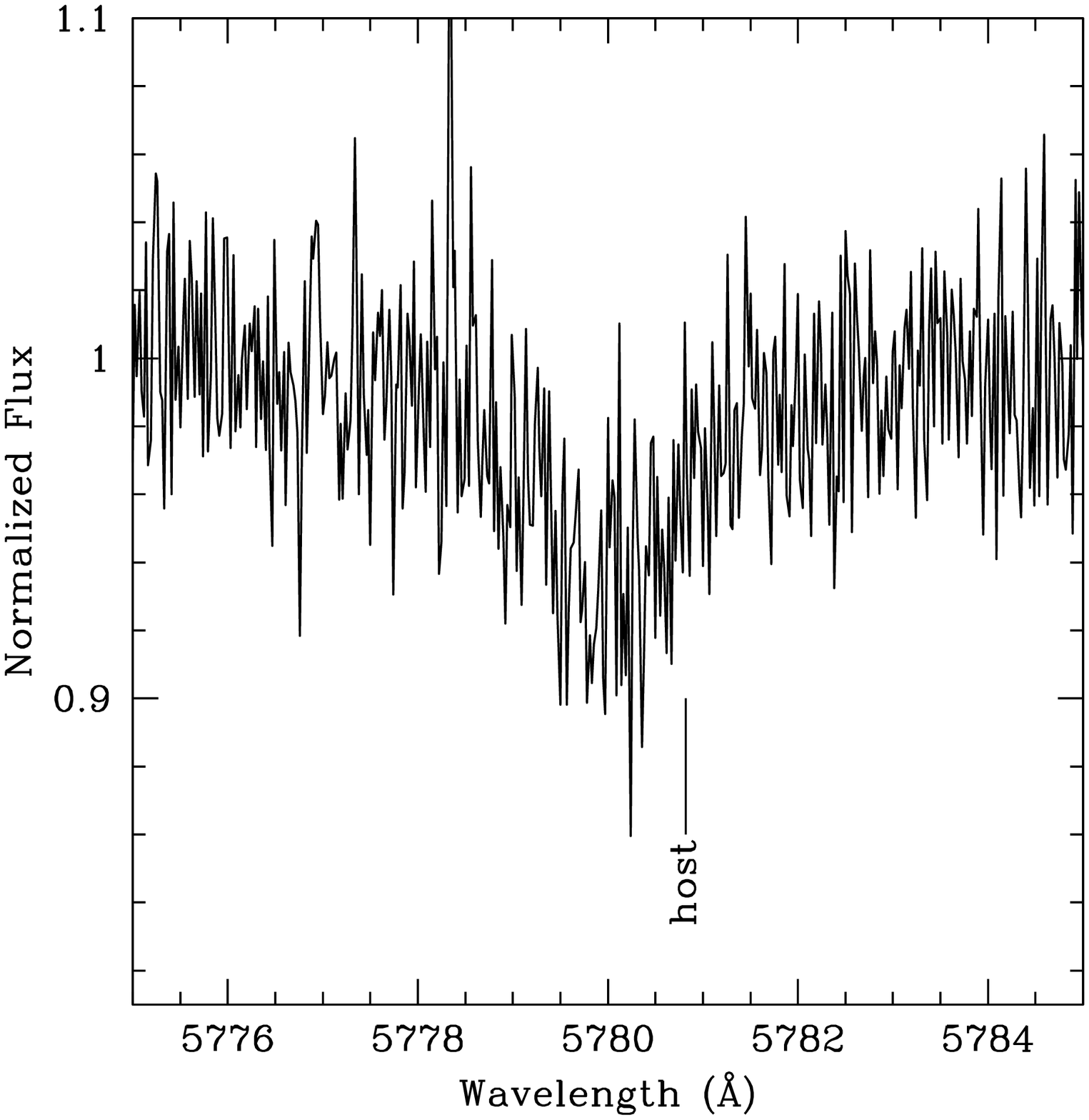}
\caption{Portions of the spectrum of SN 2017eaw obtained with HIRES on the Keck I telescope on 2017 June 2, 
focusing on the Na~{\sc i}~D lines (left panel) and the DIB $\lambda$5780 feature (right panel). The locations 
of these features at the host-galaxy redshift are indicated (by ``host'') in both panels.\label{fighires}}
\end{figure}

What is noticeable in the spectral sequence is the presence of a strong Na~{\sc i} D feature, consistent
with the large Galactic extinction we would expect for a host galaxy at a Galactic latitude of
only $b=11{\fdg}7$. Of course, at low resolution the Na~{\sc i} feature could also include a contribution
to the extinction internal to the host, given its low redshift.
As mentioned in Section~\ref{specobs}, we obtained a Keck HIRES spectrum, with the aim of
isolating both the Na~{\sc i}~D and 5780 \AA\ diffuse
interstellar band (DIB) features, to assess the amount of extinction to the SN.
We show the HIRES spectrum in Figure~\ref{fighires} for these two portions of the overall coverage.
As one can see, the Na~{\sc i}~D1 and D2 lines are essentially saturated; however, these are both at wavelengths
corresponding essentially to zero redshift (i.e., to the Galactic component of extinction), whereas at
wavelengths corresponding to NGC 6946, no clear sign of either feature exists.
\citet{Phillips+2013} cautioned that Na~{\sc i}~D is a rather poor measure of the value of the 
extinction and recommended use of the DIB $\lambda$5780 feature instead.
One can see in Figure~\ref{fighires} that a strong DIB feature is evident; however, again this corresponds
to the foreground extinction component, whereas any contribution from the host cannot be distinguished from 
the broad Galactic feature.
Hence, we conclude that there is little evidence for significant host-galaxy extinction.
Hereinafter we therefore assume the Galactic foreground contribution toward SN 2017eaw from 
\citet{Schlafly+2011} (via NED), $A_V=0.941$ mag, as the total visual interstellar extinction.
The uncertainty in the extinction is likely $\lesssim 0.1$ mag.

\section{Distance to SN 2017\lowercase{eaw}}\label{distance}

Not surprisingly for such a famous and well-studied galaxy as NGC 6946, NED lists 32 
redshift-independent distances. 
These include Tully-Fisher estimates of 5.0--5.3 Mpc by \citet{Bottinelli+1984,Bottinelli+1986}, 5.4--5.5 
Mpc by \citet{Schoniger+1994}, and 5.5 Mpc by \citet{Pierce1994}.
A number of SN-based distances have also been estimated, including early measurements with the expanding 
photosphere method (EPM) applied to SN 1980K by \citet{Schmidt+1992} (7.2 Mpc) and 
\citet{Schmidt+1994} (5.7 Mpc), with more recent attempts using SN 2004et by 
\citet{Takats+2012} (4.8 Mpc) and \citet{Bose+2014} (4.0--5.9 Mpc). The standardized candle method (SCM)
has also been applied to SN 2004et by \citet{Olivares+2010} (4.3--5.4 Mpc) and \citet{Pejcha+2015} (4.5 Mpc),
and applied with a modified version of this method by \citet{Poznanski+2009} to SN 2002hh (6.0 Mpc) and SN 
2004et (4.7 Mpc). Thus, most of these distance estimates to NGC 6946 have been quite 
``short,'' at approximately 5 Mpc.
\citet{Karachentsev+2000} estimated 6.8 Mpc using the brightest blue stars in a galaxy group 
containing NGC 6946, and \citet{Tikhonov2014} measured a tip-of-the-red-giant-branch (TRGB) distance to NGC 
6946 of $6.72 \pm 0.15$ Mpc using archival {\sl HST\/} data.

\begin{figure}
\plotone{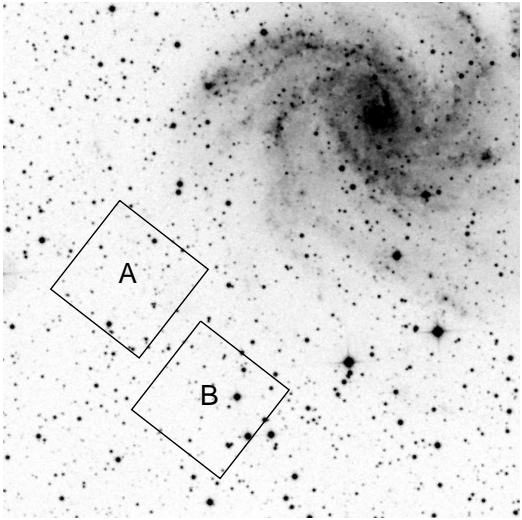}
\caption{Digitized Sky Survey image of the host galaxy, NGC 6946, showing the footprints of the two
WFC3/UVIS fields, A and B, that we analyzed when determining the TRGB distance to the galaxy.
Field A is at a galactic nuclear offset of $6{\farcm}9$, while Field B is at $7{\farcm}9$.
The latter field had been previously analyzed by \citet{Murphy+2018}. Both fields were also
analyzed by \citet{Anand+2018}.\label{figfootprint}}
\end{figure}

We have independently estimated a TRGB distance using more recent {\sl HST\/} data. We downloaded publicly
available deep archival WFC3/UVIS images in F606W (total exposure 5470 s) and F814W (5548 s) for a parallel 
field $6{\farcm}9$ from the galaxy center, obtained on 2016 October 28 by GO-14786 (PI B.~Williams). 
We were well underway analyzing this field when the work by \citet{Murphy+2018} appeared. 
These authors analyzed a different field, also observed by GO-14786 in the same bands to approximately the same depth, 
but at a nuclear offset of $7{\farcm}9$. \citeauthor{Murphy+2018}~inferred an
astonishingly larger distance of $7.83 \pm 0.29$ Mpc (i.e., a distance modulus of $29.47 \pm 0.08$ mag), 
much greater than previous estimates.
To confirm this measurement and to potentially bolster confidence in the measurement from our chosen field,
we decided to undertake analysis of both fields, shown in Figure~\ref{figfootprint}.
We note that yet another TRGB analysis, by \citet{Anand+2018}, appeared, also while our analysis was
underway.

Both fields have comparatively fewer contaminants, e.g., younger supergiants, in NGC 6946 than other 
archival {\sl HST\/} pointings in these bands and likely better probe the desired older halo population. 
We processed the data with \texttt{Dolphot} \citep{Dolphin2000,Dolphin2016}, with the same input parameters 
as described in Section~\ref{progenitorid} after first running the individual frames corrected for 
charge-transfer-efficiency (CTE) through \texttt{AstroDrizzle}, to flag cosmic-ray hits.
We further culled out objects from the output photometry list that are most probably ``good stars,'' by 
imposing cuts on the \texttt{Dolphot} parameters object type ($=1$), and sharpness $s$ 
($-0.3 \le s \le 0.3$), crowding ($<0.5$ mag), quality flag ($=0$), and $\chi$ ($<2$ at F606W and $<3$ at 
F814W), following, e.g., \citet{Mager+2008}. 
Even with these cuts we cannot rule out that source crowding and
confusion are affecting the photometry.
The photometry for these selected objects was then corrected for the appropriate 
Galactic foreground extinction in each band \citep{Schlafly+2011} toward each of the two fields.
Field B is less extinguished than is Field A. The extinction varies up to 0.01 mag in both $A_V$ and $A_I$ 
for Field A, and to 0.03 and 0.02 mag in $A_V$ and $A_I$, respectively, for Field B.

\begin{figure}
\plotone{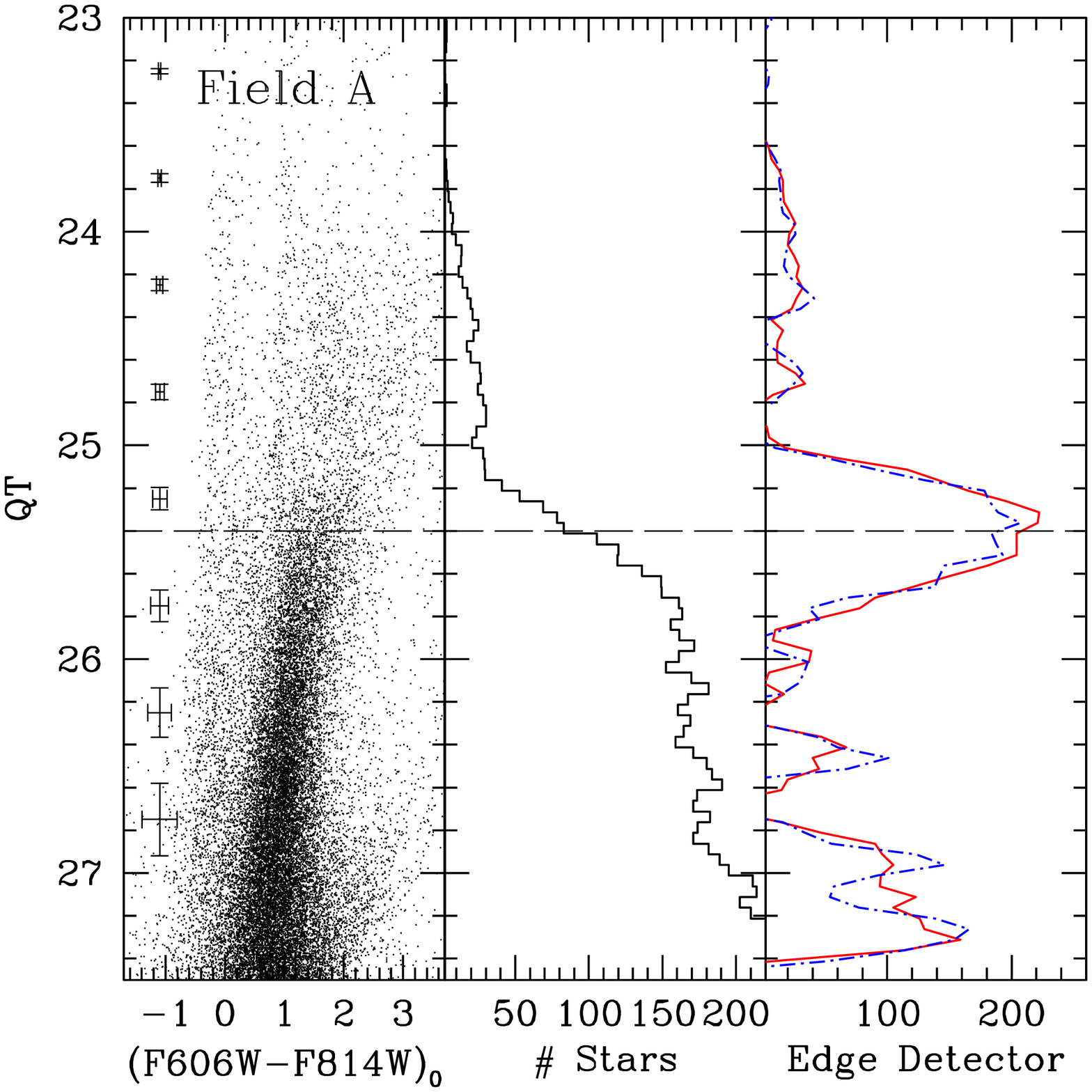}
\caption{{\it Left}: CMD of the color-dependence-corrected brightness $QT$ 
\citep{Jang+2017} versus the reddening-corrected $(F606W-F814W)$ color for stars in Field A (see 
Figure~\ref{figfootprint}), which is $6{\farcm}9$ from the center of NGC 6946. Representative 
uncertainties in the photometry are also shown. A well-developed red giant branch can clearly be seen. 
{\it Middle}: histogram of $QT$, binned by 0.05 mag, for stars in the color range $(F606W-F814W)_0=1.1$--2.0 mag.
{\it Right}: the Sobel edge detector responses, from \citet[][ solid red curve]{Madore+2009} and
\citet[][ dotted-dashed blue curve]{Jang+2017}, that we applied to the histogram in the middle panel. A dashed line indicates the 
TRGB at $QT=25.40$ mag, which is at the peak of a Gaussian fit to the edge detector responses. The width 
of that Gaussian is 0.23 mag.\label{figtrgb1}}
\end{figure}

\begin{figure}
\plotone{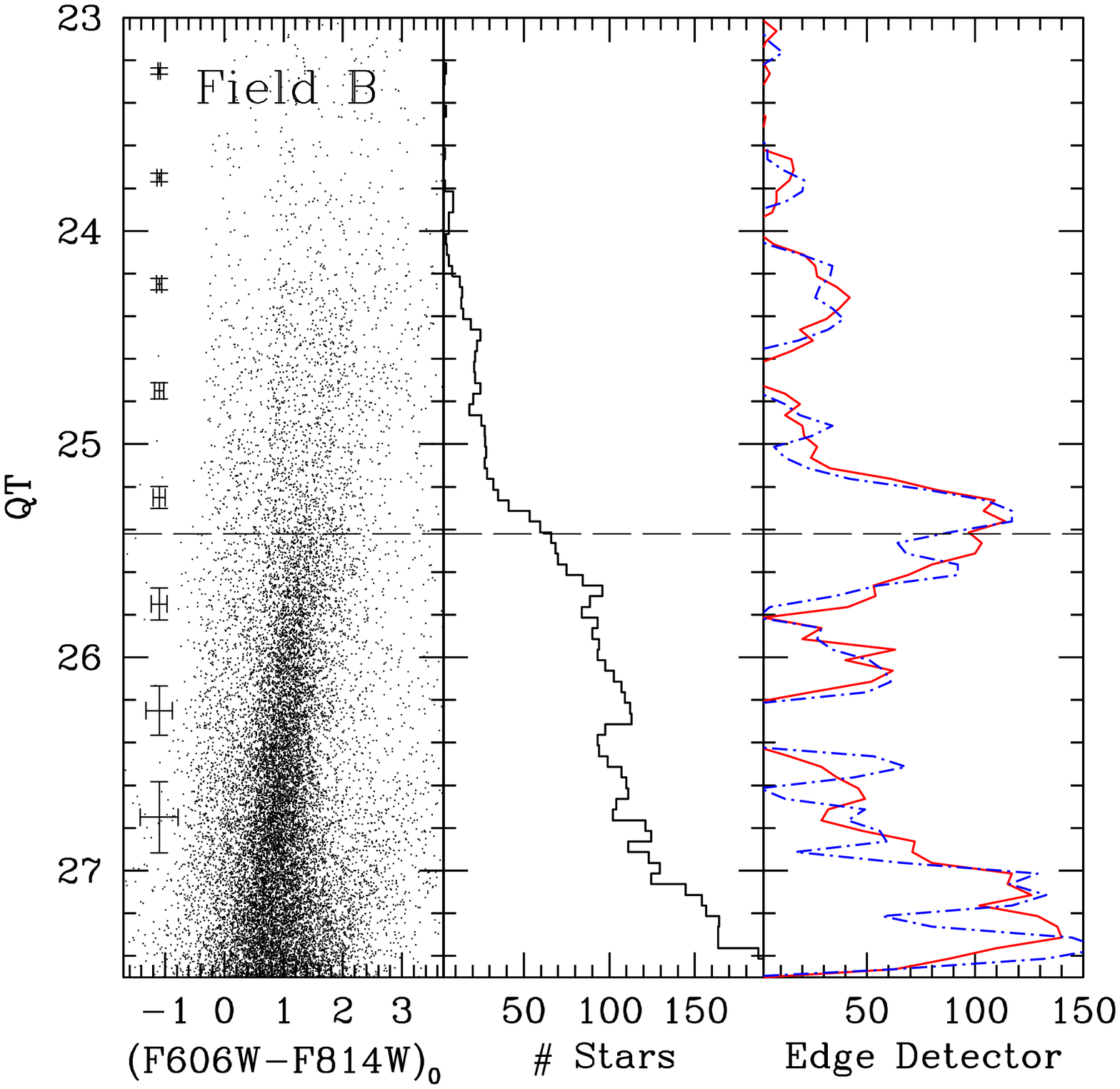}
\caption{{\it Left}: CMD of the color-dependence-corrected brightness $QT$ 
\citep{Jang+2017} versus the reddening-corrected $(F606W-F814W)$ color for stars in Field B (see 
Figure~\ref{figfootprint}), which is $7{\farcm}9$ from the center of NGC 6946. Representative 
uncertainties in the photometry are also shown. A well-developed red giant branch can clearly 
be seen. 
{\it Middle}: histogram of $QT$, binned by 0.05 mag, for stars in the color range $(F606W-F814W)_0=1.1$--2.0 mag.
{\it Right}: the Sobel edge detector responses, from \citet[][ solid red curve]{Madore+2009} 
and \citet[][ dotted-dashed blue curve]{Jang+2017}, that we applied to to the histogram in the middle panel. 
A dashed line indicates the TRGB at $QT=25.42$ mag, 
which is at the peak of a Gaussian fit to the edge detector responses. The width of that Gaussian is 0.20 
mag. Note that \citet{Murphy+2018} had previously performed this analysis for this field and found 
essentially the same value for the TRGB brightness. \citet{Anand+2018} also found a similar 
value.\label{figtrgb2}}
\end{figure}

We transformed the reddening-corrected photometry at F814W to the color-dependence-corrected brightness 
$QT$, following \citet{Jang+2017} for the {\sl HST\/} appropriate camera and filter combination. 
The resulting color-magnitude diagrams (CMDs) for the two fields are shown in 
Figures~\ref{figtrgb1} and \ref{figtrgb2} for A and B, respectively.
A prominent red giant branch can be clearly seen in both of the figures.
We applied two different Sobel edge detectors, the [$-1$, $-1$, $-1$, 0, 0, 0, $+1$, $+1$, $+1$] kernel 
from \citet{Madore+2009} and the [$-1$, $-2$, $-1$, 0, $+1$, $+2$, $+1$] kernel from \citet{Jang+2017}, 
to 0.05-mag-binned histograms of $QT$ in the color range $1.1 \le (F606W-F814W)_0 \le 2.0$ mag,
to mitigate against contamination from the likely RSGs
near $(F606W-F814W)_0 \approx 1.0$ mag and to locate the TRGB. 
We show the resulting edge detector responses in Figures~\ref{figtrgb1} and \ref{figtrgb2} as well. 
By trial and error we had found that these two edge detectors provided the clearest
discrimination of the TRGB; as can be seen in the figures, their results are quite similar.

From our analysis the TRGB appears to be at 25.40 and 25.42 mag for Fields A and B, respectively.
The RGB is more populated in Field A than in Field B.
For the uncertainties in these two estimates we assumed the width of a Gaussian fit to the edge detector 
response \citep{Sakai+1999,Sakai+2000} for the two fields, which are 0.23 and 0.20 mag, respectively. 
Taking the uncertainty-weighted mean of these two estimates, and assuming the luminosity of the TRGB is 
$QT=-4.031$ mag \citep[][ this is the value for the similar late-type spiral galaxy NGC 4258]{Jang+2017}, 
we found a reddening-corrected distance modulus to NGC 6946 of $29.44 \pm 0.21$ mag, or 
a distance of $7.73 \pm 0.78$ Mpc. 

Note that we arrive at the same value, to within the uncertainties, for both the TRGB apparent brightness 
and the distances as did \citet{Murphy+2018} and 
\citet[][ $7.72 \pm 0.32$ Mpc, from analysis of both Fields A and B]{Anand+2018}, via somewhat less sophisticated pathways 
to the TRGB estimate (we used the Sobel edge detection, whereas the other two studies used a Bayesian maximum likelihood 
technique; see their studies for details).

\begin{figure}
\plotone{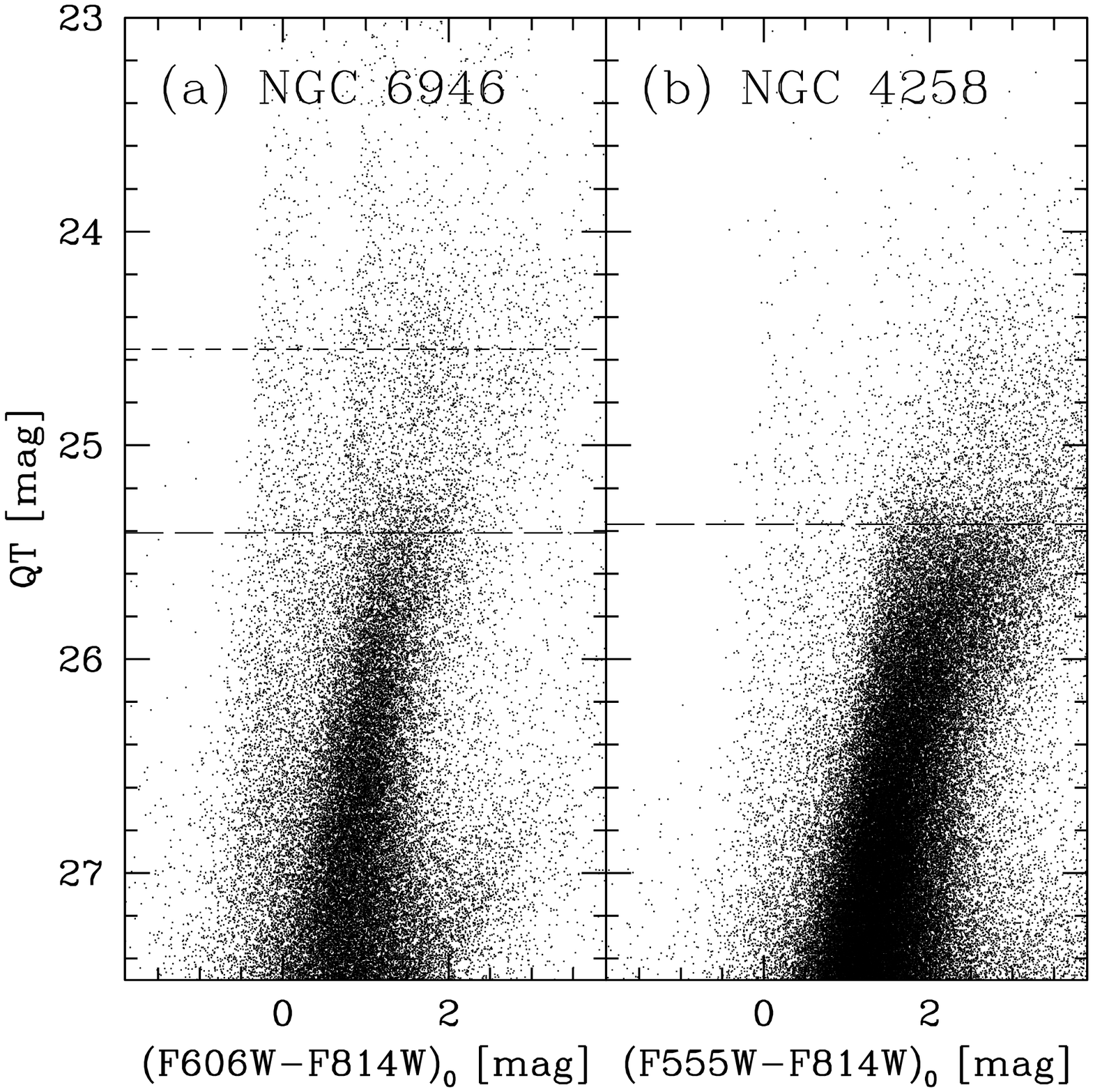} 
\caption{
{\it Left}: CMD of the color-dependence-corrected brightness $QT$ 
\citep{Jang+2017} vs.~the reddening-corrected $(F606W-F814W)$ color for stars in the combined Fields A
and B in NGC 6946 (see Figure~\ref{figfootprint}). {\it Right}: a similar CMD for stars in the halo of the
late-type spiral galaxy NGC 4258 (see \citealt{Jang+2017}). The long-dashed line in both panels indicates the
detected level of the TRGB in each galaxy. The short-dashed line in the left panel indicates where the TRGB would be if the
distance to NGC 6946 were 5.2 Mpc.\label{n4258trgb}}
\end{figure}

To demonstrate that the TRGB distance estimate for NGC 6946 is reasonable, we compare in Figure~\ref{n4258trgb} the 
CMD for the combined fields A and B with a CMD of a field in NGC 4258, considered a distance anchor by \citet{Jang+2017}. 
The TRGB distance for the latter galaxy \citep[$7.18 \pm 0.40$ Mpc;][]{Mager+2008} is consistent with distance estimates obtained using 
that galaxy's nuclear water megamaser \citep[$7.54 \pm 0.17$ Mpc; e.g.,][]{Riess+2016}. It is evident from the figure that the TRGB 
distance to NGC 6946 is comparable to that of NGC 4258, with the former galaxy
being slightly more distant than the latter. The TRGB distance to NGC 6946 is inconsistent with the previous shorter SN-based
distances. We note that general comparisons of Cepheid and TRGB distance estimates have been in excellent agreement
(e.g., \citealt{Jang+2018}).

Hereinafter we adopt our estimate, above, of the distance to NGC 6946. As a cross-check, we also computed SCM and
EPM distances to SN 2017eaw in Section~\ref{epmscm}.

\section{Metallicity of the SN 2017\lowercase{eaw} Site}\label{metallicity}

One means of estimating the metallicity at the site of a SN is from an abundance gradient for the host
galaxy, if it has been established. Since the oxygen abundance is often adopted as a proxy for metallicity, 
the gradient considered is usually that of oxygen. First, we deprojected an image of NGC 6946 assuming the 
relevant galaxy parameters (inclination, position angle) from \citet{Jarrett+2003}.
From the absolute positions of SN 2017eaw and the NGC 6946 nucleus, 
the nuclear offset of the SN site is then $\sim 221\arcsec$. At the assumed host distance, this 
corresponds to 7.3 kpc.
Using the oxygen abundance gradient from 
\citet[][ although they assumed a distance of 5.9 Mpc to NGC 6946]{Belley+1992}, we
estimate that the abundance at the SN 2017eaw site is $12 + \log({\rm O/H})=8.71 \pm 0.05$.
With the Sun's oxygen abundance assumed to be $12 + \log({\rm O/H})=8.69 \pm 0.05$\citep[$Z=0.014$;][]{Asplund+2009}, 
this would imply a solar-like metallicity at the SN site.

Another, potentially more accurate, way of estimating the SN site metallicity is from the O abundances of 
nearby emission regions, if available \citep[e.g.,][]{Modjaz+2011}.
Examination of the archival {\sl HST\/} F658N image reveals that no such regions exist in the SN's
immediate environment. Fortunately, \citet{Gusev+2013} measured the O abundance for three H~{\sc ii} regions 
nearest the SN site, \#2, \#22, and \#23 (all $\sim 73\arcsec$ to the northwest), 
to be $12 + \log({\rm O/H})=8.46 \pm 0.01$, $8.56 \pm 0.02$, and $8.46 \pm 0.01$, respectively; 
the next closest, \#6, $\sim 92\arcsec$ to the southeast, has $12 + \log({\rm O/H})=8.58 \pm 0.02$.
All of these measurements would imply that the SN 2017eaw site is somewhat subsolar in metallicity.
Given that this method is likely a more direct means of estimating the metallicity, we adopt hereinafter
a subsolar metallicity in the range of 
$Z=0.009$ ([Fe/H] = $-$0.2) to $Z=0.011$ ([Fe/H] = $-$0.1).

\section{Analysis of SN 2017\lowercase{eaw}}\label{photanal}

\subsection{Date of Explosion}\label{explosion}

Not many observational constraints exist on the explosion epoch of SN 2017eaw.
For instance, the KAIT SN search did not obtain its first image of the host (independent of the multiband 
SN 2017eaw monitoring) until 2017 June 12.99, nearly a month after discovery, owing to hour-angle limitations 
imposed for the search. \citet{Steele+2017} did not detect the SN on 2017 May 6.18 to $R>21.2$ mag, eight 
days prior to discovery. \citet{Wiggins2017} reported that nothing was detected at the SN position
to an unfiltered mag ${>19}$ on May 12.20, $\sim 2$ days before his discovery. 
Interestingly, \citet{Drake+2017} reported detection of a source at the SN position in the Catalina 
Real-Time Transient Survey (CRTS) data at $V \approx 19.8$ mag on 2017 May 7.43, 
slightly more than one day later than the deep upper limit by \citeauthor{Steele+2017}.

Patrick Wiggins kindly provided his unfiltered images both of the discovery and of the pre-discovery upper 
limit. We analyzed the images with \texttt{DAOPHOT}, using our photometric sequence at $R$ from 
Table~\ref{tabseq} for calibration. We confirmed that the SN was discovered at 12.8 mag. However, we found 
that the nondetection limit on 2017 May 12 was $>16.8$ mag, rather than $>19$ mag.

Additionally, Andrew Drake graciously provided us with four nearly contemporaneous CRTS exposures from 2017 
May 7, which were the basis of their report \citep{Drake+2017}. We analyzed these images
both individually and as a coaddition. 
In short, we could not convince ourselves that the SN had been detected on that date.
In more detail, we compared the coadded CRTS image
astrometrically, first with a good-quality KAIT image of the SN, and subsequently with the archival 
{\sl HST\/} images in which the SN progenitor
is detected (see Section~\ref{progenitorid}), using 25 and 13 stars in common between the two image
datasets, respectively. In the coadded image there is indeed what appears to be a source within 1 pixel of 
the nominal SN position.  
However, \texttt{DAOPHOT} did not detect this source, and therefore we were not able to measure 
its brightness. Additionally, just a hint of this source appears only in one of the four individual exposures. 

\begin{figure}
\plotone{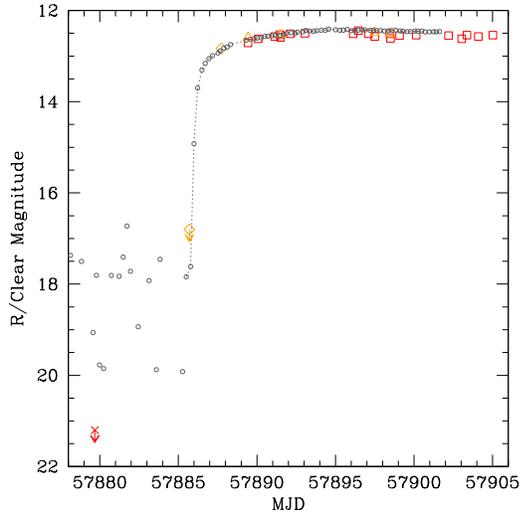} 
\caption{The rise of SN 2017eaw compared to that of KSN 2011a \citep{Garnavich+2016}. Shown are the $R$-band
(red open squares) and KAIT unfiltered (orange open triangles) light curves for SN 2017eaw. The discovery and pre-discovery
nondetection by \citet{Wiggins2017} are shown as orange open diamonds; the $R$ nondetection by \citet{Steele+2017} is
shown as a red cross. The very early KSN 2011a light curve is represented by open gray circles.\label{risetime}}
\end{figure}

Nevertheless, with the Wiggins discovery and pre-discovery nondetection, in particular, we can employ 
knowledge of the rise time of SNe II-P to place a further constraint on the explosion date. Specifically, we draw upon
the beautifully defined rise of KSN 2011a \citep{Garnavich+2016}, which appears to be a relatively normal SN II-P, albeit at higher redshift
than SN 2017eaw. In Figure~\ref{risetime} we show a comparison of the very early KSN 2011a light curve with the $R$-band
and unfiltered SN 2017eaw light curves. We note, of course, that the {\sl Kepler\/} bandpass used to detect KSN 2011a is
far broader than $R$, although comparable in breadth to the unfiltered KAIT CCD response function \citep{Riess+1999}.
Assuming that the rise of the latter was similar to that of the former (and 
the comparison tends to imply that this is the case), we can infer that the explosion date for SN 2017eaw was
on about JD 2,457,885.7 (May 12.2). The Wiggins nondetection constrains this date to about $\pm 0.1$ day, which
is remarkable. The implied rapid rise time of SN 2017eaw would also rule out the CRTS ``pre-discovery,'' some five 
days prior to the Wiggins discovery. Our assumed explosion date is almost exactly a day earlier than the date assumed by \citet{Rui+2019}.

\subsection{Absolute Light and Color Curves}\label{photabs}

\begin{figure}
\plotone{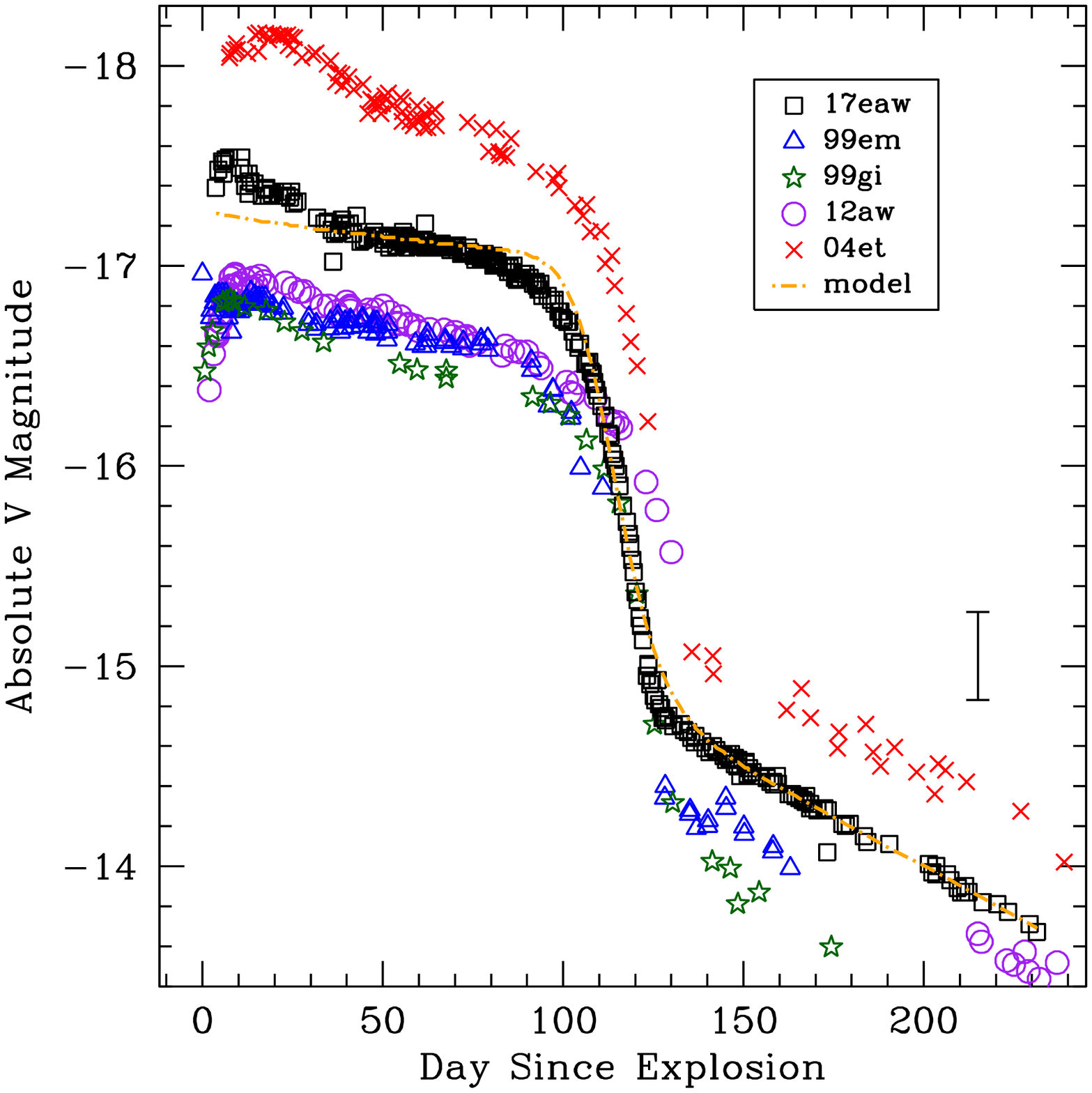} 
\caption{Absolute $V$ light curve of SN 2017eaw (black open squares). For comparison we show the absolute 
curves of SN 1999em \citep[][ blue open triangles]{Hamuy+2001,Leonard+2002a,Elmhamdi+2003b}, SN 1999gi 
\citep[][ dark green open stars]{Leonard+2002b}, SN 2004et \citep[][ red crosses]{Sahu+2006,Maguire+2010}, and SN 2012aw 
\citep[][ purple open circles]{Bose+2013,DallOra+2014}, all adjusted by the distances and reddenings in the literature, 
although SN 2004et was adjusted to our assumed distance to NGC 6946. 
The displayed error bar is representative of the average uncertainty in the 
SN 2017eaw curve, with the predominant source of error being the uncertainty in the adopted distance 
modulus to NGC 6946. Also shown is a simple model for the behavior of the light curve, following 
\citet[][ yellow dotted-dashed curve]{Elmhamdi+2003b}.\label{figlcabs}}
\end{figure}

We corrected the observed {\em BVRI\/} light curves of SN 2017eaw for the assumed Galactic foreground 
extinction and then adjusted them to be absolute light curves with our adopted distance modulus to NGC 6946.
As an illustration we show the absolute $V$ curve in Figure~\ref{figlcabs}.
At maximum $V=-17.57$ mag,
SN 2017eaw is generally more luminous than SN 1999em \citep{Hamuy+2001,Leonard+2002a,Elmhamdi+2003b}, SN 
1999gi \citep{Leonard+2002b}, and SN 2012aw \citep{Bose+2013,DallOra+2014}. 
However, it is intermediate in luminosity between these SNe and SN 
2004et (\citealt{Sahu+2006,Maguire+2010}; when adjusted to our assumed distance to NGC 6946), which appears to be 
significantly more luminous than SN 2017eaw.
(See also the comparison of SN 2017eaw with SN 2004et by \citealt{Tsvetkov+2018}.)
The bump in the light curve near maximum brightness
(at $\sim 8$ days after explosion) appears more prominent for SN 2017eaw, and not nearly as pronounced for the other SNe, 
although \citet{morozova:17a} required $\lesssim 0.3\ M_{\odot}$ of dense circumstellar matter (CSM) to account for enhanced 
emission in the early-time curves of SN 1999em, SN 2004et, and SN 2012aw.
(SN 2004et may have exhibited a less prominent, more extended bump peaking at about 20 days post-explosion.)
We have also fit via a minimum $\chi^2$ the simple model from \citet{Elmhamdi+2003b}, shown in the 
figure, for the behavior of the light curve. See Section~\ref{nickel} for the ramifications of this fit.

\begin{figure}
\plotone{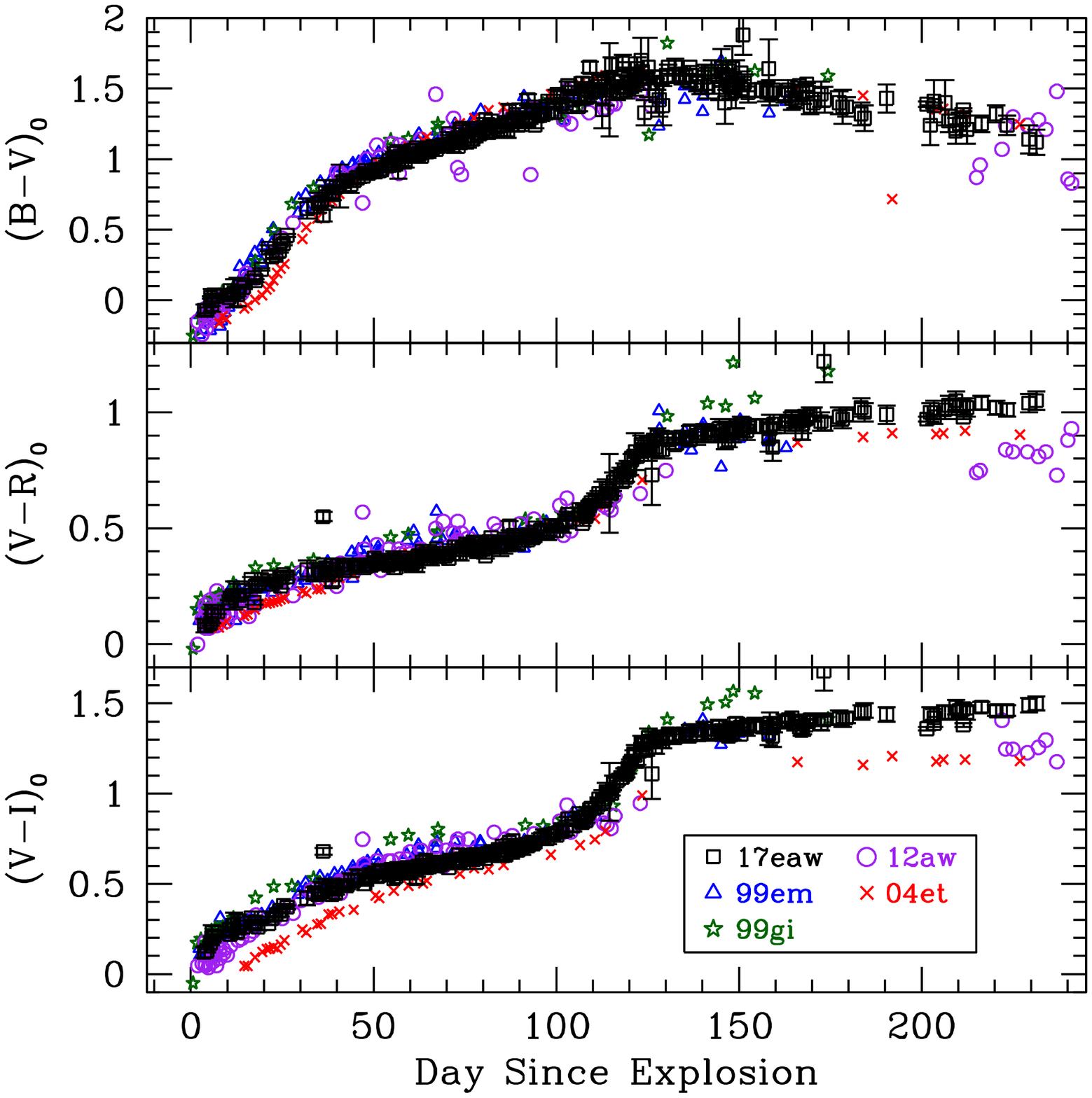} 
\caption{$B-V$, $V-R$, and $V-I$ color curves (black open squares) for SN 2017eaw, corrected for our 
assumed value of the reddening to the SN. For comparison we show the color curves for the SNe II-P 
1999em \citep[][ blue open triangles]{Hamuy+2001,Leonard+2002a,Elmhamdi+2003b}, 
1999gi \citep[][ dark green open stars]{Leonard+2002b}, 
2004et \citep[][ red crosses]{Sahu+2006}, and 
2012aw \citep[][ purple open circles]{Bose+2013,DallOra+2014}, all corrected by the reddening assumed by 
those studies and adjusted in time to match approximately the curves for SN 2017eaw.\label{figcolor}}
\end{figure}

The color evolution of SN 2017eaw is shown in Figure~\ref{figcolor}. Generally, reasonably close agreement 
exists between this SN and other SNe II-P, such as SN 1999em, SN 1999gi, SN 2004et, and SN 2012aw. SN 
1999gi may have been somewhat redder than SN 2017eaw in all colors, while SN 2004et appears to have been 
bluer in all of our observed colors, especially at early and later times.

\subsection{Bolometric Light Curve}\label{photbolo}

With an absence of photometry both shortward and longward of {\em BVRI}, 
we transformed the absolute observed light curves to a bolometric light curve via several different methods.
One method was to assume bolometric corrections to the broadband photometry derived from the
modeling of SNe~II-P by \citet{Pejcha+2015}. 
Similarly, we also used the bolometric corrections from \citet{Bersten+2009}.
Another was to assume bolometric corrections derived 
empirically and more generally for SNe II by \citet{Lyman+2014}, and more specifically for SN 2004et and SN 1999em
relative to the $R$ band by \citet{Maguire+2010}.
Finally, we produced a bolometric light curve from extrapolations of  
blackbody fits to the observed {\em BVRI\/} datapoints, using the routine \texttt{superbol} 
\citep{Nicholl2018}\footnote{https://github.com/mnicholl/superbol}.
The fitting with this routine was initially set relative to the observed $V$-band maximum, which was on
JD 2,457,893.06. 
We note that we just missed the actual $V$ maximum: fitting a Gaussian function approximately 
to the early bump in the observed light curve, we estimate that the time of $V$ maximum likely occurred 
around JD 2,457,893.7.

The ensemble of results from these various methods indicates a general trend for the bolometric light curve, and so we 
computed an average from all of these various results. However, the earliest portion of the curve ($\lesssim 25$ days) 
was established only from the average of the bolometric corrections from \citet{Bersten+2009} and \citet{Pejcha+2015} and the 
\texttt{superbol} fit, since the former two corrections tend to agree with the early blackbody fitting when the SN was still hot 
($T>8000$ K) and relatively free from emission lines. We consider it less likely that the early behavior of the curves resulting from 
the corrections from \citet{Maguire+2010} and \citet{Lyman+2014} adequately represents the actual early-time bolometric 
evolution of SN 2017eaw. In a forthcoming paper (V.~Morozova et al.~2019, in preparation), we will demonstrate that the 
bolometric light curve --- particularly the early peak --- is consistent with the presence of 
dense CSM immediately adjacent to the progenitor at the time of explosion. Such CSM is strongly suspected to be present
for a number of SNe II-P \citep{moriya:17,moriya:18,morozova:17,morozova:17a,foerster:18,paxton:18} and could be related to
pre-explosion outbursts during the late stages of RSG nuclear burning \citep[e.g.,][]{quataert:12,shiode:14,fuller:17}.
The behavior of the curve on the exponential tail is likely more consistent with that resulting from the bolometric correction from 
\citet{Lyman+2014} and the bolometric correction for SN 1999em from \citet{Maguire+2010} than the
\texttt{superbol} blackbody extrapolation, which is affected by the presence of strong spectral emission lines during this phase
and overpredicts the luminosity on the tail. The uncertainty in our average curve conservatively includes the individual 
uncertainties in the \texttt{superbol} fit and in the individual bolometric corrections.

For
comparison we also show the bolometric light curves for SN 2004et (\citealt{Maguire+2010,Faran+2018}; after adjusting 
their published curves from their respective assumed distances to the distance of NGC 6946 that we assume)
and SN 1987A \citep{Suntzeff+1990}. The former SN, again, appears to have been more luminous than SN 2017eaw,
except possibly at peak.

\begin{figure}
\plotone{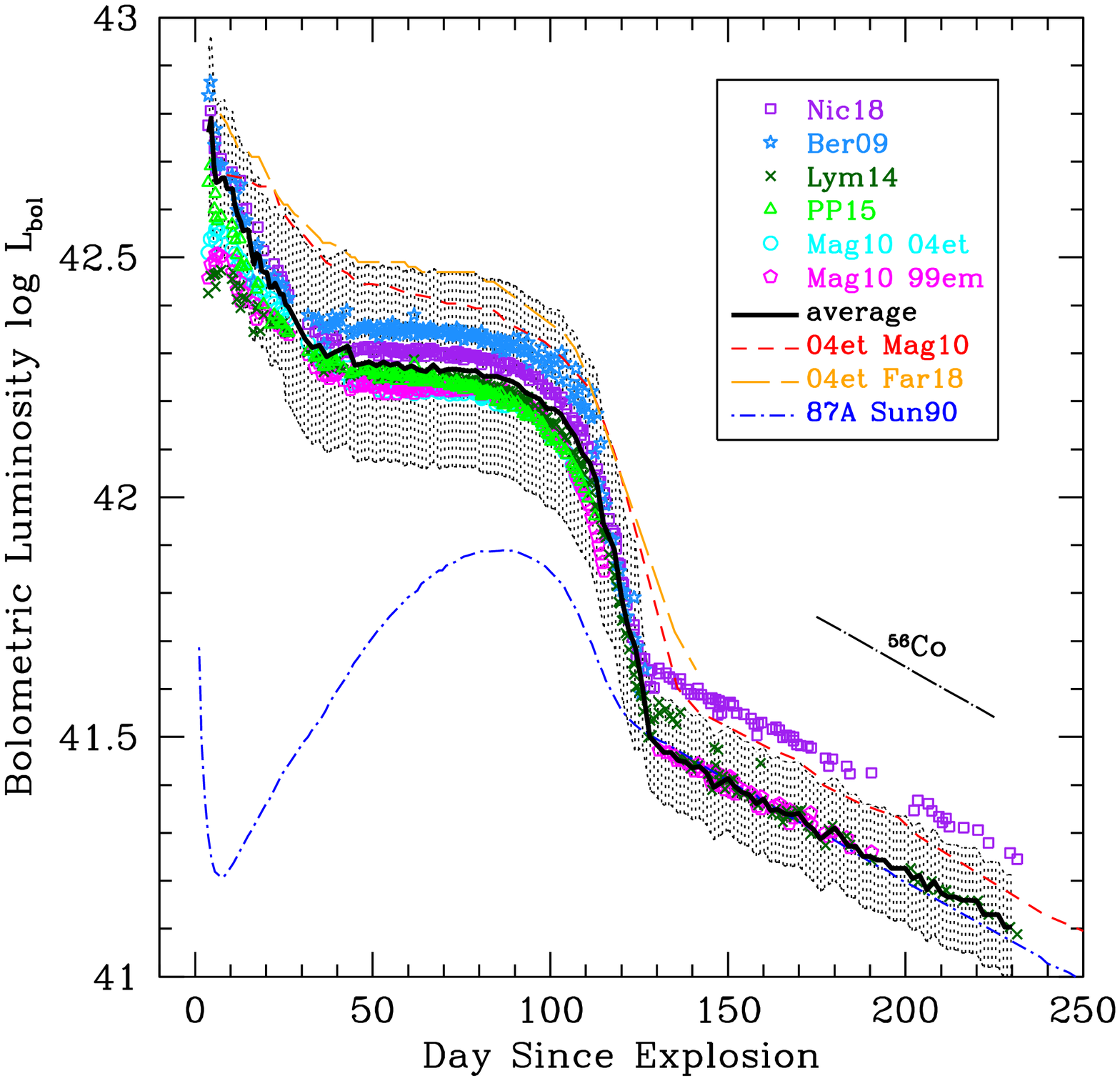} 
\caption{Bolometric light curve of SN 2017eaw, which we adopt as the average (solid curve) of various methods: The 
blackbody-fitting routine \texttt{superbol} \citep[][ Nic18, purple open squares]{Nicholl2018} and bolometric corrections from 
\citet[][ Ber09, light blue stars]{Bersten+2009}, 
\citet[][ Lym14, darkgreen crosses]{Lyman+2014}, \citet[][ PP15, green open triangles]{Pejcha+2015},
and \citet[][ Mag10, cyan open circles for SN 2004et and magenta open pentagons for SN 1999em, both
relative to $R$-band]{Maguire+2010}. The hashed region indicates the conservative 1$\sigma$ uncertainty in our adopted curve.
Also shown for comparison are bolometric light curves for SN 2004et from 
\citet[][ red short-dashed curve, adjusted to our assumed distance from their adopted distance of $5.9 \pm 0.4$ Mpc]{Maguire+2010}
and \citet[][ Far18, orange long-dashed curve, adjusted from their adopted distance of $4.81 \pm 0.16$ Mpc]{Faran+2018}, and of 
SN 1987A \citep[][ Sun90, dark-blue short-dashed-dotted curve]{Suntzeff+1990}. Also indicated is the expected decline rate for
the light-curve tail if it is powered primarily by the decay of $^{56}$Co (long-dashed-dotted curve).\label{figbolo}}
\end{figure}

\subsection{Estimate of the Nickel Mass}\label{nickel}

We can estimate the mass, $M({^{56}{\rm Ni}})$, of radioactive nickel, whose daughter is $^{56}$Co, the decay
of which powers the exponential tail of SN II-P light curves via deposition and trapping of $\gamma$-rays
released by the decay. For a given SN II-P, assuming that the $\gamma$-ray thermalization is equally
efficient, the comparison is usually made to SN 1987A,
the bolometric light curve of which \citep{Suntzeff+1990} we show in Figure~\ref{figbolo}.
The luminosities of the light-curve tails for the two 
SNe after about day 135 are comparable, if not the same, to within the uncertainties. The nickel mass for SN 1987A was 
estimated at $M_{\rm Ni}=0.075 \pm 0.015\ M_{\odot}$ \citep{Arnett+1989}, so it is likely safe to assume that $M_{\rm Ni}$ for
SN 2017eaw is essentially the same, based on this comparison. Another means of estimating $M_{\rm Ni}$ is
via a ``steepness'' parameter,
$S$, introduced by \citet{Elmhamdi+2003a} for the $V$-band light curve. We show a best fit of 
their simple analytical model for the behavior of the $V$ curve in Figure~\ref{figlcabs}. From that model
fit we find that $S=0.089$ mag day$^{-1}$ at an ``inflection point'' of day 113.6. 
From \citet[][ their Equation (3)]{Elmhamdi+2003a} we then estimate that $M_{\rm Ni}=0.04\ M_{\odot}$, which is
about a factor of two less than the SN 1987A comparison. \citet{Rho+2018} found that dust had started forming as early as 
day 124, so there may be an increase in extinction local to the
SN that could decrease the luminosity of the exponential tail.
However, we note that \citeauthor{Rho+2018} arrived at 
satisfactory model fits of their near-infrared nebular spectra
assuming that $M_{\rm Ni}= 0.084\ M_{\odot}$, which is consistent with the SN 1987A-based value.
The uncertainty in $M_{\rm Ni}$ for SN 2017eaw, based on our bolometric light curve, is $\sim \pm 0.03\ M_{\odot}$.
An estimate of $\sim 0.07$--$0.08\ M_{\odot}$ is consistent with the trend that \citet{Valenti+2016} (their Figure
22) found for SNe II, given $M_V=-17.13$ at day 50 for SN 2017eaw.

\subsection{Spectral Analysis}\label{specanal}

We compared our spectra of SN 2017eaw with the template spectra in both \texttt{SNID} \citep{Blondin+2007} and 
\texttt{GELATO} \citep{Harutyunyan+2008}.
With \texttt{SNID} the SN 2017eaw spectra compared best with other SNe II-P, such as SN 2006bp, SN 1999em, 
SN 2005dz, SN 2004fx, SN 2004et, and (at later times) even with SN 1987A.
For the \texttt{GELATO} comparison the best matches were with SN 1999gi and, most often, SN 2004et.

\begin{figure}
\plotone{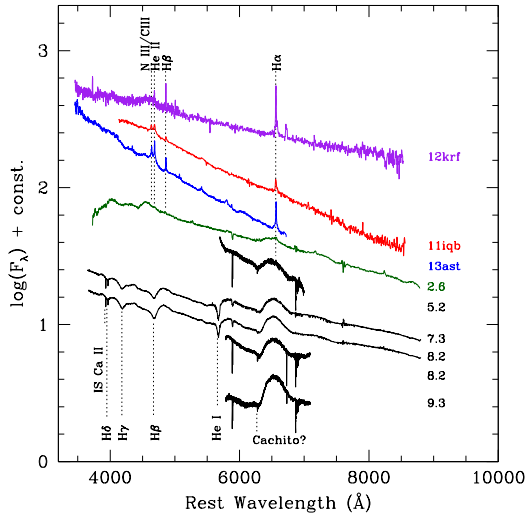} 
\caption{Spectra of SN 2017eaw in the first $\sim9$ days after explosion. For comparison we show the 
spectra of three SNe~II that showed flash-ionization features, indicative of the presence of 
CSM --- PTF12krf \citep{Khazov+2016}, PTF11iqb \citep{Smith+2015,Khazov+2016}, and 
iPTF13ast \citep[SN 2013cu; ][]{Gal-Yam+2014,Khazov+2016}. The ages, in days, of the SN 2017eaw spectra are 
labeled. The spectra of PTF12krf, PTF11iqb, and iPTF13ast are from days 4$^{+1}_{-1}$, 2.1$^{+0}_{-1.1}$, and
3$^{+0}_{-0.2}$, respectively. Various emission and absorption features are indicated.\label{figspecearly}}
\end{figure}

At very early times ($\le 10$ days) some 
SNe II-P have shown ``flash'' features in their spectra indicative of an explosion within dense CSM.
The features arise from recombination of the CSM ionized by the UV/X-ray flash from shock breakout
\citep{Khazov+2016}.
(We mention that \citealt{kochanek:18}, however, has explained the flash spectral features via a collision
interface formed between the regular stellar winds in a binary system.)
In Figure~\ref{figspecearly} we show our SN 2017eaw spectra within the first nine days or so after explosion.
We have also included here the earliest available spectrum of SN 2017eaw of which we are aware, obtained 
by \citet{Xiang+2017}\footnote{Posted on the Transient Name Server, https://wis-tns.weizmann.ac.il/object/2017eaw.} 
and shown by \citet{Rui+2019}.
Xiaofeng Wang graciously permitted us to present this spectrum in the figure.
We compare these spectra with some examples that have exhibited prominent flash features\footnote{These 
spectra have been obtained from WISEReP, https://wiserep.weizmann.ac.il/ \citep{Yaron+2012}.}, the SNe II 
PTF12krf and PTF11iqb (\citealt{Khazov+2016}; also \citealt{Smith+2015}), as well as the SN IIb iPTF13ast 
\citep[SN 2013cu; ][]{Gal-Yam+2014,Khazov+2016}, which is among the first known instances of this
phenomenon.
The spectra of PTF12krf, PTF11iqb, and iPTF13ast are from days 4$^{+1}_{-1}$, 2.1$^{+0}_{-1.1}$, and
3$^{+0}_{-0.2}$, respectively.
Various emission features --- H$\alpha$, H$\beta$, He~{\sc ii} $\lambda$4686, and an
N~{\sc iii} $\lambda\lambda$4634, 4640/C~{\sc iii} $\lambda\lambda$4647, 4650, 4651
blend --- are all strong in the early spectra of the latter three SNe.
However, for SN 2017eaw there may be only quite weak indications of these features in the 
spectrum by \citeauthor{Xiang+2017}~and in our earliest MMT spectrum from day 5.2, in particular at H$\alpha$. 
\citet{Rui+2019} argued that in the day 2.6 spectrum the weak H$\alpha$ feature is indicative of slow-moving
(163 km s$^{-1}$) CSM, although we see no evidence of this in our day 5.2 spectrum, and instead only weak H$\alpha$ near zero
rest-frame velocity.
By the day 7.3 spectrum any traces of these features have vanished.

It is possible therefore that either spectra were not obtained sufficiently early and the features were 
missed, or these flash features were just intrinsically weaker for SN 2017eaw than for the notable cases 
that do show them. If the latter is the case, then, as noted earlier, an implication is that any CSM 
in immediate proximity to SN 2017eaw may have been too dense and massive for these features to form.
Alternatively, narrow lines might have been hidden if the CSM was asymmetric, as in the case of PTF11iqb, 
which had early SN~IIn-like signatures present and an enhanced early luminosity peak that declined to a 
plateau \citep{Smith+2015}. In order to explain the rapid disappearance of the narrow lines, while the excess 
CSM interaction luminosity remained, \citet{Smith+2015} proposed that a highly asymmetric CSM geometry, such 
as a disk, could allow the opaque SN ejecta to envelop the CSM interaction occurring in a disk.  In that 
scenario, CSM interaction would continue to generate luminosity, which would heat the SN ejecta and produce a 
larger emergent luminosity, but it would be buried inside the opaque ejecta, so that the narrow lines are 
hidden from view. This type of asymmetric CSM interaction might also help to explain the lack of narrow lines 
during the initial peak of SN 2017eaw.

\begin{figure}
\plotone{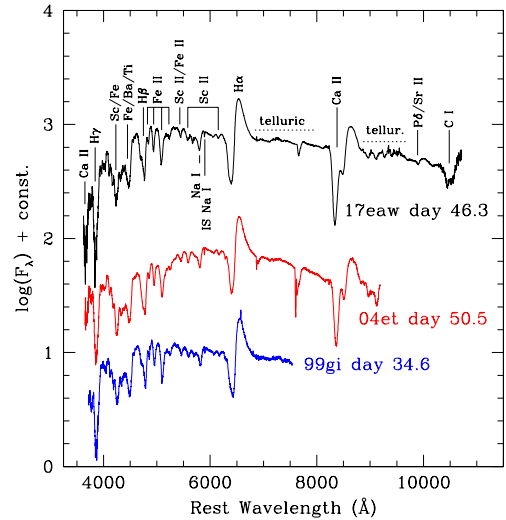}  
\caption{Comparison of the day 46.3 spectrum of SN 2017eaw with spectra of two SNe~II-P, SN 2004et 
\citep{Sahu+2006}, and SN 1999gi \citep{Leonard+2002b}, at comparable ages during the plateau phase. 
Various spectral lines and features are indicated. Telluric lines have not been removed from the
spectrum of SN 2004et, and may be weakly present in the spectrum of SN 2017eaw.\label{figspeccomp}}
\end{figure}

\begin{figure}
\plotone{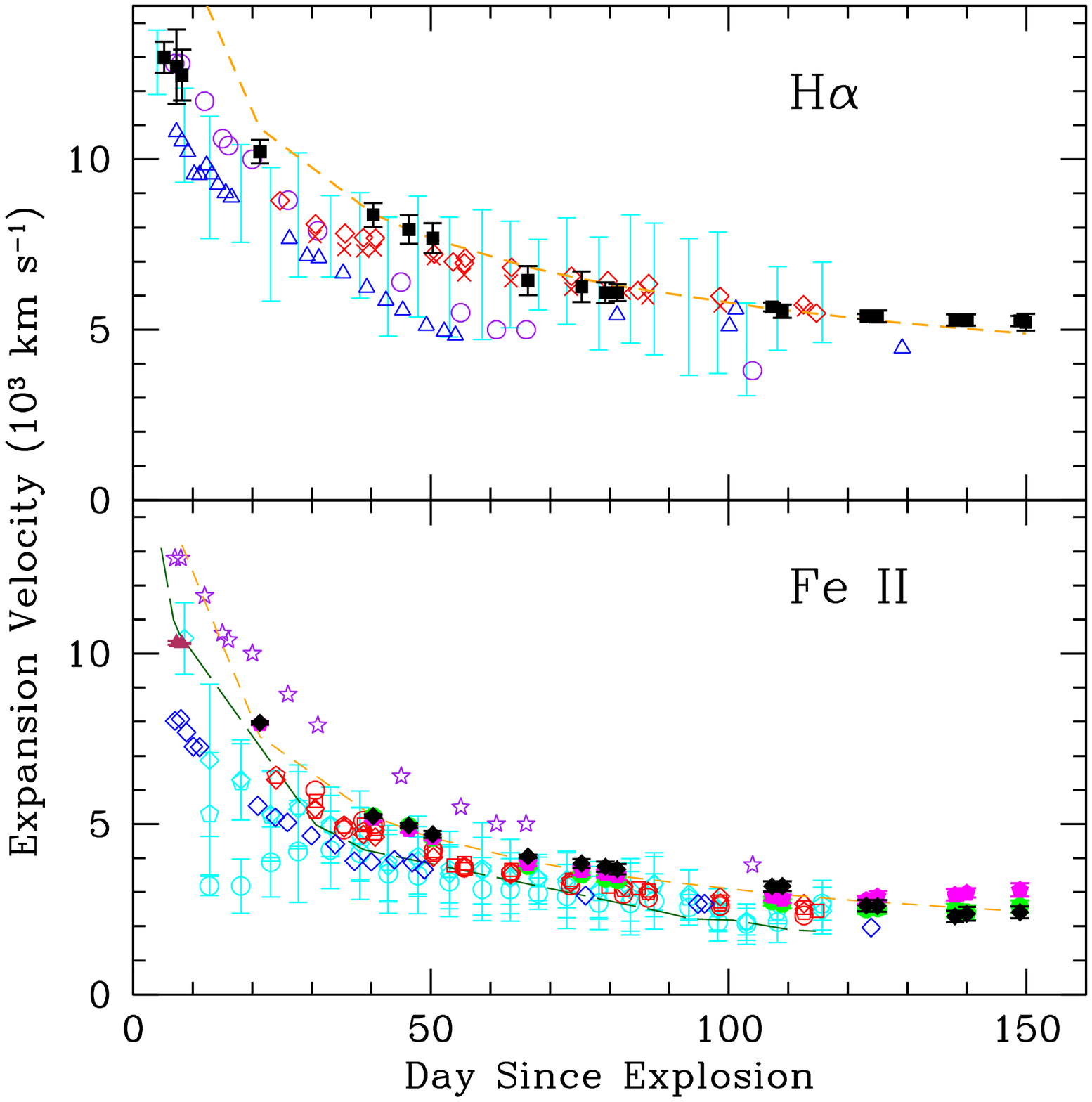} 
\caption{Evolution of the expansion velocity of SN 2017eaw. {\it Top panel}: From H$\alpha$
absorption, solid black points; data for SN 1999em (\citealt{Leonard+2002a}, open blue triangles), SN 2012aw
(\citealt{Bose+2013}, open purple circles), SN 2004et (\citealt{Sahu+2006}, red crosses; \citealt{Maguire+2010}, open
red diamonds), and the range of the sample of 96 SNe II-P from \citet{Gutierrez+2017} (cyan error bars) shown for comparison. 
 {\it Bottom panel}: from Fe~{\sc ii} line absorption, solid green circles ($\lambda$4924), 
solid magenta pentagons ($\lambda$5018), and solid black diamonds ($\lambda$5169); data for SN 1999em (\citealt{Leonard+2002a}, 
$\lambda$5169, open blue diamonds), SN 1999gi (\citealt{Leonard+2002b}, photospheric velocity, dark green long-dashed line), 
SN 2012aw (\citealt{Bose+2013}, all spectral lines, open purple stars), SN 2004et (\citealt{Sahu+2006}, $\lambda$4924, red open circles,
$\lambda$5018, red open pentagons, $\lambda$5169, red open diamonds; \citealt{Maguire+2010}, average, red open squares),
and the sample of \citet{Gutierrez+2017} ($\lambda$4924, cyan open circles, $\lambda$5018, cyan open pentagons, $\lambda$5169, cyan 
open diamonds). The measurements for SN 2017eaw from He~{\sc i} $\lambda$5876 absorption are shown as maroon solid triangles.
Also shown in both panels are the trends for SNe II-P found by \citet{Faran+2014} (orange short-dashed line).\label{figspecvel}}
\end{figure}

\begin{figure}
\plotone{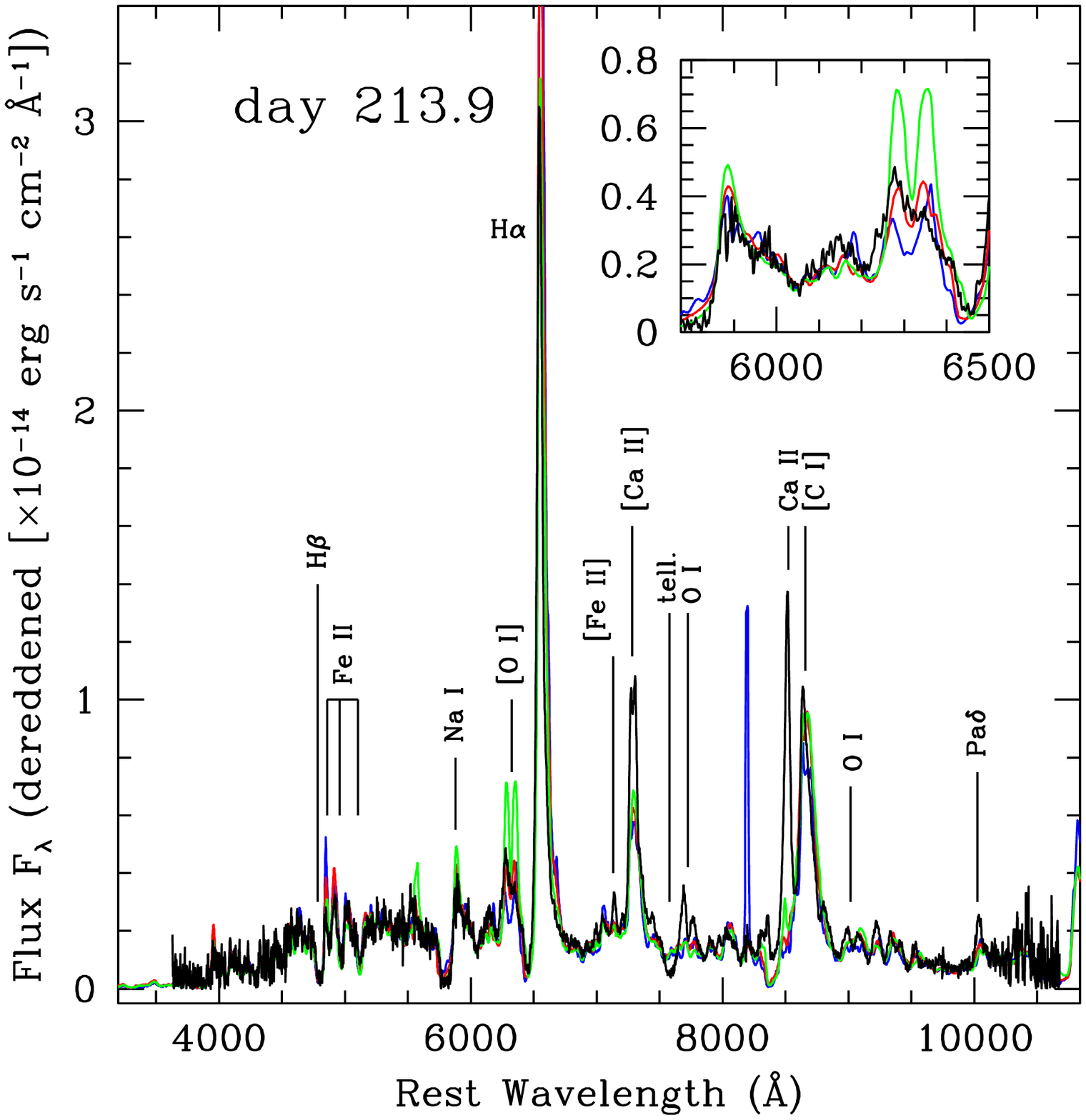}
\caption{Lick/Kast low-resolution spectrum of SN 2017eaw obtained on day 213.9 (black curve); see 
Section~\ref{specobs}. The spectrum has been dereddened following our assumed value for the reddening to 
the SN (Section~\ref{reddening}).
We compare the SN 2017eaw spectrum with model spectra on day 212 from \citet{Jerkstrand+2012}, assuming a progenitor 
at $M_{\rm ZAMS}=12\ M_{\odot}$ (blue curve), 
$15\ M_{\odot}$ (red curve), and $19\ M_{\odot}$ (green curve). The model spectra have been scaled in flux to
the observed spectrum. Various spectral lines and 
features are indicated. The inset in the figure focuses on the spectral region containing the Na~{\sc i}~D 
$\lambda\lambda$5890, 5896 and [O~{\sc i}] $\lambda\lambda$6300, 6364 lines.\label{figspeclate1}}
\end{figure}

\begin{figure}
\plotone{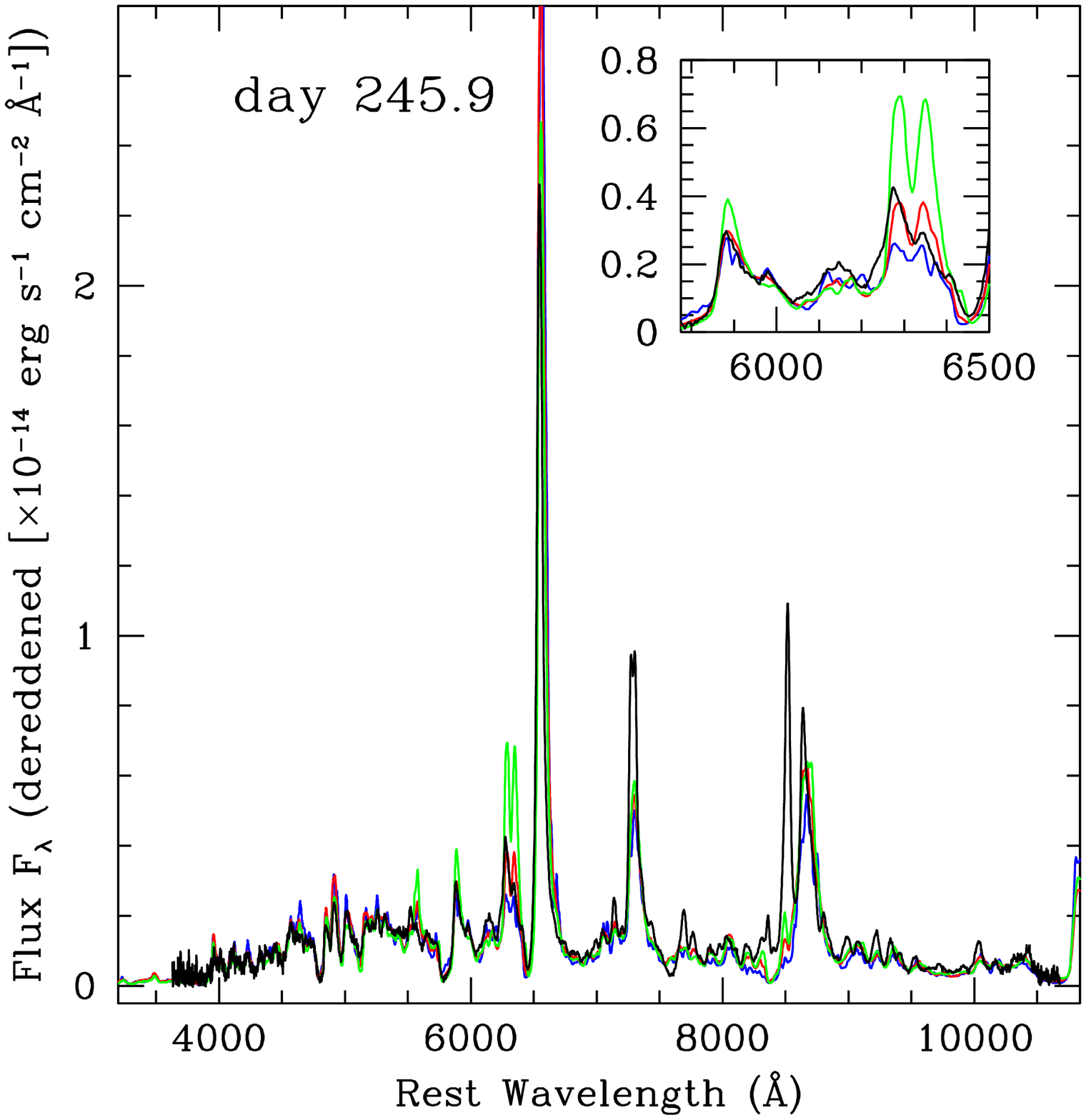} 
\caption{Lick/Kast low-resolution spectrum of SN 2017eaw obtained on day 245.9 (black curve); see 
Section~\ref{specobs}. The spectrum has been dereddened following our assumed value for the reddening to 
the SN (Section~\ref{reddening}). 
We compare the SN 2017eaw spectrum with model spectra on day 249 from \citet{Jerkstrand+2012}, 
assuming a progenitor at $M_{\rm ZAMS}=12\ M_{\odot}$ (blue curve), 
$15\ M_{\odot}$ (red curve), and $19\ M_{\odot}$ (green curve). The model spectra have been scaled in flux to
the observed spectrum.
Similar spectral features are seen as in Figure~\ref{figspeclate1}.
The inset in the figure focuses on the spectral region containing the Na~{\sc i}~D $\lambda\lambda$5890, 
5896 and [O~{\sc i}] $\lambda\lambda$6300, 6364 lines.\label{figspeclate2}}
\end{figure}

\begin{figure}
\plotone{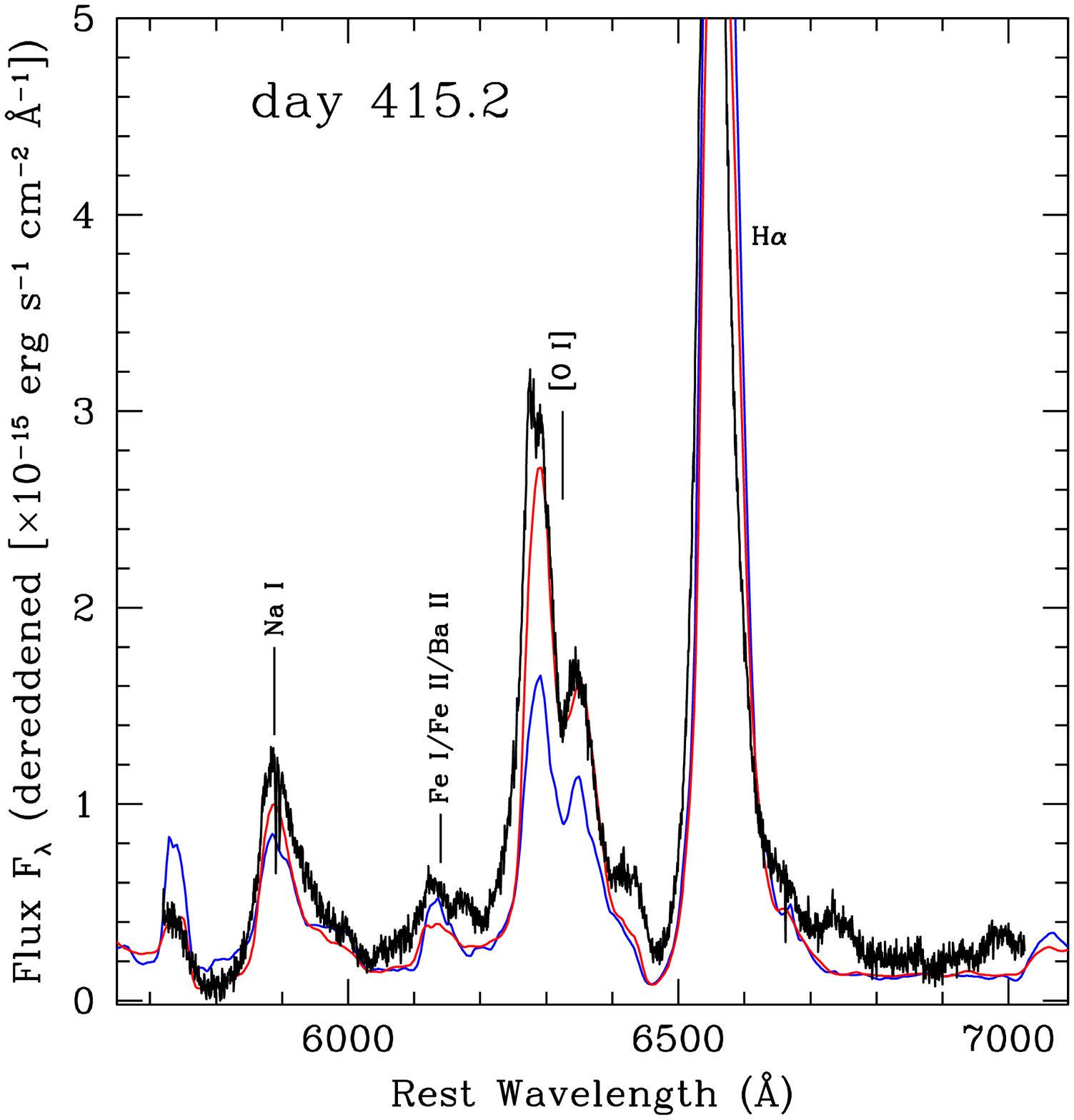} 
\caption{MMT moderate-resolution spectrum of SN 2017eaw obtained on day 415.2 (black curve); see 
Section~\ref{specobs}. 
The spectrum has been dereddened following our assumed value for the reddening to the SN 
(Section~\ref{reddening}).
This spectrum only covers the region containing the Na~{\sc i}~D $\lambda\lambda$5890, 5896 and [O~{\sc i}] 
$\lambda\lambda$6300, 6364 lines (plus H$\alpha$).
We compare the SN 2017eaw spectrum with model spectra generated by \citet{Jerkstrand+2012} 
at progenitor mass $M_{\rm ZAMS}=12\ M_{\odot}$ (blue curve) and 
$15\ M_{\odot}$ (red curve) on day 400. (No model is available at $M_{\rm ZAMS}=19\ M_{\odot}$ for this 
epoch.) The model spectra have been scaled in flux to the observed spectrum. Various spectral lines and 
features are indicated.\label{figspeclater}}
\end{figure}

The primary absorption features in the early-time spectra are the Balmer lines, He~{\sc i} $\lambda$5876, 
the strong Na~{\sc i}~D line (discussed above), and what is likely interstellar Ca~{\sc ii}, also due to 
the Galactic foreground.
The Balmer profiles continue to develop their characteristic P-Cygni-like appearance over time, and 
He~{\sc i} is no longer visible by or before day 39.
It is possible that in our moderate-resolution data we see (weakly) the ``Cachito'' feature discussed by 
\citet{Gutierrez+2017} to the blue side of H$\alpha$. However, the feature as observed in SN\,2017eaw is 
comprised not of one wide absorption, but of two narrow (FWHM $\approx 100$ km\,s$^{-1}$)
absorption minima that evolve very little over time, with respect to both profile substructure and wavelength position. The 
position overlaps with a grouping of O$_2$ absorption lines centered near 6280 \AA, so some telluric 
contamination is possible. 
If the feature is associated with H$\alpha$, then its velocity in the earliest spectrum is $-$13,300 
km\,s$^{-1}$.  The feature is not discernible in our spectra beyond day 47.9. We note that this feature does 
bear striking resemblance to persistent absorptions blueshifted with respect to H$\alpha$ that are sometimes observed 
in SNe~IIb \citep{Milisavljevic+2013}. The origin of these high-velocity features is unclear, and 
explanations involving ejecta asymmetry, a mixture with Fe~{\sc ii} or Si~{\sc ii} lines, and interaction 
with CSM have been put forward \citep{Baron+1994,ZW1996,JB2010,Milisavljevic+2013}. 

At later times during the plateau phase, the spectra evolved gradually. We show an example spectrum in
Figure~\ref{figspeccomp} from day 46.3. 
When we compare this spectrum using \texttt{SNID} we obtain a best match with SN 2004et \citep{Sahu+2006}. 
Using \texttt{GELATO} the best comparison is with SN 1999gi \citep{Leonard+2002b}. 
We show 
the spectra for these two other SNe~II-P in the figure for comparison\footnote{Again, obtained from 
WISEReP.}. The comparison shows that, spectroscopically, SN 2017eaw is evidently quite 
normal.
We have indicated various features in these spectra (including a few telluric lines primarily in
the spectrum of SN 2004et), following \citet{Gutierrez+2017}: the
Balmer profiles, Na~{\sc i}, Ca~{\sc ii}, and various metal lines of Fe~{\sc ii}, Sc~{\sc ii}, and blends.
We may also see evidence for the C~{\sc i} $\lambda$10,691 line.

We measured the evolution of the expansion velocity of SN 2017eaw from our multi-epoch spectra, starting with 
our earliest spectrum on day 5.2 through well off the plateau on day 151.9. Velocities were estimated from 
the minimum of the absorption features of the H$\alpha$, He~{\sc i} 
$\lambda$5876, and Fe~{\sc ii} $\lambda\lambda$4924, 5018, 5169 lines.
Multiple measurements were performed to estimate uncertainties. See Figure~\ref{figspecvel}.
We compared the evolution of the 
expansion velocity of SN 2017eaw to the sample of 96 SNe~II from \citet{Gutierrez+2017}, as well as to other 
well-studied SNe II, including SN 1999em \citep{Leonard+2002a}, SN 1999gi \citep{Leonard+2002b},
SN 2004et \citep{Sahu+2006,Maguire+2010}, and SN 2012aw \citep{Bose+2013}.
We also show the trends in velocity evolution from \citet{Faran+2014}.
The expansion velocities of SN 2017eaw  are generally within the mean of all SNe II-P, although the expansion 
velocities from Fe~{\sc ii} $\lambda$5169 are somewhat higher, particularly at the earliest epochs.

We can also analyze the spectra at ages that are among the latest available (213.9, 245.9, and 415.2 days) in light 
of the modeling of the nebular spectra of SN 2004et by \citet{Jerkstrand+2012} at comparable 
ages\footnote{We do not analyze the 482.1 day spectrum, since the closest model in time is from day 451.}.
(These models were also applied to the nebular spectra of SN 2012aw; \citealt{Jerkstrand+2014}.)
We show these three observed spectra once again in Figures~\ref{figspeclate1}, \ref{figspeclate2}, and 
\ref{figspeclater}, after dereddening them 
(see Section~\ref{reddening}; assuming the reddening law of \citealt{Cardelli+1989} with $R_V=3.1$), 
along with the model 
spectra\footnote{Obtained from https://star.pst.qub.ac.uk/webdav/public/\\ ajerkstrand/Models/Jerkstrand+2014/.} 
for initial masses $M_{\rm ZAMS}=12$, 15, and $19\ M_{\odot}$ at days 212, 250, and 400, 
respectively (a $19\ M_{\odot}$ model on day 400 is not available).
For each figure we have photometrically scaled the model spectra to the observed one, via \texttt{pysynphot}, using the light-curve
data closest in time to the observed spectrum on day 213.9; for the day 245.9 and day 415.2 spectra, we had to linearly extrapolate 
the light curves at $V$ and $R$, respectively.
We especially focus in the figure on the spectral region containing the Na~{\sc i}~D 
$\lambda\lambda$5890, 5896 and [O~{\sc i}] $\lambda\lambda$6300, 6364 lines, whose strengths (as
\citealt{Jerkstrand+2012} pointed out) are most influenced by the assumed progenitor initial mass.
One can see that the observed spectra at all
three epochs are best matched by the $15\ M_{\odot}$ model, with the $12\ M_{\odot}$ model generally underpredicting and the 
$19\ M_{\odot}$ model overpredicting the line strengths (on days 213.9 and 245.9), particularly for [O~{\sc i}]. In total, the 
implication of this comparison is that the progenitor initial mass of SN 2017eaw is closest to $15\ M_{\odot}$.

\subsection{SN-based Distance Estimates}\label{epmscm}

\begin{deluxetable*}{cccccccccc}                    
\tablewidth{0pt}
\tablecolumns{10}
\tablenum{4}
\tablecaption{Quantities Derived from the EPM Analysis of SN 2017eaw\tablenotemark{a}\label{tabepm}}
\tablehead{
\colhead{Age} & \colhead{$\theta_{BV}$} & \colhead{$T_{BV}$} & \colhead{$\zeta_{BV}$} 
& \colhead{$\theta_{BVI}$} & \colhead{$T_{BVI}$} & \colhead{$\zeta_{BVI}$}
& \colhead{$\theta_{VI}$} & \colhead{$T_{VI}$} & \colhead{$\zeta_{VI}$} \\
\colhead{(days)} & \colhead{($10^8$ km Mpc$^{-1}$)} & \colhead{(K)} & \colhead{} 
& \colhead{($10^8$ km Mpc$^{-1}$)} & \colhead{(K)} & \colhead{}
& \colhead{($10^8$ km Mpc$^{-1}$)} & \colhead{(K)} & \colhead{}}
\startdata
21.3 & 20.67(0.40) & 7473(189) & 0.716(0.021) & 21.47(0.40) & 8041(96)   & 0.595(0.005) & 21.52(0.36) & 8578(133) & 0.543(0.003) \\
40.3 & 26.27(3.33) & 5032(627) & 1.256(0.243) & 25.55(2.71) & 6056(339) & 0.774(0.051) & 24.84(0.88) & 7227(352) & 0.594(0.021) \\
46.3 & 26.54(0.84) & 4818(123) & 1.350(0.058) & 26.22(0.92) & 5730(80)   & 0.828(0.015) & 26.04(0.56) & 6739(163) & 0.629(0.014) \\
50.3 & 26.94(0.52) & 4684(73)   & 1.416(0.038) & 26.51(0.55) & 5582(47)   & 0.857(0.010) & 26.40(0.27) & 6573(76)   & 0.643(0.007) \\
\enddata
\tablenotetext{a}{Uncertainties are in parentheses.}
\end{deluxetable*}

Primarily to provide a check on our TRGB distance estimate to the host galaxy (see Section~\ref{distance}), 
we also estimated distances to SN 2017eaw itself through the SCM \citep{Hamuy+2002,Nugent+2006} and the
EPM \citep{Kirshner+1974,Eastman+1989,Schmidt+1992}.
To compute the SCM distance, we closely followed the technique of \citet{Polshaw+2015} and input the values
directly drawn from our data into their Equation 1: the $I$ brightness on day 50, $I_{50} = 12.29 \pm 0.01$ mag; 
the extinction at $I$ to SN 2017eaw, $A_I = 0.517$ mag (with a very conservative uncertainty of 0.1 mag); and, 
the expansion velocity on day 50, $v_{50} = 4697 \pm 103$ km s$^{-1}$, which we estimated from the Fe~{\sc ii} 
$\lambda$5169 absorption line in our spectrum on day 50.3 ($\sim$ day 50; see Figure~\ref{figspecvel}).  
For all other constants and their uncertainties, 
we adopted the values provided by \citeauthor{Polshaw+2015}
Our resulting distance estimate from SCM is then $7.32 \pm 0.60$ Mpc.

For the EPM distance estimate, we only considered spectral epochs earlier than 60 days (i.e., 21.3, 40.8, 46.7, and 50.2 days) 
after explosion that included the Fe {\sc ii} $\lambda$5169 absorption feature. We adopted the
velocities measured from that spectral feature as representing the photospheric velocity, $v_{\rm phot}$.
We considered the photometry data of the SN that were as contemporaneous as possible with the spectral epochs. We assumed our
estimate of the extinction to the SN (Section~\ref{reddening}).
We also adopted the dilution factors from \citet{Dessart+2005}. 
(We note that dilution factors have also been generated more recently by \citealt{Vogl+2019} and are in good agreement 
with those of \citeauthor{Dessart+2005}.)
To implement the EPM
to estimate the theoretical angular size, $\theta$, of the photosphere, we closely followed the procedure detailed by
\citet{Leonard+2002b}. We carried out this procedure for each of three bandpass combinations ($BV$, $BVI$, and
$VI$) to determine the photospheric temperatures ($T$) and appropriate dilution factors ($\zeta$); see Table~\ref{tabepm}.

\begin{figure}
\plotone{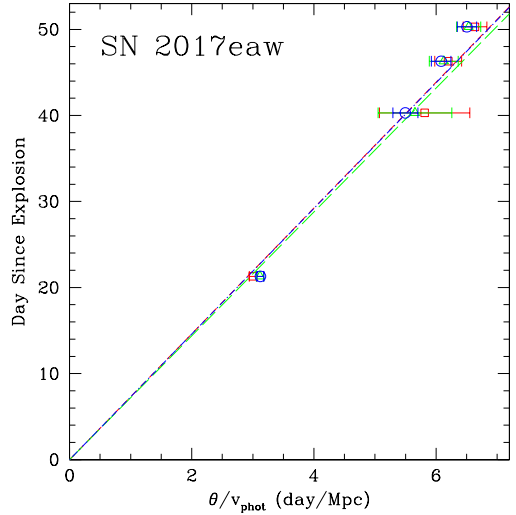} 
\caption{EPM fitting for SN 2017eaw for three combinations of bandpasses, $BV$ (red open squares, short-dashed line),
$BVI$ (green open triangles, long-dashed line), and $VI$ (blue open circles, short-dashed-dotted line). We have fixed the
explosion time, $t_0$, to our adopted date of JD 2,457,885.7 (Section~\ref{explosion}). See text for further details. \label{figepm}}
\end{figure}

To estimate the distances using each of the bandpass combinations, we fixed the explosion date ($t_0$) to be
our adopted value, JD 2,457,885.7 (see Section~\ref{explosion}). We feel confident that we can do this, given how
well the rise time of the SN appears to be constrained. We then merely measured the slope of the best-fitting line
(via weighted least-squares) describing the relation between the day since explosion and the ratio
$\theta/v_{\rm phot}$ for each of the three combinations; see Figure~\ref{figepm}. 
Uncertainties in each of the three slopes were established by determining the 1$\sigma$ dispersion in model slopes
generated from 1000 simulated datasets characterized by the parameter values and uncertainties in 
Table~\ref{tabepm}, together with the uncertainty in the assumed reddening and a flat likelihood of $\pm 1$~day uncertainty
around $t_0$ (which contributed $\sim 0.25$ Mpc to the total uncertainty). We then calculated
the unweighted mean of the three slopes (which were all quite similar in value, $\sim 7.3$ Mpc, as can be seen in the
figure) as our final EPM distance estimate. Since the three distance values are not
independent, we conservatively report the final uncertainty as the  sum in quadrature of the largest uncertainty in an 
individual distance and the 1$\sigma$ dispersion in the three individual distances.
Our distance estimate from EPM is then $7.27 \pm 0.42$ Mpc.

We note that both of the uncertainties given for the SCM and EPM estimates are purely statistical in nature and do not 
reflect potential systematics inherent in both techniques. Nevertheless, these estimates appear to corroborate the
TRGB distance we have estimated for NGC 6946 (see Section~\ref{distance}). 

\section{The SN Progenitor}

\subsection{Identification of the Progenitor}\label{progenitorid}

\begin{deluxetable}{cccc}
\tablewidth{0pt}
\tablecolumns{4}
\tablenum{5} 
\tablecaption{Photometry of the SN 2017eaw Progenitor\label{tabprog}}
\tablehead{
\colhead{Instrument} & \colhead{Band} & \colhead{Obs.~Date} &\colhead{Magnitude}}
\startdata
{\sl HST\/} ACS/WFC & F606W & 16-10-26 & 26.40(05) \\
{\sl HST\/} ACS/WFC & F658N & 04-07-29 & $>$24.6 \\
{\sl HST\/} ACS/WFC & F814W & 04-07-29 & 22.60(04) \\
{\sl HST\/} ACS/WFC & F814W & 16-10-26 & 22.87(01) \\
{\sl HST\/} WFC3/IR & F110W & 16-02-09 & 20.32(01) \\
{\sl HST\/} WFC3/IR & F128N & 16-02-09 & 19.69(02) \\
{\sl HST\/} WFC3/IR & F160W & 16-10-24 & 19.36(01) \\
{\sl Spitzer\/} IRAC & 3.6 $\mu$m & 15-12-23 -- 17-03-31\tablenotemark{a} & 18.01(03) \\
{\sl Spitzer\/} IRAC & 4.5 $\mu$m & 15-12-23 -- 17-03-31\tablenotemark{a} & 17.80(04) \\
{\sl Spitzer\/} IRAC & 5.8 $\mu$m & 04-06-10 -- 08-07-18\tablenotemark{a} & $>$16.15 \\
{\sl Spitzer\/} IRAC & 8.0 $\mu$m & 04-06-10 -- 08-01-27\tablenotemark{a} & $>$15.47 \\
{\sl Spitzer\/} MIPS & 24 $\mu$m & 04-07-09 -- 04-07-11\tablenotemark{a} & $>$10.39 \\
\enddata
\tablenotetext{a}{Consists of a coaddition of data obtained in this band during the indicated date range.}
\end{deluxetable}

\begin{figure*}[htp] 
\centering
\includegraphics[width=0.37\textwidth]{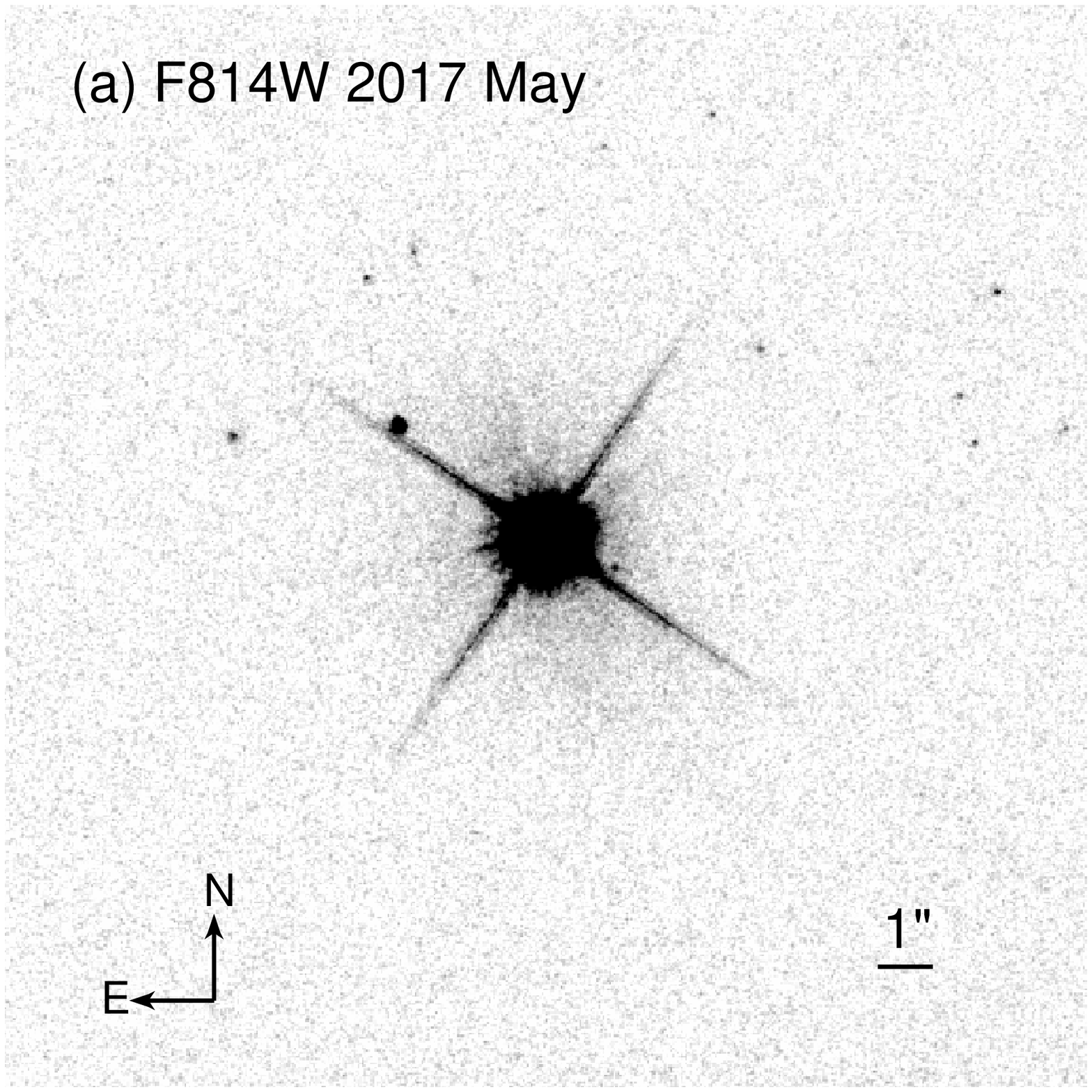}\quad
\includegraphics[width=0.37\textwidth]{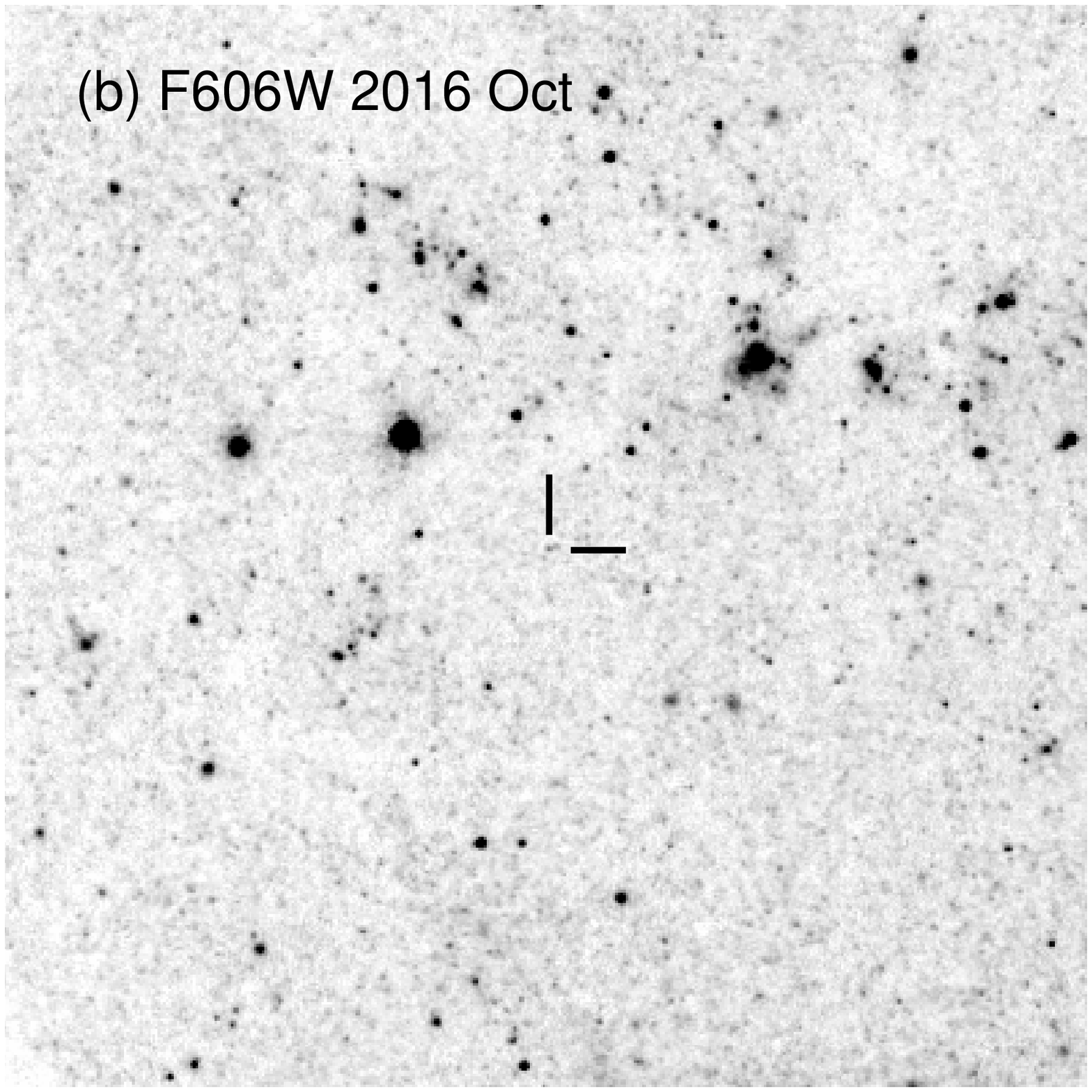}
\medskip
\includegraphics[width=0.37\textwidth]{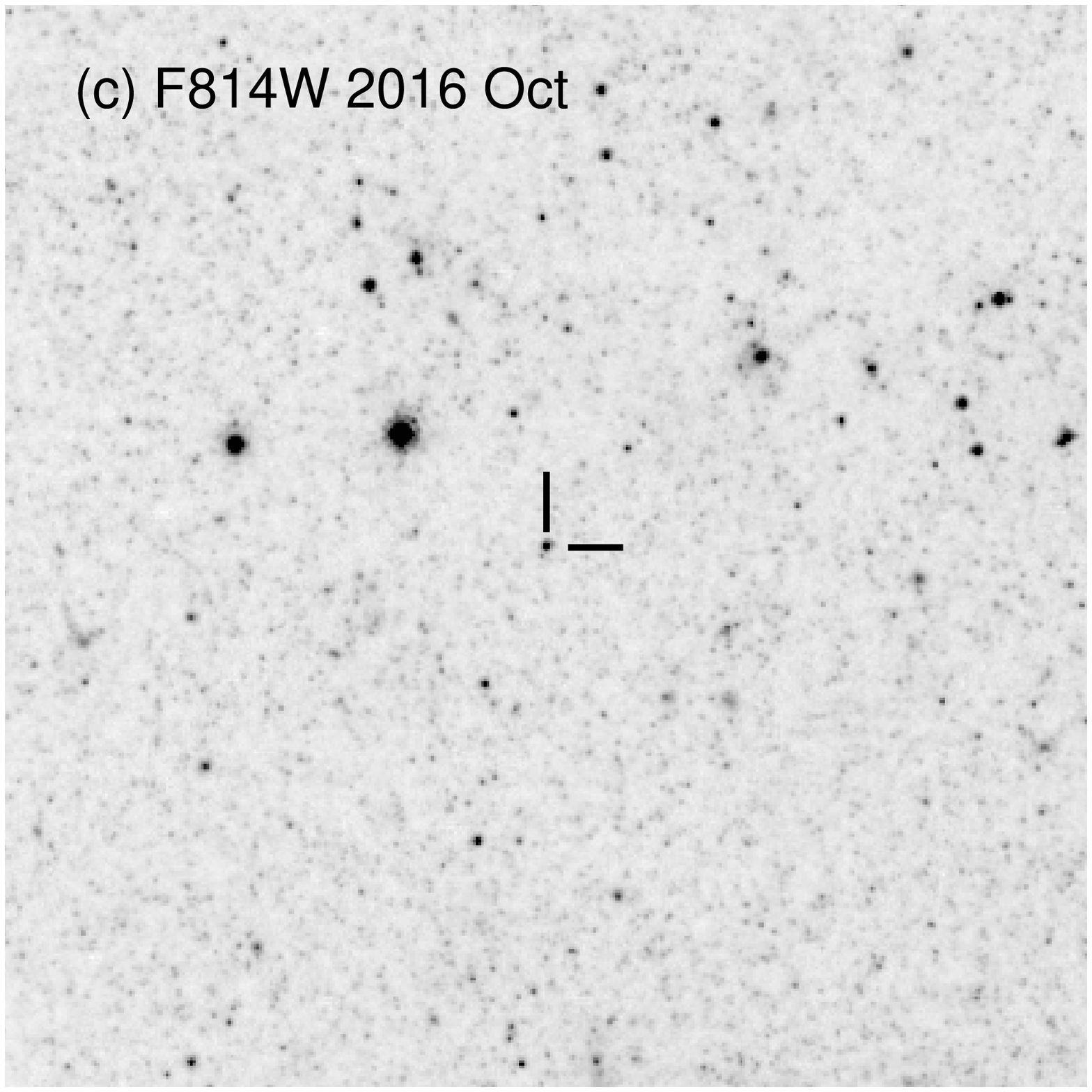}\quad
\includegraphics[width=0.37\textwidth]{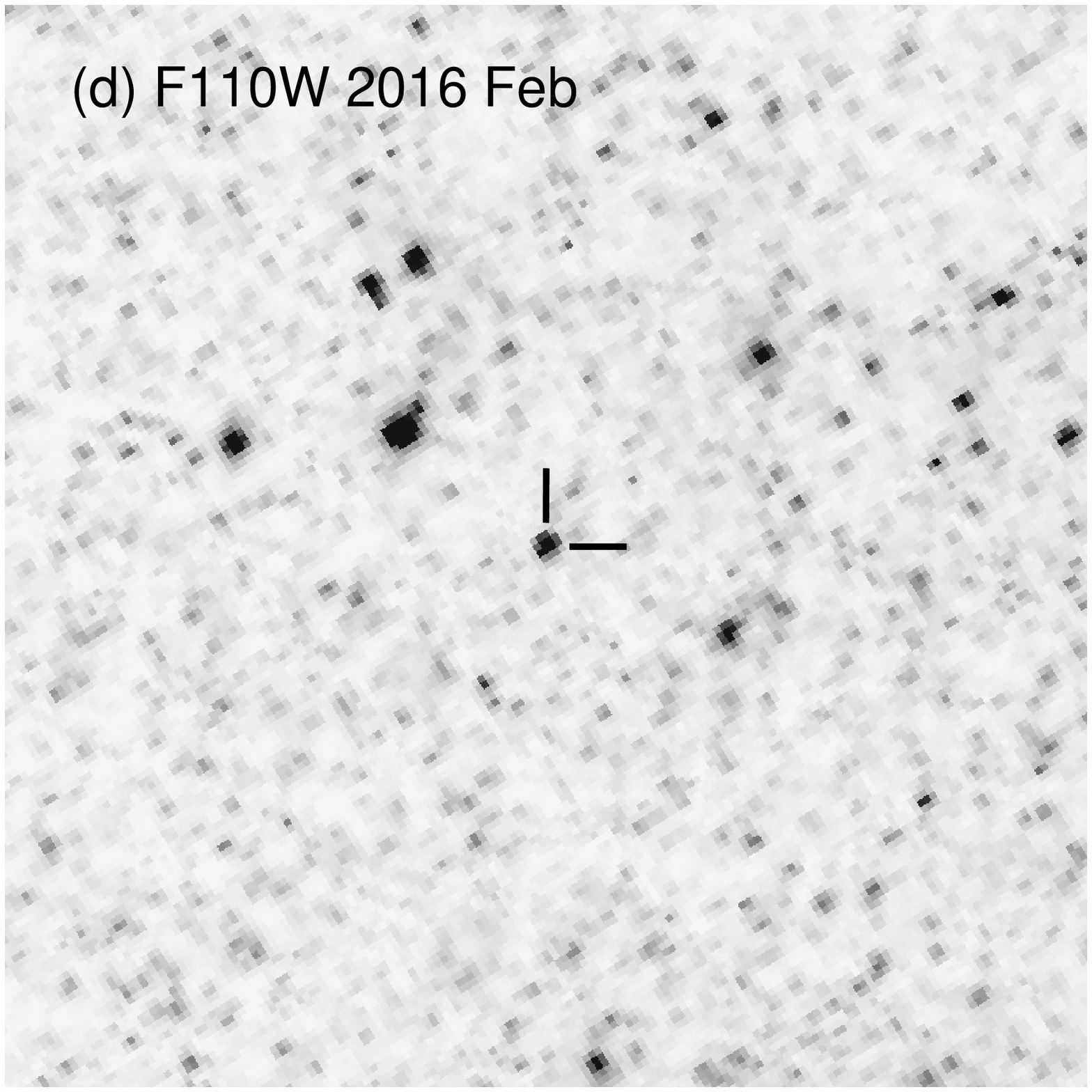}
\medskip
\includegraphics[width=0.37\textwidth]{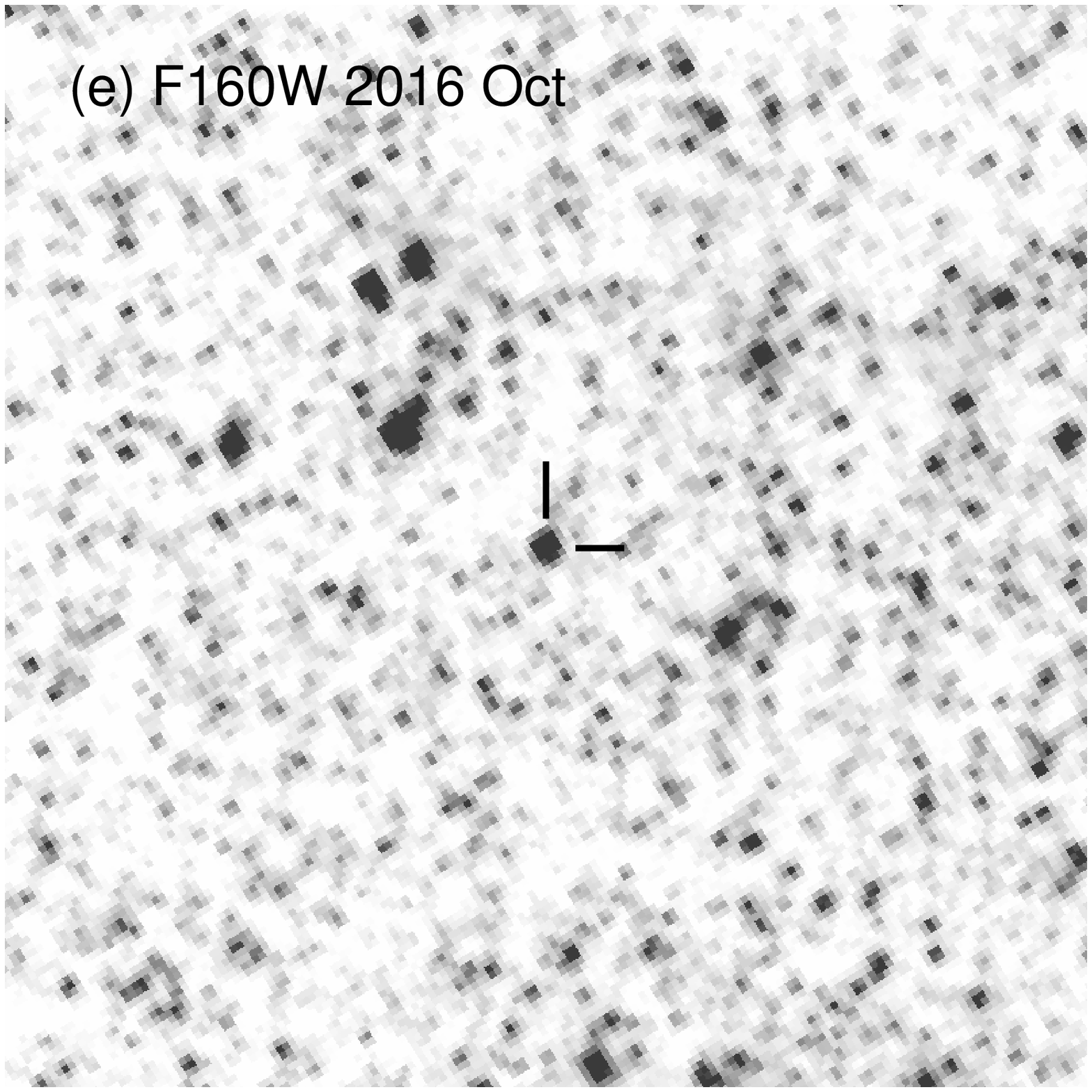}\quad
\includegraphics[width=0.37\textwidth]{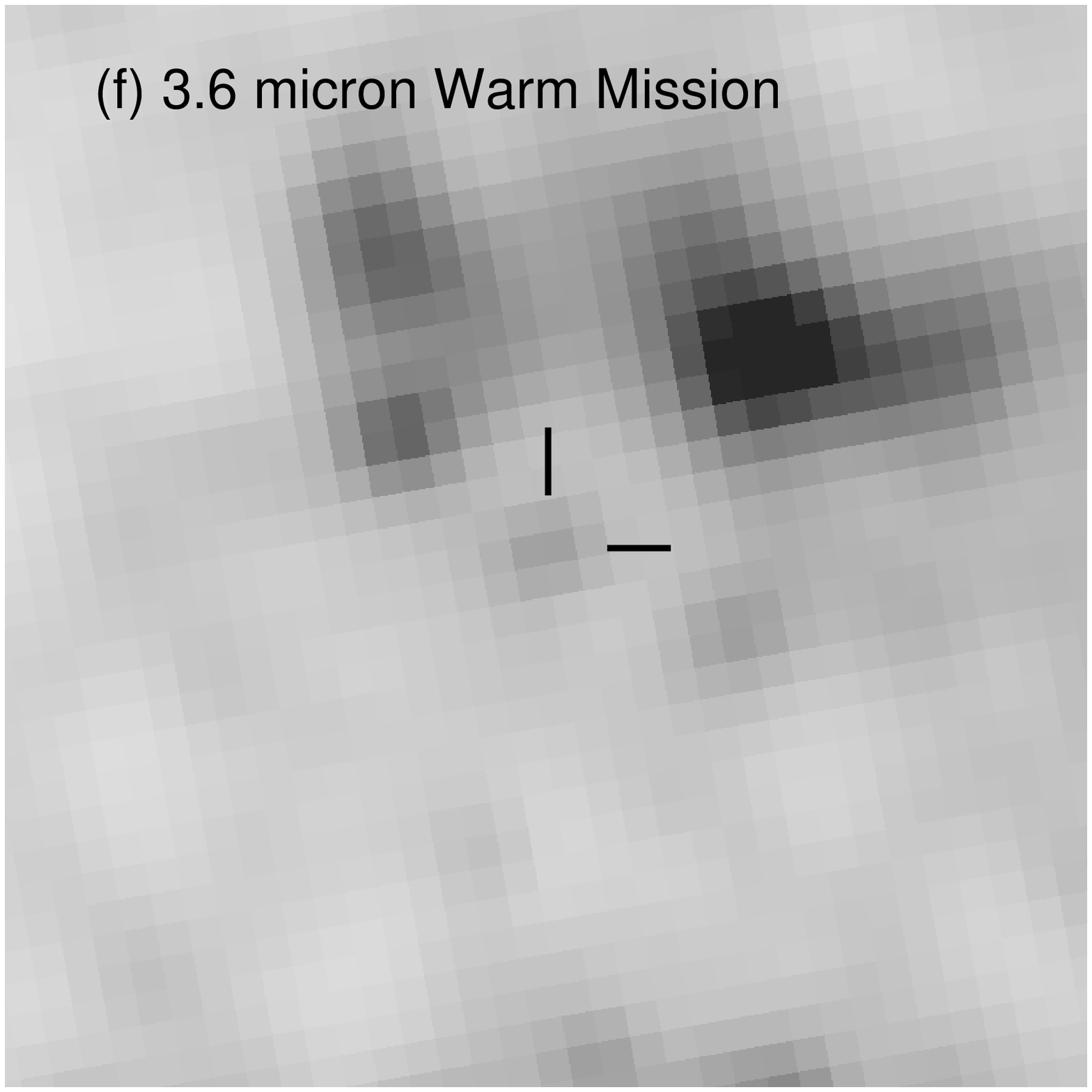}
\caption{(a) Portion of the {\sl HST\/} WFC3/UVIS image mosaic of SN 2017eaw in F814W the SN 2017eaw 
obtained on 2017 May 29; the SN is the brightest object in the image. (b) Portion of the pre-explosion
ACS/WFC image mosaic in F606W from 2016 October 26, with the likely candidate for the SN progenitor 
indicated with tick marks. 
(c) Same as for (b), but at F814W. (d) Same as for (b), but for WFC3/IR F110W from 2016 February 9.
(e) Same as for (b), but for WFC3/IR F160W from 2016 October 24.
(f) Same as for (b), but in the {\sl Spitzer\/} IRAC 3.6 $\mu$m band during the ``Warm'' (post-cryogenic)
Mission (data from 2009 August 6 through 2017 March 31).
All panels are shown to the same scale and orientation. North is up and east is to the left.\label{figprog}}
\end{figure*}

We \citep{VanDyk+2017} were the first initially to identify a stellar object in the optical/near-infrared 
with characteristics indicative of an RSG and not far from the nominal position of the SN in a 
ground-based image, when matched to the pre-SN {\sl HST\/} data.
However, we noted then that high-spatial-resolution imaging of the SN was required to confirm this candidate.
We astrometrically registered our 2017 WFC3 ToO image mosaic at F814W to the 2004 ACS/WFC F814W exposure 
(the 2004 image was publicly available in 2017 May, whereas the 2016 F814W data were not). Using 
25 stars in common between the two datasets, we registered the images to an rms uncertainty of 
0.37 ACS/WFC pixel (18.5 mas). The difference between the transformed centroid of the SN and the centroid
of the candidate object is 0.31 pixel, within the 1$\sigma$ astrometric uncertainty.
We therefore consider it highly likely that this object is the progenitor of SN 2017eaw.
We show the ACS and WFC3 images at the same scale and orientation in Figure~\ref{figprog}. The progenitor
candidate is indicated in the figure. 
We note that \citet{Johnson+2018}, \citet{Kilpatrick+2018}, and \citet{Rui+2019} also identified this object as 
the SN progenitor in their studies.

The pre-SN {\sl HST\/} images obtained from the {\sl HST\/} archive had been pre-processed with the
standard pipeline at STScI.
We measured photometry from all of the pre-SN {\sl HST\/} images with \texttt{Dolphot}.
First, we processed the individual CTE-corrected frames 
through \texttt{AstroDrizzle} to flag cosmic-ray hits.
We then set the Dolphot parameters FitSky=3, RAper=8, and InterpPSFlib=1 and used the default 
\texttt{TinyTim} model PSFs.
We list the resulting (Vegamag) photometry in Table~\ref{tabprog}.
The progenitor candidate was detected in all bands, except F658N, for which we provide a 3$\sigma$
upper limit estimate of the star's brightness.
Our measurements in these bands roughly agree with those presented by \citet{Kilpatrick+2018}
and \citet{Rui+2019}, although both of these studies included a measurement in the F164N band from {\sl HST\/} 
program GO-14786, whereas we considered that this would contribute little additional insight into the progenitor SED, since 
the bandpass of the F164N filter is encompassed by that of F160W, and therefore did not include it.
As both \citet{Kilpatrick+2018} and \citet{Rui+2019} noted, the brightness of the progenitor appears to have 
dimmed somewhat at F814W between 2004 and 2016.

Our progenitor detections in the {\sl HST\/} data are consistent with the conservative upper limit 
($R>22.4$ mag) that \citet{Steele+2017} placed on detection.

\subsection{Analysis of the {\sl Spitzer\/} Data}\label{spitzeranal}

As \cite{Khan2017} pointed out, the progenitor candidate is clearly detected at 3.6 and 4.5 $\mu$m;
see Figure~\ref{figprog}.
We know that RSGs, particularly of high luminosity, experience variability in the optical 
\citep[e.g.,][]{Soraisam+2018}, although
\citet{Johnson+2018} appeared to have ruled out significant variability at the $\gtrsim 5$--10\% level
for the SN 2017eaw progenitor within the decade prior to explosion.
\citet{Tinyanont+2019} have also ruled out any variability at $K_s$ greater than 6\% lasting longer
than 200 days from 1 yr to 1 day before explosion.
Given the large amount of available {\sl Spitzer\/} data, we were able to explore possible variability of the 
progenitor at 3.6 and 4.5 $\mu$m, during both the cryogenic and ``Warm'' (post-cryogenic) missions.
We note that \citet{Kilpatrick+2018} performed a similar analysis.
For each of the observation dates listed in Table~\ref{tabspitzer} 
we analyzed the {\sl Spitzer\/} individual artifact-corrected basic calibrated data (cBCDs) with 
\texttt{MOPEX} and \texttt{APEX} \citep{Makovoz+2005a,Makovoz+2005b,Makovoz+2006}.
We performed point-response-function (PRF) fitting photometry with the \texttt{APEX} User List 
Multiframe module, forcing the model PRF to find and fit the progenitor at its absolute position. 
In addition to the progenitor, we also forced the PRF fitting on two objects of similar brightness, both
southwest of the progenitor, one at $3{\farcs}5$ (it can be seen in Figure~\ref{figprog}(f)) and the other 
at $13{\farcs}2$.

To successfully avoid oversubtracting the PRF from the progenitor and the two comparison objects
against the complex galactic background, within the Extract Med Filter module of \texttt{APEX} we adjusted 
(decreased) the values of Window X, Window Y, and Outliers per Window from the default values and
visually inspected the residual image created with the task \texttt{APEX\_QA}.
The photometry was executed in exactly the same way for all three objects.
For each epoch we created with \texttt{MOPEX} an array-location-dependent photometric correction mosaic 
for each of the two bands, the correction factors from which we applied to the photometry.
(This correction is very close to unity, in general.)
The photometry was further aperture- and 
pixel-phase-corrected\footnote{http://irsa.ipac.caltech.edu/data/SPITZER/docs/irac/\\ iracinstrumenthandbook/79/\#{\textunderscore}Toc410728413, Table C.1}; the progenitor photometry was further color-corrected\footnote{http://irsa.ipac.caltech.edu/data/SPITZER/docs/irac/\\ iracinstrumenthandbook/18/\#{\textunderscore}Toc410728306, Table 4.4} 
assuming a 3500 K blackbody appropriate for an RSG (this correction was also very close to unity).

\begin{figure}
\plotone{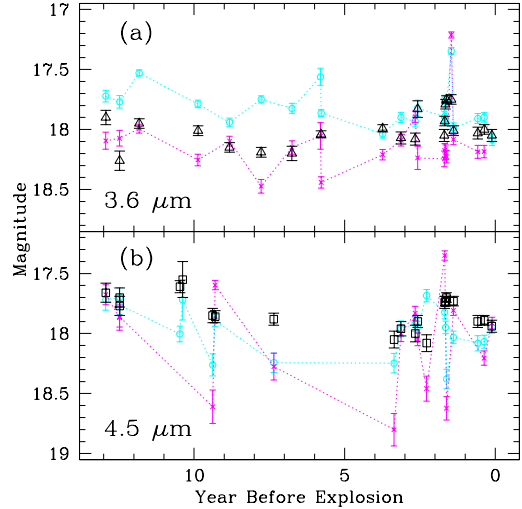} 
\caption{Apparent brightness of the SN 2017eaw progenitor in the {\sl Spitzer\/}
(a) 3.6 $\mu$m (open triangles) and (b) 4.5 $\mu$m (open squares) 
bands over $\sim12.9$ yr prior to explosion. We also show measurements for two objects in the {\sl Spitzer\/}
data with comparable brightness in the vicinity of the progenitor, one at $3{\farcs}5$ (open cyan circles) and the
other $13{\farcs}2$ (magenta crosses), both to the southwest of the progenitor.\label{figspitzer}}
\end{figure}

We show the multi-epoch photometry at 3.6 and 4.5 $\mu$m for all three objects in Figure~\ref{figspitzer}.
One can readily see that the scatter in the datapoints for all three objects far exceeds the formal \texttt{APEX}
uncertainties, and the scatter may, in fact, be somewhat correlated for all three, indicating that the photometry of objects
at this brightness level in the {\sl Spitzer\/} data may be dependent both on the image quality and on the photometric
extraction technique. Although the objects were formally detected at an appreciable S/N in the
majority of epochs in both bands, the objects, including the progenitor, are still quite faint relative to the background and are 
in close proximity to brighter objects; assessing the degree of oversubtraction or undersubtraction of the PRF model was 
therefore quite subjective, and the formal photometric uncertainties almost certainly underrepresent the 
actual uncertainties.

From this analysis we can rule out any detectable variability in the progenitor at these wavelengths at the 
$\sim 0.5$--0.6 mag level over the nearly 13 yr prior to explosion. One could potentially better investigate the existence
of any variability below this level in the two {\sl Spitzer\/} bands through the template-subtraction technique. One would have 
to wait, though, ostensibly until the SN has faded sufficiently at 3.6 and 4.5 $\mu$m to provide for an adequate template. 
As of the end of 
2018, SN 2017eaw is still quite bright as seen by {\sl Spitzer} \citep{Tinyanont+2019}, and given the limited lifetime of the 
{\sl Spitzer\/} mission, it may not be possible to acquire the desired images in time. We may need to turn to the 
{\sl James Webb Space Telescope\/} ({\sl JWST}) to obtain the templates, degrading the resolution to match the existing 
{\sl Spitzer\/} data before subtraction.

\subsection{Modeling of the Progenitor SED}\label{progmodel}

\begin{figure*}[htp] 
\centering
\includegraphics[width=0.6\textwidth]{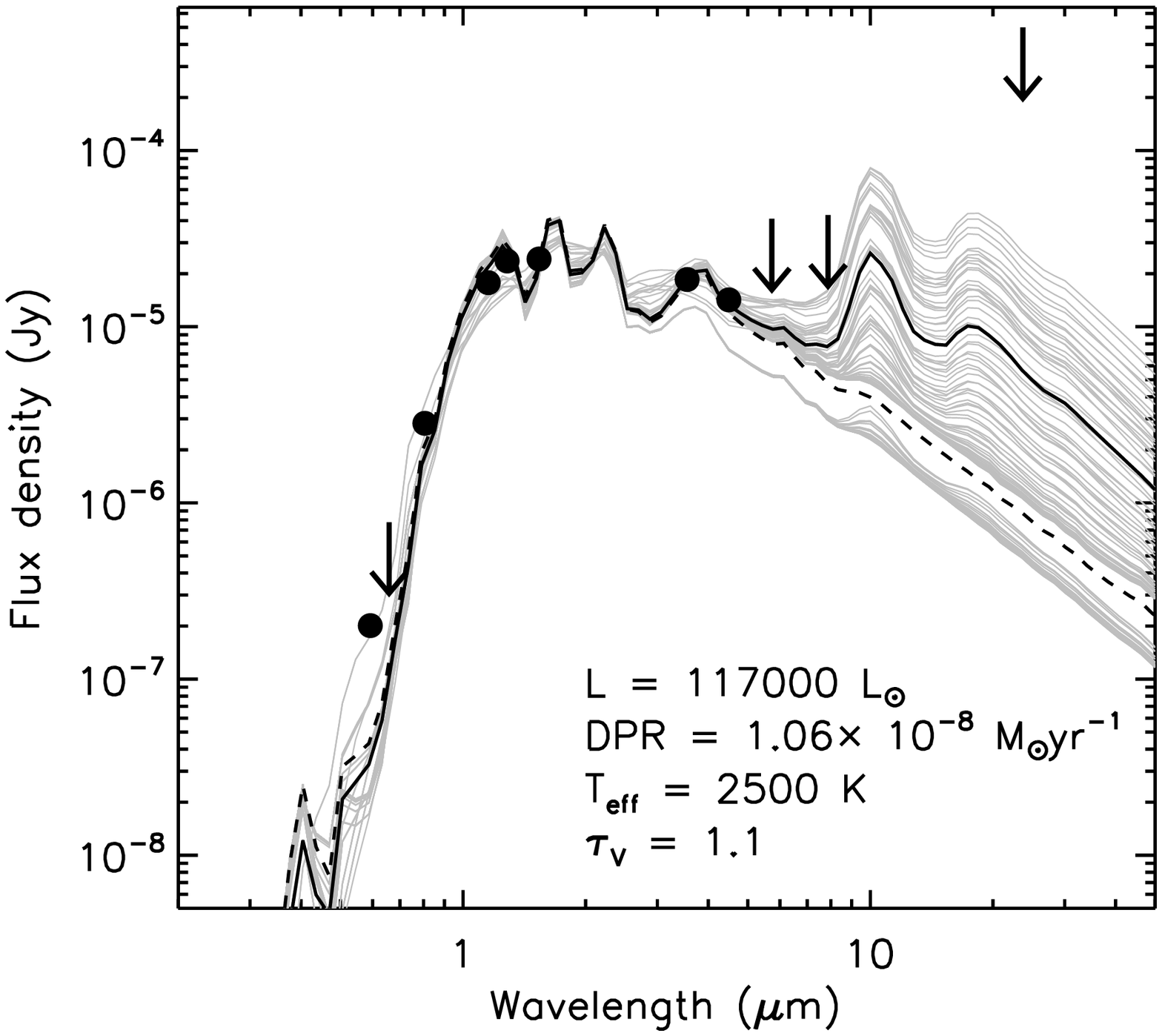}\quad
\includegraphics[width=0.6\textwidth]{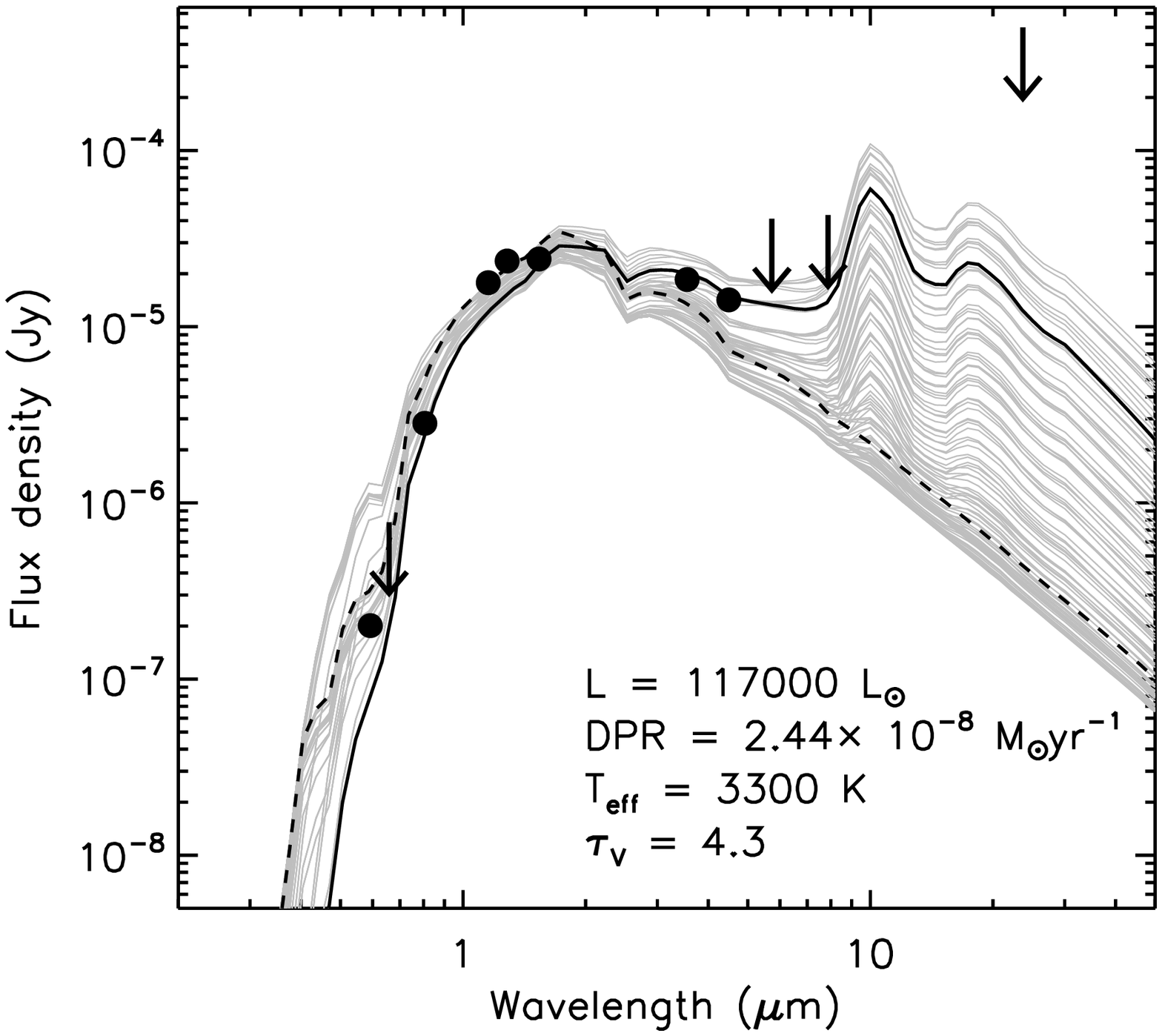}
\caption{Reddening-corrected observed SED for the SN 2017eaw 
progenitor (solid points and upper limits). The measurements are from pre-explosion
{\sl HST\/} and {\sl Spitzer\/} images (see Table~\ref{tabprog}). 
The solid curves are the GRAMS models \citep{Sargent+2011,Srinivasan+2011}, with the black curve being the best-fit 
model, and the gray curves being the family of 100 models with acceptable fit quality, when $T_{\rm eff}$ is allowed to be a 
free parameter (top panel) and when it is constrained to the range 3300--3600 K (bottom panel). The dashed curve in both 
panels is the input RSG model 
photosphere. In both panels we indicate the best-fit luminosity, DPR, $T_{\rm eff}$, and $\tau_V$.\label{figsed}}
\end{figure*}

To attempt to mitigate against the existence of {\em any\/} variability of the progenitor, 
we considered only those {\sl Spitzer\/} measurements at 3.6 and 4.5 $\mu$m --- that is, 
those from 2015 December through 2017 March --- that bracket the {\sl HST\/} progenitor brightness from 2016 February through October.
We note how fortunate we are in this case to have as much temporal and wavelength coverage, both to
be able to so exquisitely characterize the progenitor SED and also to minimize the impact of variability
(see, e.g., \citealt{Soraisam+2018}). This is usually not the case when determining the nature of 
detected SN progenitors!
Although the profile of the progenitor could also include other, fainter objects within it (and we
will not be able to assess this until after the SN has long since vanished),
we possessed far less trepidation over including the {\sl Spitzer\/} data in the progenitor SED than did
\citet{Rui+2019}, since the object's FWHM was essentially the same as that of brighter stars in the mosaics 
($\sim 2.9$ pixels) and we produced residual images during the PRF fitting, with the progenitor cleanly
subtracted away.
We computed an uncertainty-weighted mean of those {\sl Spitzer\/} measurements in each band.
The star is not detected at longer {\sl Spitzer\/} wavelengths (4.5, 8.0, and 24 $\mu$m)
as a result of the relative lack of sensitivity in those bands, and therefore we estimated upper 
limits to its detection.
We provide the final photometry of the progenitor, adopting the IRAC and MIPS 24 $\mu$m zero-points, in Table~\ref{tabprog}.

Next, we corrected the photometry presented in Table~\ref{tabprog} for Galactic foreground reddening
(the reddening law adopted for the {\sl Spitzer\/} data is from \citealt{Indebetouw+2005}, assuming the 
value of $A_K$ for the SN from \citealt{Schlafly+2011}) and then adjusted the corrected photometry
by our assumed distance modulus.
We have assumed that the interstellar reddening toward the progenitor is the same as toward the SN.
We show the resulting SED for the progenitor in Figure~\ref{figsed}.
We found that this SED is totally inconsistent with a cool supergiant photosphere at any effective 
temperature, $T_{\rm eff}$.
In particular, excess flux in the infrared clearly exists, as represented by the two shortest wavelength 
{\sl Spitzer\/} bands, relative to a bare photosphere.
We concluded that the SED had to be fit by models of RSGs which included additional circumstellar dust immediately surrounding the star.

We therefore fit the observed SED of the progenitor with O-rich models from the 
Grid of RSG and AGB Models \citep[GRAMS;][]{Sargent+2011,Srinivasan+2011}. GRAMS is a precomputed grid of radiative-transfer models 
for circumstellar dust shells around hydrostatic photosphere models with two fixed prescriptions for the 
dust properties, one each for O-rich and C-rich dust.
The radiative transfer for the GRAMS models is performed using
the \texttt{2DUST} code \citep{Ueta+2003}.

For these RSG models we input a PHOENIX model photosphere \citep{Kucinskas+2005}, around which varying amounts 
of silicate dust \citep{Ossenkopf+1992} are placed in spherically symmetric shells.
Both \citet{Kilpatrick+2018} and \citet{Rui+2019} employed MARCS model atmospheres \citep{Gustafsson+2008}
in their modeling, and \citeauthor{Kilpatrick+2018} used the circumstellar dust prescriptions from \citet{Kochanek+2012},
which assume dust according to \citet{Draine+1984} consistent with {\em interstellar\/} properties.
We computed two sets of fits: one for which $T_{\rm eff}$ was allowed to attain any value available in the 
grid, and another for which $T_{\rm eff}$ was limited to the range 3300--3600 K. The results for the unconstrained and 
constrained cases 
are shown in Figure~\ref{figsed} in the top and bottom panels, respectively. The fit procedure was as follows. We computed a 
$\chi^2$ per datapoint ($\chi^2$ divided by the number of bands with valid flux measurements) for every 
O-rich model in the grid, using the model with the lowest value to obtain the best-fit bolometric luminosity, $L_{\rm bol}$, and 
dust production rate (DPR) or, equivalently, the optical depth, $\tau_V$. Data with upper limits were also included in the fit, following the method
described by \citet{Sawicki2012}. We set the uncertainty in each parameter to the 
median absolute deviation from the median of that parameter, computed using the 100 models with the 
lowest $\chi^2$ per point (gray curves in Figure~\ref{figsed}). The uncertainty in distance is incorporated into the uncertainties in 
the $L_{\rm bol}$ and DPR.

The best-fit temperature for the case where $T_{\rm eff}$ was a free parameter is 2500 K. For this case, the best-fit luminosity and DPR are
$(1.2\pm 0.2) \times 10^5\ L_\odot$ and $(1.1 \pm 0.1) \times 10^{-8}\ M_\odot$ yr$^{-1}$, respectively. 
The fit tends to follow the observed datapoints reasonably well, although with a somewhat irregularly shaped model SED.
The luminosity is effectively the same in the case in which $T_{\rm eff}$ was constrained. 
In this case, the best-fit $T_{\rm eff}$ is a somewhat warmer
3300 K, and the DPR is $(2.4 \pm 0.3) \times 10^{-8}$ $M_\odot$ yr$^{-1}$. 
With the constrained, somewhat higher $T_{\rm eff}$, the overall fit is not quite as good as the
unconstrained fit. In both cases the fits are significantly driven by the two {\sl Spitzer\/} datapoints
and far less so by the {\sl HST\/} optical data.
The luminosity of the progenitor RSG is quite high based on this modeling; however, even simply ``eyeballing'' the 
absolute brightness of the star at $K$ from the dereddened observed SED and applying the bolometric correction for that single 
band from \citet{Levesque+2005}, one would already arrive at a luminosity of $\sim 10^5\ L_\odot$, indicating that a high
luminosity is certainly plausible.

For both best fits the flux at bluer wavelengths is redistributed into the mid-infrared by the presence of the
circumstellar dust. The upper limits at the longer {\sl Spitzer\/} wavelength, particularly at 24 $\mu$m,
provide comparatively poorer constraints on this mid-infrared emission and the overall model fit.
It is interesting to note that in both cases, constrained and unconstrained $T_{\rm eff}$, the best fits
tend toward the lowest possible temperatures. Through their modeling, \citet{Kilpatrick+2018} also found
a fit at $T_{\rm eff} \approx 3350$ K (\citealt{Rui+2019} arrived at a somewhat warmer $\sim 3550$ K).
These temperatures are at the low end of the RSG $T_{\rm eff}$ scale of 
\citet{Levesque+2005} and much cooler than the scale of \citet{Davies+2013}.
The best-fit stellar radii in the modeling are 1828 and $1049\,R_{\odot}$ for the unconstrained and
constrained $T_{\rm eff}$, respectively. This radius estimate for the 
constrained case is essentially the same as the effective radius we compute from the best-fit luminosity and 
temperature (see Section~\ref{sec:initial_mass}).
For both cases the corresponding inner radius of the dust shell is at $15\,R_{\star}$ and the outer radius is $10^4$
times the inner radius.

It should be mentioned that in the current version of the GRAMS model grid the luminosity resolution is 
limited at the highest luminosities. Additionally, 
the fractional uncertainty in the luminosity, according to the suite of model fits, is about 17\%.
The $V$ optical depth is $\tau_V = 1.1$ ($A_V=1.2$ mag) for the unconstrained fit and 
a significantly higher $\tau_V=4.3$ ($A_V=4.7$ mag) for the constrained fit.
It is not surprising that the warmer models would require more circumstellar dust to achieve a good fit.
We show in each panel of Figure~\ref{figsed} the best-fit $L_{\rm bol}$, $T_{\rm eff}$, DPR, and 
$\tau_V$ corresponding to this DPR.

We note that the PHOENIX photospheric models 
used here are at solar metallicity, whereas the 
SN 2017eaw site is likely at subsolar metallicity (see Section~\ref{metallicity}). However, we do not 
consider this to be an issue since, while metallicity will affect the UV/optical/near-infrared
photospheric spectrum and, for a given luminosity, the stellar parameters, such as mass and surface gravity
($\log g$), for a dusty source we should not be able to distinguish between solar and subsolar metallicity 
models based on photometry alone. 
Furthermore, the DPR is directly proportional to the assumed expansion speed in the shell. GRAMS models are 
computed for $v_{\rm exp} = 10$ km s$^{-1}$. However, expansion velocities measured for Galactic RSGs are 
larger than this value, and can be as high as $\sim45$ km s$^{-1}$ \citep{deBeck+2010}. 
The expansion velocity and therefore the DPR may depend on metallicity, but this dependence is not well 
calibrated and nonetheless appears to be quite weak \citep[e.g.,][]{vanLoon2006}.

Note that this modeling of the progenitor SED does not depend on invoking application of bolometric 
corrections to the observed broadband measurements, which for RSGs can be plagued by uncertainties as a 
result of both intrinsic photometric variability and changes in spectral type at the latest evolutionary 
phases (\citealt{Soraisam+2018,Davies+2018}).

\subsection{Initial Mass of the Progenitor}\label{sec:initial_mass}

\begin{figure}
\plotone{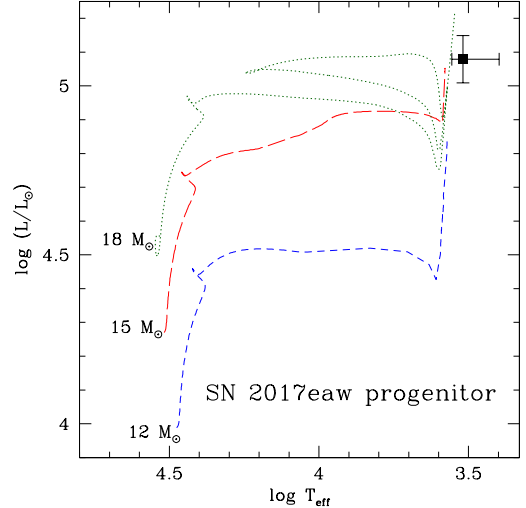} 
\caption{Hertzsprung-Russell diagram showing the locus of the SN 2017eaw progenitor (solid point).
Shown for comparison are single-star evolutionary tracks from the Geneva group at subsolar
metallicity $Z=0.006$ (with rotation at $\Omega/\Omega_{\rm crit}=0.3$; \citealt{Georgy+2013}) 
for 12 $M_{\odot}$ (blue short-dashed line) and 15 $M_{\odot}$ (red long-dashed line),
and from PARSEC \citep{Bressan+2012,Chen+2015} at $Z=0.010$ for 18 $M_{\odot}$ (dark green
dotted line).\label{fighrd}}
\end{figure}

Finally, given $T_{\rm eff}$ and $L_{\rm bol}$ from our modeling in the previous section, we can now place
the SN 2017eaw progenitor in a Hertzsprung-Russell (H-R) diagram and make comparisons with theoretical 
stellar evolutionary tracks for massive stars, in order
to estimate the initial mass, $M_{\rm ini}$, of the star. We indicate the SN progenitor in the H-R diagram in 
Figure~\ref{fighrd}. The uncertainties in both $T_{\rm eff}$ and $L_{\rm bol}$ shown in the figure
arise from the theoretical modeling of the star's SED in Section~\ref{progmodel}.
For comparison we show theoretical single massive-star evolutionary tracks from the Geneva group 
\citep{Georgy+2013} at subsolar metallicity $Z=0.006$ for 12 and 15 $M_{\odot}$, with rotation at 
$\Omega/\Omega_{\rm crit}=0.3$. 
We also show a PARSEC track \citep{Bressan+2012,Chen+2015} at $Z=0.010$ for 18 $M_{\odot}$ (subsolar Geneva tracks for 
$M> 15\ M_{\odot}$ have not been published).

None of these tracks provides a particularly satisfactory comparison with the locus for the progenitor.
The tracks all appear to terminate (at the initiation of carbon burning for PARSEC, at the end of C burning 
for Geneva) at warmer $T_{\rm eff}$ than what we infer for the star itself.
Conversely, the results from our modeling of the progenitor SED in Section~\ref{progmodel} might also be too 
cool, although, as mentioned before, to within the uncertainties of the modeling the trend of the best fits 
is definitely toward cooler $T_{\rm eff}$.
Nevertheless, as one can see, the $T_{\rm eff}$ and $L_{\rm bol}$ for the progenitor are most consistent, to 
within the 1$\sigma$ uncertainties, with the endpoint of the RSG phase for the 15 $M_{\odot}$ track.
At the 1$\sigma$ level the tracks with $M_{\rm ini}=12\ M_{\odot}$ and with $M_{\rm ini}=18\ M_{\odot}$ appear to be ruled 
out. We therefore conclude that the SN 2017eaw progenitor likely arose from a star with $M_{\rm ini} \approx 15\ M_{\odot}$.
Note that this is larger than the $13\ M_{\odot}$ that \citet{VanDyk+2017} estimated, based on a cursory
comparison with an RSG photospheric SED, without the inclusion of circumstellar dust, and also assuming a 
shorter distance to the host galaxy.
Assuming $T_{\rm eff}=3300$ K, with $L_{\rm bol}$ we further estimate that the effective radius of the RSG 
progenitor within approximately a year of explosion was $R_{\rm eff}=1046 \pm 90\ R_{\odot}$.
This radius, for instance, is larger than what \citet{Dessart+2013} derived for the radius of
the Type II-P SN 1999em.

\section{Discussion and Conclusions}

We have presented extensive optical photometric and spectroscopic monitoring of SN 2017eaw
from about 4 to 482 days after explosion. 
We also independently confirmed the TRGB distance estimate to the host galaxy, NGC 6946, 
$7.73{\pm}0.78$ Mpc (see also \citealt{Murphy+2018} and \citealt{Anand+2018}).
This distance is corroborated by both our SCM and EPM distance estimates based on SN 2017eaw --- $7.32 \pm 0.60$ 
and $7.27 \pm 0.42$ Mpc, respectively.
\citet{Eldridge+2019} have also endorsed this larger distance to NGC 6946.
The extinction to SN 2017eaw appears to be primarily the result of appreciable 
Galactic foreground dust, consistent with expectations for the host galaxy being at low Galactic latitude.
SN 2017eaw is a normal  SN II-P, possibly intermediate in both photometric and spectral properties 
between other SNe II-P, such as SN 1999em, SN 1999gi, and SN 2012aw, and SN 2004et, which also occurred in NGC 6946.
We estimated that a nickel mass $M_{\rm Ni} \approx 0.07\ M_{\odot}$ was synthesized in the explosion.
Also, the metallicity at the SN site is likely to be subsolar.

We concluded that SN 2017eaw arose from a luminous ($L_{\rm bol} \approx 10^{5.1}\ L_{\odot}$),
massive, and cool ($T_{\rm eff} \approx 2500$--3300 K) RSG. 
The progenitor was surrounded by extended CSM with substantial dust that was established by mass loss during 
previous stages of stellar evolution, especially during the RSG phase.
From detailed, realistic modeling of the observed SED for the progenitor star in 2016 (derived from 
combined {\sl HST\/} and {\sl Spitzer\/} data) and comparison of the inferred location of the progenitor
in the H-R diagram with recent, state-of-the-art theoretical massive-star evolutionary tracks, 
we found that the star is consistent with an initial mass of $15\ M_{\odot}$.

Such a high mass for the progenitor is also supported by comparison of the late-time spectra of SN 2017eaw
with existing models produced to analyze SN 2004et \citep{Jerkstrand+2012}: these nebular spectra are more
consistent with the model for an $M_{\rm ini}=15\ M_{\odot}$ progenitor than a $12\ M_{\odot}$ one, whereas
a model with even higher mass, at $19\ M_{\odot}$, also appears to be ruled out.
We note that both \citet{Kilpatrick+2018} and \citet{Rui+2019} concluded that the progenitor initial mass
was nearer 12--$13\ M_{\odot}$, although both studies had assumed a shorter distance to the host galaxy.

This case of SN 2017eaw represents an unprecedented photometric characterization, in terms of wavelength 
coverage, for the progenitor of an SN of any type, especially at this distance. Such coverage has allowed us
to model the progenitor SED with the sort of detail normally reserved for RSGs within (for example) the Local
Group. It is especially rare to have datapoints in the {\sl Spitzer\/} bands.
It is by virtue of the star's high luminosity and infrared excess that this was possible.
If we had had, as is usually the situation, only {\sl HST\/} F606W and F814W data available, we
would have attempted to fit a 3300 K model atmosphere with spherical geometry for a $15\,M_{\odot}$ star
from (for example) MARCS \citep{Gustafsson+2008} --- without any knowledge of the need for additional CSM 
extinction --- which would have overpredicted somewhat the brightness at F606W. (No atmosphere models at a 
cooler 3200 K are available from MARCS for a $15\,M_{\odot}$ star, only for a $5\,M_{\odot}$ one.)
We would have then assumed the bolometric correction for this temperature at solar metallicity 
from \citet[][ $BC_V=-3.66$ mag]{Levesque+2005} and applied that to the $V$ magnitude inferred from the
model 3300 K atmosphere --- 25.90 mag, when fixing the model SED to the dereddened observed F814W 
magnitude. This would have resulted in a lower overall luminosity $L_{\rm bol} \approx 10^{4.8}\,L_{\odot}$.
Assuming the lower-metallicity $BC_V$ from \citet[][ $-3.26$ mag]{Levesque+2006}, the luminosity would be
an even lower $L_{\rm bol} \approx 10^{4.6}\,L_{\odot}$. Even assuming the $BC$ to F814W from 
\citet[][ 0.0 mag]{Davies+2018}, we would have arrived at $L_{\rm bol} \approx 10^{4.7}\,L_{\odot}$.
All of these would have greatly underestimated the actual likely $L_{\rm bol}$ of the progenitor, and
subsequently its initial mass: based on the Geneva models alone, $M_{\rm ini}$ would only be 
$\sim 10$--$11\,M_{\odot}$.
This should provide a cautionary tale with respect to deriving the initial masses of SN II-P RSG 
progenitors, especially of nominally higher luminosity, based on the typically limited datasets available
so far for most cases. It would, of course, be inordinately valuable to obtain better wavelength coverage
in the future of nearby potential SN hosts, especially in the near- to mid-infrared.

This high range of initial mass for the SN 2017eaw progenitor pushes up against, but does not entirely 
exceed, the previously established upper limit for SN II-P progenitor initial masses
\citep{Smartt+2009,Smartt2015}. This limit remains to be challenged.
We expect such massive stars to have extensive wind-driven, dusty mass loss as an RSG
\citep[e.g.,][]{Massey+2005}, although we note that the inferred mass range for the SN 2017eaw progenitor 
is less than the threshold ($\sim 18\ M_{\odot}$) at which 
pulsationally driven superwinds are predicted to be prevalent \citep{yoon:10}.
Whether late-stage, pre-explosion outbursts occur for RSGs \citep{quataert:12,shiode:14,fuller:17}
remains to be confirmed observationally.
Such a pre-explosion outburst in the case of SN 2017eaw cannot be entirely ruled out
observationally (see, e.g., \citealt{Tinyanont+2019}).

It is interesting to
contemplate whether SN 2004et, also in NGC 6946, with its higher luminosity might have arisen from a more 
massive progenitor than SN 2017eaw (see, e.g., \citealt{Faran+2014}).
That the two are fortuitously in the same host galaxy makes such a comparison that much 
more straightforward and compelling. 
The nature of the progenitor of SN 2004et has been up for debate, with much of the uncertainty arising
from detection of the star in ground-based data with limited wavelength coverage: 
\citet{Li+2005} concluded that the progenitor had $M_{\rm ini}=15^{+5}_{-2}\ M_{\odot}$, 
whereas a revisit by \citet{Crockett+2011} resulted in an estimate of $8^{+5}_{-1}\,M_{\odot}$ (the former
study assumed a distance of 5.5 Mpc, and the latter, 5.7 Mpc); 
\citet[][ also adopting 5.5 Mpc]{Jerkstrand+2012}, based on the modeling of the nebular spectra mentioned 
above, found the progenitor to be more consistent with $15\ M_{\odot}$; and
\citet{Maund2017}, based on the SN 2004et stellar environment, estimated that 
$M_{\rm ini}=17 \pm 2\ M_{\odot}$ (with distance assumed at 4.9 Mpc).
\citet{Eldridge+2019} also argued for a larger initial mass for the SN 2004et progenitor.
Based on the comparison with SN 2017eaw it now seems more credible that the higher mass estimate applies 
for SN 2004et.

It is notable that the SN 2017eaw progenitor mass estimates by \citet{Kilpatrick+2018}, \citet{Rui+2019}, and us are 
all substantially larger than that inferred by \citet[][ $8.8^{+2.0}_{-0.2}\ M_{\odot}$]{Williams+2018} from the 
properties of the stellar environment around the SN. Such low progenitor masses have been generally found
for subluminous SNe II-P, and no indication exists that SN 2017eaw is subluminous; in fact, just the 
opposite seems more plausible.
The low mass estimate based on the stellar environment is consistent with the progenitor being relatively
spatially isolated (see Figure~\ref{figprog}). The origin of this isolation for such a massive star is 
curious.

It is, of course, necessary to revisit SN 2017eaw when it has significantly faded, particularly at F814W
and the WFC3/IR bands with {\sl HST\/} or with {\sl JWST}, 
to determine whether the
candidate progenitor has vanished. One could also reimage the site with {\sl Spitzer}, although if ejecta 
dust is forming, as \citealt{Rho+2018} have found, the SN may not be fading as rapidly; also, opportunities
to observe with {\sl Spitzer\/} are drawing to a close.
We cannot rule out that the light curves may flatten at late times, as a result of SN ejecta-CSM 
interaction or a light echo off pre-existing circumstellar dust, as in the case of SN 2004et
\citep{Kotak+2009,Fabbri+2011}.
Thus, one of these possibilities may, in fact, be in effect, in which case our wait time may be extended for several years.
Nonetheless, we already consider it very unlikely that a less luminous RSG, or even a 
bluer star, may have been hidden in the glare of the candidate progenitor and be responsible for the SN. 
None of the observed properties
of SN 2017eaw is consistent with an origin as a low-luminosity RSG, or as a blue or yellow star.
We therefore fully expect that the candidate progenitor will no longer be at the SN site.
(However, we certainly believe it would be worthwhile to place constraints on the existence of a less
luminous companion.)
We remain confident that the star we have identified will turn out to have been the SN 2017eaw
progenitor.
Finally, given the inferred lack of interstellar dust internal to the host galaxy at the SN site, we 
consider it unlikely that SN 2017eaw will have resulted in a detectable extended, interstellar light echo 
\citep[as is the case, e.g., for SN 2012aw;][]{VanDyk+2015}.

\acknowledgements

We are grateful to Patrick Wiggins for providing his discovery and pre-discovery images and to Andrew Drake 
for providing the CRTS pre-discovery images.
Daniel Huber kindly donated some observing time for us to obtain the Keck HIRES spectrum.
We appreciate useful discussions with Shoko Sakai,
Insung Jang, and Myung Gyoon Lee regarding TRGB distance estimates, Matt Nicholl regarding use of \texttt{superbol}, 
Jim Fuller about pre-SN outbursts, and Luc Dessart concerning EPM distance estimates.
We also appreciate Xiaofeng Wang giving us permission to include the spectrum their team posted to TNS.
Andrew G. Halle, Costas Q. Soler, Nick Choksi, Kevin Tang, Jeffrey D. Molloy, 
Goni Halevy, and Ben Stahl helped with some of the Lick observations.
We thank the Lick and Keck Observatory staffs for their expert assistance. 

This work is based in part on observations made with the NASA/ESA
{\it Hubble Space Telescope}, obtained from the Data Archive at the Space Telescope Science Institute
(STScI), which is operated by the Association of Universities for Research in Astronomy (AURA), Inc.,
under National Aeronautics
and Space Administration (NASA) contract NAS5-26555.
Support for programs GO-14645 and GO-15166 was provided by NASA through grants from STScI. 
This work is based in part on archival data obtained with the {\it Spitzer Space Telescope}, which is operated by the Jet Propulsion Laboratory, 
California Institute of Technology, under a contract with NASA. Support for this work was provided by an award issued by JPL/Caltech.
This research has made use of NED, which is operated by the Jet Propulsion Laboratory, California Institute of Technology, under contract with NASA.

The research of J.R.M. is supported through a Royal Society University Research Fellowship.
S.S. acknowledges grant MOST104-2628-M-001-004-MY3 from the Taiwan Ministry of Science and Technology.
Partial support for N.S.'s supernova and transient research group at the University of Arizona was provided by NSF grant AST-1515559.
M.R.K. acknowledges support from the NSF Graduate Research Fellowship, grant No. DGE 1339067.
Support for A.V.F.'s supernova research group at U.C. Berkeley has            
been provided by NSF grant AST-1211916, the TABASGO Foundation,          
the Christopher R. Redlich Fund, Gary and Cynthia Bengier, 
and the Miller Institute for Basic Research in Science (U.C. Berkeley),
and NASA/HST grant AR-14295 from STScI. 
A.L.P. acknowledges financial support for this research from a Scialog award made by the Research Corporation for Science Advancement.
PyRAF is a product of the Space Telescope Science Institute, which is operated by AURA for NASA.

KAIT and its ongoing operation were made possible by donations from Sun 
Microsystems, Inc., the Hewlett-Packard Company, AutoScope Corporation, 
Lick Observatory, the NSF, the University of California, the Sylvia \& Jim 
Katzman Foundation, and the TABASGO Foundation.  A major upgrade of the 
Kast spectrograph on the Shane 3~m telescope at Lick Observatory was made 
possible through generous gifts from the Heising-Simons Foundation, as well 
as William and Marina Kast. Research at Lick Observatory is partially 
supported by a generous gift
from Google. We also greatly appreciate contributions from numerous
individuals, including Charles Baxter and Jinee Tao,
Firmin Berta, Marc and Cristina Bensadoun, Frank and Roberta Bliss, Eliza Brown and Hal Candee,
Kathy Burck and Gilbert Montoya, Alan and Jane Chew, David and Linda Cornfield, Michael Danylchuk,
Jim and Hildy DeFrisco, William and Phyllis Draper, Luke Ellis and Laura Sawczuk, Jim Erbs and Shan Atkins,
Alan Eustace and Kathy Kwan, David Friedberg, Harvey Glasser, Charles and Gretchen Gooding,
Alan Gould and Diane Tokugawa, Thomas and Dana Grogan, Alan and Gladys Hoefer, Charles and Patricia Hunt,
Stephen and Catherine Imbler, Adam and Rita Kablanian, Roger and Jody Lawler, Kenneth and Gloria Levy,
Peter Maier, DuBose and Nancy Montgomery, Rand Morimoto and Ana Henderson, Sunil Nagaraj and Mary Katherine Stimmler,
Peter and Kristan Norvig, James and Marie O'Brient, Emilie and Doug Ogden, Paul and Sandra Otellini,
Jeanne and Sanford Robertson, Stanley and Miriam Schiffman, Thomas and Alison Schneider,
Ajay Shah and Lata Krishnan, Alex and Irina Shubat, the Silicon Valley Community Foundation, 
Mary-Lou Smulders and Nicholas Hodson, Hans Spiller, Alan and Janet Stanford, the Hugh Stuart Center Charitable Trust,
Clark and Sharon Winslow, Weldon and Ruth Wood, and many others.

The pt5m is a collaborative effort between the Universities of Durham and Sheffield. The telescope is kindly hosted by the Isaac 
Newton Group of Telescopes, La Palma. The Roque de los Muchachos Observatory is operated by the Instituto de Astrof\'isica de
Canarias. Financial contributions from the University of Sheffield Alumni Foundation are gratefully acknowledged. We also thank 
the Science and Technology Facilities Council for financial support in the form of grant ST/P000541/1. Some of the data 
presented herein were obtained at the W. M. Keck Observatory, which is operated as a scientific partnership among the California 
Institute of Technology, the University of California, and NASA. The Observatory was made possible by the generous financial 
support of the W. M. Keck Foundation. The authors wish to recognize and acknowledge the very significant cultural role and 
reverence that the summit of Maunakea has always had within the indigenous Hawaiian community.  We are most fortunate to 
have the opportunity to conduct observations from this mountain. 

\vspace{5mm}
\facilities{{\it HST\/} (ACS, WFC3), {\it Spitzer\/} (IRAC, MIPS), KAIT, Nickel, Shane (Kast Spectrograph), MMT (Blue Channel spectrograph), Keck:I (HIRES), MAST.}


\end{document}